\documentclass[%
 reprint,%
 amssymb, %
 prd,longbibliography,%
 aip,cha,%
]{revtex4-1}

\usepackage{cancel}

\usepackage{ucs}
\usepackage[utf8x]{inputenc}
\usepackage{fontenc}
\usepackage{graphicx}

\usepackage[dvipsnames]{xcolor}

\usepackage{epstopdf}
\usepackage{caption}

\usepackage{amsmath}
\usepackage{ulem}

\usepackage[titletoc]{appendix}
\usepackage{bm}%
\usepackage[colorlinks=true,linkcolor=blue]{hyperref}%
\expandafter\ifx\csname package@font\endcsname\relax\else
 \expandafter\expandafter
 \expandafter\usepackage
 \expandafter\expandafter
 \expandafter{\csname package@font\endcsname}%
\fi
\begin{document}
%%% user defined
%\newcommand{\aka}[1]{\textcolor{red}{#1}}
%\newcommand{\awo}[1]{\textcolor{blue}{#1}}
\newcommand{\chk}[1]{\textcolor{purple}{#1}}
%%%%
%\title{Author's Guide to AIP Substyles for \revtex~4.1}%
\title{Benchmark of the Local Drift-kinetic Models for Neoclassical Transport Simulation in Helical Plasmas}%

\author{B. Huang}%
\email{huang.botsz@nifs.ac.jp}
\affiliation{Sokendai (The Graduate University for Advanced Studies), Toki, Gifu, Japan}%

\author{S. Satake}%
\author{R. Kanno}
\author{H. Sugama}
\affiliation{Sokendai (The Graduate University for Advanced Studies), 509-5292, Toki, Gifu, Japan}%
\affiliation{National Institute for Fusion Science, National Institutes of National Sciences, 509-5292, Toki, Gifu, Japan}%

\author{S. Matsuoka}
\affiliation{Japan Atomic Energy Agency, 277-0871, Kashiwa, Chiba, Japan}%

\date{Oct. 2016}%

\begin{abstract}
The benchmarks of the neoclassical transport codes based on the several local drift-kinetic models are reported here.
Here, the drift-kinetic models are ZOW, ZMD, DKES-like, and global, as classified in [Matsuoka et al., Physics of Plasmas 22, 072511 (2015)].
The magnetic geometries of HSX, LHD, and W7-X are employed in the benchmarks.
It is found that the assumption of $ \boldsymbol E \times \boldsymbol B $ incompressibility causes discrepancy of neoclassical radial flux and parallel flow among the models when $ \boldsymbol E \times \boldsymbol B $ is sufficiently large compared to the magnetic drift velocities. For example, $ \mathcal{M}_p \leq 0.4$ where $\mathcal{M}_p$ is the poloidal Mach number.
On the other hand, when $\boldsymbol E \times \boldsymbol B$ and the magnetic drift velocities are comparable, the  tangential magnetic drift, which is included in both the global and ZOW models, fills the role of suppressing unphysical peaking of neoclassical radial-fluxes found in the other local models at $E_r \simeq 0$.
In low collisionality plasmas, in particular, the tangential drift effect works well to suppress such unphysical behavior of the radial transport caused in the simulations. 
It is demonstrated that the ZOW model has the advantage of mitigating the unphysical behavior in the several magnetic geometries, and that it also implements evaluation of bootstrap current in LHD with the low computation cost compared to the global model.
\end{abstract}

\maketitle

\section{Introduction}
The magnetic field geometry of a fusion device is given by the external coil system and the plasma current. 
One of the advantages of stellarator/heliotron configuration compared to axisymmetric tokamaks is that plasma current is not necessary to sustain the confinement magnetic field.
However, owing to the geometry, the helical ripple enhances the neoclassical radial particle and energy transport. 
Therefore, optimization of the field geometry is required for minimizing the neoclassical transport together with stabilizing the magnetohydrodynamics (MHD) equilibrium and improving the fast particle confinement.\cite{Grieger1992} 
The future fusion device will be operated in the higher-beta and higher-temperature condition compared to that in the present experimental devices.
In such a collisionless and high pressure gradient plasma, the neoclassical bootstrap current is supposed to increase enough to interact with the MHD equilibrium.
A self-consistent algorithm is required to investigate both the optimization of the neoclassical transport and the MHD equilibrium. 
The algorithm must satisfy both the efficiency and the accuracy in order to evaluate a quantitative study for the design of a fusion reactor.

From this viewpoint, neoclassical transport in helical plasmas has been investigated by transport codes based on several local approximations, for example, DKES\cite{Rij_1989} \cite{Hirshman1986}, GSRAKE\cite{Beidler_2001}, EUTERPE\cite{JMGarcia_PPCF55_2013_074008}, and NEO-2\cite{Kernbichler_2008} et al.
A comprehensive cross-benchmark of several local codes has been presented in Ref. \cite{Beidler_nf2011}
In the local models, the tangential grad-B and curvature drift on the flux surfaces is often assumed to be negligibly small compared to the parallel motion and $\boldsymbol E \times \boldsymbol B$ drift. 
Further, the mono-energy assumption is sometimes employed in the local neoclassical codes, in which the momentum conservation of the collision operator is broken because the Lorentz pitch-angle scattering operator is adopted.
Note that momentum correction techniques by Taguchi\cite{Taguchi_1992}, Sugama-Nishimura\cite{penta_theory_2002}\cite{penta_code_2005}\cite{spong_2005}, and Maa\ss berg\cite{Maassberg_2009} 
have been devised to recover the parallel momentum balance.
Several benchmarks have shown that the momentum correction affects the quantitative accuracy of neoclassical transport calculations in helical plasmas\cite{Maassberg_2009}\cite{Tribaldos_2011}, especially in the quasi-axisymmetric HSX plasma.\cite{Lore_2010}\cite{Briesemeister_PPCF_55_014002_2013}
 
The recent studies indicate the contribution of the magnetic tangential drift\cite{Matsuoka2015}\cite{Landreman_2014} \cite{sugama_2016} in the evaluation of radial neoclassical transport when the $\boldsymbol E\times \boldsymbol B$ drift velocity is slower than the magnetic drift.
Matsuoka\cite{Matsuoka2015} has devised a way to include the tangential magnetic drift in the local drift-kinetic equation solver.
There are also some global neoclassical codes which treat the full 3-dimensional guiding-center motion including both the radial and tangential magnetic drift term. However, only a few global neoclassical codes have been applied on helical plasmas.\cite{Satake_2008}\cite{Murakami_gnet_nf2000}\cite{Tribaldos_PPCF_2005_545}
Compared with the local codes, the global codes are stricter solutions to evaluate the drift-kinetic equation with the finite magnetic drift effect, but it takes more computational resources than the local codes.
Therefore, it is almost impossible to utilize the global codes to investigate the interaction between bootstrap current and MHD equilibrium because it requires iterations between neoclassical transport and MHD simulations.
The local approximations are appropriate for the purpose, but this has not been thoroughly verified among the neoclassical local models with global ones to guarantee the quantitative reliability of the neoclassical radial flux and parallel flow obtained from these local drift-kinetic models.

In this paper, following the previous study by Matsuoka\cite{Matsuoka2015}, the neoclassical transport is examined with four types of neoclassical transport codes in Large Helical Device(LHD), Helically Symmetric Experiment(HSX), and Wendelstein 7-X(W7-X).
The series of numerical simulations are carried out by the $\delta$-f drift-kinetic equation solver FORTEC-3D\cite{Satake_2008}. 
In the beginning, FORTEC-3D was developed as a global neoclassical transport code; 
recently, it has been extended to treat several types of the local drift-kinetic models\cite{Matsuoka2015}.
The following approximations are employed to evaluate the neoclassical transport. 
(a) The global model takes the minimum assumption, which considers both the tangential and radial magnetic drift on the convective derivative term on the perturbed distribution, $\boldsymbol{v}_{m} \cdot\nabla \delta f$. 
The global model solves the drift-kinetic equation in 5-dimensional phase space.
(b) The zero orbit width model (ZOW) excludes the radial component of magnetic drift, and it becomes a local neoclassical model. The magnetic drift term is treated as
\begin{equation} \label{eq:hat_v_m}
 \hat{\boldsymbol{v}}_{m} \equiv \boldsymbol{v}_{m} - ( \boldsymbol{v}_{m} \cdot \nabla \psi ) \boldsymbol{e}_{\psi},
\end{equation}
where $\psi$ is a flux-surface label and $\boldsymbol{e}_{\psi} \equiv \partial \boldsymbol{X} / \partial \psi $.
The {\it local} indicates the neglect of radial drift in the guiding-center equation of motion. 
Therefore, the ZOW model becomes a 4-dimensional model and reduces computational resources. 
However, the ZOW model breaks Liouville's theorem in the phase space. 
The ZOW model requires a modification in the delta-f method to solve the model properly as will be explained in Sec. \ref{sec:delta_f_scheme}. 
(c) The zero magnetic drift (ZMD) model takes a further approximation. 
The ZMD model ignores not only the radial magnetic drift but also the tangential magnetic drift from a particle orbit.
Then, the magnetic drift term in the drift-kinetic equation is treated as  
$\boldsymbol{v}_{m}\cdot\nabla\delta f = 0$.
Liouville's theorem is satisfied in the ZOW. 
(d) The DKES model further employs mono-energetic assumption, i.e., 
$\dot{v}(\partial \delta f/\partial v ) =0$, and the incompressible $\boldsymbol E \times \boldsymbol B$ drift approximation.\cite{Rij_1989}\cite{Hirshman1986}
With the Lorentz pitch-angle scattering operator, the drift-kinetic equation in DKES model reduces to a 3-dimensional model.

The remainder of this paper is organized as follows.
In Sec.\ref{sec:Local_Drift_kinetic_Models},the drift kinetic equations based on global, ZOW, ZMD and DKES models are described. The conservation properties of the phase-space volume of each model is also discussed in this section.
Then, the numerical scheme of the $\delta f$ method is explained briefly in Sec.\ref{sec:delta_f_scheme}. 
The particle, parallel momentum, and energy balance equations in each drift-kinetic model are examined in Sec.\ref{section:Drift-kinetic_Equation}.
In Sec.\ref{sec:result}, the simulation results are presented. The drift-kinetic models are benchmarked by the neoclassical fluxes such as the radial particle flux, radial energy flux, and flux-surface average parallel mean flow.
The effect of $\boldsymbol E \times \boldsymbol B$, the effect of magnetic drift, and the electron neoclassical transport are analyzed.
Finally, the bootstrap current is presented. A summary is given in Sec.\ref{sec:summary}. 
In Appendix \ref{AppendixA}, the property of the source/sink term is presented. 
In Appendix \ref{AppendixB}, the derivations of the second-order viscosity tersors $\boldsymbol{\Pi}_{2}$ for the local models are presented.

\section{Local Drift-kinetic Models} \label{sec:Local_Drift_kinetic_Models}
The neoclassical transport simulations are carried out by the $\delta f$ method under the following transport ordering assumptions.
The gyro-radius $\rho$ is small compared with the typical scale length $L$, i.e.,  $\rho/L\sim \mathcal{O} (\delta)$, where $\delta$ represents a small ordering parameter.
It is assumed that the plasma time evolution is slow,   
\begin{equation*}
 \frac{\partial}{\partial t} \sim \mathcal{O} \left(\delta^2 \frac{v_{th}}{L}\right)
\end{equation*}
where $v_{th}=\sqrt{2T/m}$ is thermal velocity.
The order of magnitude of the $\boldsymbol E\times \boldsymbol B$ drift velocity is assumed as
 \begin{equation*}
  \frac{v_{E} }{v_{th}} \sim \mathcal{O} \bigg ( \frac{ \rho }{L} \bigg ) \sim \mathcal{O} ( \delta),
 \end{equation*}
where the $\boldsymbol E \times \boldsymbol B$ drift velocity is given as
\begin{equation*}
v_{E} \equiv \frac{ | \boldsymbol E \times \boldsymbol B | }{ B^2 }.
\end{equation*}
If the radial electric field satisfies the ambipolar conditions, its magnitude is assumed as 
\begin{equation}\label{eq:Mp}
\mathcal{M}_p \equiv \frac{ v_{E}}{v_{th} } \frac{ B }{ B_p } 
 \sim \frac{ E_r }{ v_{th}  B_{ax} } \frac{ q R_{ax} }{ r } 
%\sim \mathcal {O} ( \delta )
\end{equation}
where $B_p$ and  $B_{ax}$ are the poloidal magnetic field strength and the magnetic field strength on the magnetic axis, respectively.
$r$, $R_{ax}$, and $q$ denote the minor radius, the major radius of the magnetic axis, and the safety factor, respectively.
In the present work, the order of magnitude $\mathcal{M}_p \sim 1$ for ions is allowed on the local drift-kinetic 
simulations because (a) the ion thermal velocity is much slower than the electron and
(b) the order of the poloidal magnetic field magnitude is approximately
\begin{equation*}
 B_p \sim \frac{ r }{ q R_{ax} }B_{ax} \sim \mathcal{O} ( \delta B).
\end{equation*}
Even though $\mathcal{M}_p\sim 1$ is allowed, it still assumes that the slow-flow ordering is valid, $v_{E} / v_{th} \ll 1$.

The guiding-center distribution function of species $a$ is denoted as $f_{a}(\boldsymbol{Z}, t)$. 
The guiding-center variable $\boldsymbol{Z}$ is chosen as $\boldsymbol{Z} \equiv (\boldsymbol{X}, v, \xi; t )$ 
with the guiding-center position $\boldsymbol{X}$, guiding-center velocity $v$, and the cosine component of parallel velocity pitch-angle  
$\xi \equiv v_{\parallel} / v$. 
The parallel velocity $v_{\parallel}$ is defined as $v_{\parallel} \equiv \boldsymbol{v} \cdot \boldsymbol{b}$ where $\boldsymbol{b} \equiv \boldsymbol{B} / | \boldsymbol{B} |$ is a unit vector of the magnetic field.
In Boozer coordinates, the position vector $\boldsymbol{X}$ is assigned as $\boldsymbol{X} \equiv (\psi, \theta, \zeta)$,
where $\psi$, $\theta$, and $\zeta$ are toroidal magnetic flux, poloidal angle, and toroidal angles, respectively. 
The magnetic field $\boldsymbol{B}$ is given as 
\begin{align*}
\boldsymbol{B} &= \nabla \psi \times \nabla \theta + \iota (\psi) \nabla \zeta \times \nabla \psi                                                                                                                                                                                                                                                                 \\         &= I(\psi) \nabla \theta + G(\psi) \nabla \zeta + {\beta}^{*} ( \psi, \theta , \zeta )\nabla \psi.
\end{align*}
where $\iota (\psi)$ is defined as rotational transform.
The radial covariant component ${\beta}^{*} ( \psi, \theta , \zeta )$ is assumed to be negligible because it does not influence the drift equation of motion up to the standard drift ordering $ \mathcal{O} ({\rho}/ L)$.   

The guiding center drift-kinetic equation of species $ a$ is given by
\begin{equation}\label{eq:d-k-1}
  \frac{ \partial f_{a} }{\partial t}
                        + \frac{ d {Z}_{i} }{ d t } \frac {\partial f_{a} } { \partial {Z}_{i} }  
= \mathcal{C}_a + \mathcal{S}_a,
\end{equation}
where $\mathcal{C}_a$ is Coulomb collision operator and $\mathcal{S}_a$ is a source/sink term. 
The conservation law in the phase-space  or the Liouville's theorem is presented as
\begin{equation}\label{eq:j-variable}
  \frac{ \partial  \mathcal{J}   }{\partial t}
 +\frac{ \partial }{ \partial {Z}_{i} }\bigg( \mathcal{J} \frac{ d {Z}_{i} }{ d t } \bigg)
 = \mathcal{J} \mathcal{G}.
\end{equation}
Here, $\mathcal{J}$ represents the Jacobian of the phase space.
$\mathcal{G} = 0$ if the trajectory follows the guiding center Hamiltonian. 
As the recent studies showed, the local drift-kinetic models are derived from approximation of the guiding center motion but do not satisfy the Hamiltonian. 
Therefore, $\mathcal{G} = 0$ is not guaranteed in general.
For some local neoclassical models, the approximated guiding-center equations of motion $d {Z}_{i} / dt$ are chosen ingeniously to maintain $\mathcal{G} = 0$.
To consider a general case, $\mathcal{G} \neq 0$ is retained in the following derivation.
Combining Eqs.\eqref{eq:j-variable} and \eqref{eq:d-k-1}, the conservative form of drift-kinetic equation is obtained as
\begin{equation}\label{eq:d-k-1-0}
  \frac{ \partial \big(  \mathcal{J} f_{a} \big ) }{\partial t}
 +\frac {\partial  } { \partial {Z}_{i} }\bigg (  \mathcal{J} f_{a} \frac{ d {Z}_{i} }{ d t } \bigg )
 = \mathcal{J} \big [ \mathcal{C}_a + \mathcal{S}_a \big ]
   + \mathcal{J} f_{a} \mathcal{G},
\end{equation}
which is used in taking the moments of drift-kinetic equation in section \ref{section:Drift-kinetic_Equation}.

\subsection{Global Drift-kinetic Model} \label{sec:Global}%OK 

The original FORTEC-3D is a global drift-kinetic code of which guiding center motion satisfies the Hamiltonian.
FORTEC-3D treats the drift-kinetic equation for the perturbed distribution function and equation as follows:
the distribution function $f_{a} $ is decomposed into a Maxwellian $f_{a,M}$ and  perturbation $f_{a,1}$
\begin{equation} \label{eq:fff} 
 f_{a,1} (\boldsymbol{X}, v,\xi , t ) \equiv f_{a} (\boldsymbol{X}, v, \xi, t ) - f_{a,M} ( \psi, v ),
\end{equation}
where the local Maxwellian $f_{a,M}$ is defined as
\begin{align}\label{eq:f_M_local}
 f_{a,M}  &= n_{a}(\psi) \Bigg \lgroup 
           \frac{ m_{a} }{ 2 \pi T_{a}(\psi) } \Bigg \rgroup ^{3/2} 
           \cdot 
           \mathrm{exp} \Bigg \lgroup - \frac{ m_a v^{2} }{ 2 T_{a} (\psi) } \Bigg \rgroup,
\end{align}

\begin{align}\label{eq:DKE0-1}
 \bigg (  \frac{ \partial } { \partial t } & + \dot{\boldsymbol Z} \cdot \frac{ \partial  }{ \partial \boldsymbol{ Z } } \bigg ) f_{a,1}
 % -\mathcal{C} ( f_{a,1} ) 
 = \mathcal{S}_{a,0} + \mathcal{C}^L ( f_{a,1}) + \mathcal{S}_{a,1},
\end{align}
where $ \dot{\boldsymbol Z} = \frac{d}{dt} ( \boldsymbol X, v,\xi ) $ and $\mathcal{C}^{L} ( f_{a,1})$ is a linearized Fokker-Planck collision operator
\begin{align} \label{eq:collision-0}
 \mathcal{C}^L ( f_{a} ) &= \sum_{b}\mathcal{C} ( f_{a,M}, f_{b,1}) + \mathcal{C} ( f_{a,1}, f_{b,M}),
\end{align}
and the source term $\mathcal{S}_{a,0}$ is defined as
\begin{equation}\label{eq:S0_f3d}
 \mathcal{S}_{a,0} \equiv- \frac{d \boldsymbol Z }{dt} \cdot\frac{\partial}{\partial\boldsymbol Z} f_{a,M} =  - \bigg ( {\dot v} \frac{\partial}{\partial \psi} + {\dot \psi} \frac{\partial}{\partial \psi} \bigg) f_{a,M}.
\end{equation}
On the other hand, $\mathcal{S}_{a,1} $ is an additional source/sink term, which helps the numerical simulation to reach a quasi-steady state. 
The $\mathcal{S}_{a,1} $ is discussed in Sec. \ref{section:particle_flux} and \ref{sec:parall_monentum_balance}.

The guiding-center trajectory is given as follows:\cite{littlejohn1983} 
%%% Equation %%%
%\input{./text/eq_F3D-1.tex}

\begin{subequations} \label{eq:f3d-o1}
\begin{alignat}{3}
  \dot{ \boldsymbol X } =& v \xi \boldsymbol{b} + \frac{1}{ e_a B_{\parallel}^* } \boldsymbol{b} \times \bigg \{ m_a (v \xi)^2 \boldsymbol{b} \cdot \nabla \boldsymbol{b} + \mu \nabla B -  e_a\boldsymbol{E}^* \bigg \},
  \\
  \frac{d v }{dt} =& \frac{ e_a}{m_a v} \dot{ \boldsymbol X } \cdot \boldsymbol{E}^{*} 
    + \frac{ \mu }{m_a v} \frac{ \partial B }{ \partial t } ,
 \\
 \frac{d \xi }{dt}   =& - \frac{ \xi}{ v }\frac{dv}{dt} 
  - \frac{ \boldsymbol{b} }{ m_a v} \cdot ( \mu \nabla B - e_{a}\boldsymbol{E}^* ) + \xi \frac{d \boldsymbol X}{dt} \cdot
  \boldsymbol{\kappa},
\end{alignat}
\end{subequations}
where, $m_a$ and $e_a$ denote the mass and charge of the species $a$, and
\begin{subequations}
\begin{alignat}{7}
 & \mu \equiv \frac{m_a v^2}{2B} ( 1 - {\xi}^2 ),
\\
 & \boldsymbol A^{*} \equiv \boldsymbol A  + \frac {m_a v \xi}{e_a} \boldsymbol { b },
\\ \label{eq:E_ast}
 & \boldsymbol E^{*} \equiv - \frac{ \partial \boldsymbol A^{*} }{ \partial t } - \nabla \Phi,
\\
 & \boldsymbol B^{*} \equiv \nabla \times \boldsymbol A^{*},
\\ \label{eq:B_para_ast}
 &B^{*}_{\parallel} \equiv \boldsymbol { b } \cdot \boldsymbol B^{*},
 \\ \label{eq:kappa}
 & \kappa \equiv \left ( { \boldsymbol b} \cdot \nabla \right ) { \boldsymbol b}.
\end{alignat}
\end{subequations}
The trajectory is derived from Hamiltonian so that it satisfies the Liouville equation, i.e.,
\begin{equation} \label{eq:f3d-g=0}
 \mathcal{JG}=0                                                                                                                                                                                                                                                                       \end{equation}
Note that the phase-space Jacobian in Boozer coordinates is
\begin{equation}
 \mathcal{J} = \frac{ 2 \pi B^{*}_{\parallel} v^{2} }{ B } \frac{ G + \iota I }{ B^{2} }.
\end{equation}

\subsection{Zero Orbit Width(ZOW) Model} %OK
The zero orbit width (ZOW) approximation\cite{Matsuoka2015} is a local drift-kinetic model, which ignores only the radial drift $\dot \psi ~ \partial f_{1} / \partial \psi$.
The subscript of particle species is omitted here and hereafter unless it is necessary.
The drift-kinetic equation Eq.\eqref{eq:DKE0-1} becomes                                                                                                                                                                                                                                                                                                
\begin{align}\label{eq:zow-f1}
   &  
   \bigg ( \frac{ \partial } { \partial t } 
   + {\dot {\boldsymbol{Z}} }^{\text{zow}} \cdot\frac{ \partial }{ \partial \boldsymbol{Z} } \bigg ) f_{1} 
  = \mathcal{S}_{0} + \mathcal{C}^L ( f_{1}) + \mathcal{S}_{1}
\end{align}
where $\dot { \boldsymbol{Z} }^{\text{zow}} = \frac{d}{dt} ( \theta, \zeta, v, \xi ) $.
%The radial neoclassical transport is available following the local drift-kinetic equation Eq.\eqref{eq:zow-f1}.
In the present study, stationary electromagnetic field approximation is assumed
\begin{equation} \label{eq:dBdt=dphidt=0}
 \frac{ \partial B  }{ \partial t } = \frac{ \partial \Phi  }{ \partial t } = 0.
\end{equation} 
Thus, the electric field is approximated as 
\begin{equation} \label{eq:dEdt=0}
 \boldsymbol E^* \simeq - \nabla \psi \frac{d \Phi}{d \psi}
\end{equation}
where $\Phi = \Phi (\psi) $ is the electrostatic potential, which is assumed to be a flux-surface function for simplicity. 
Other approximations employed in local models are $ \boldsymbol{B}^* \cdot { \boldsymbol b } \simeq B$ and 
\begin{align}\label{eq:kappa-app}
 \boldsymbol \kappa \simeq \frac{ {\nabla}_{\perp} B }{ B }.
\end{align}
Here, the $\mathcal{O}(\delta)$ correction in $B^{*}_{\parallel}$ is neglected.
When $\beta \equiv p/(B^2/2\mu_0) $, one has
\begin{align}\label{eq:kappa-app-1}
\boldsymbol \kappa  \nonumber
 &= { \boldsymbol b} \times \left ( \frac{ \nabla B \times \boldsymbol B }{ B^2 } - \frac{ \nabla \times \boldsymbol B }{ B} \right )
 \\ \nonumber
 &= \frac{1}{B} \left( \nabla B - { \boldsymbol b} \cdot \nabla B \right)
  + \mu_{0} \frac{ \boldsymbol J \times \boldsymbol B }{ B^2 }
 \\ \nonumber
 &= \frac{ {\nabla}_{\perp} B }{ B } + \mu_{0} \frac{ \boldsymbol J \times \boldsymbol B }{ B^2 }
 \\ 
 &= \frac{ {\nabla}_{\perp} B }{ B } + \frac { \mu_{0} \nabla p } { B^2 }
 \simeq \frac{ {\nabla}_{\perp} B }{ B } + \mathcal{O} ( \beta )
\end{align}
where $p = p(\psi)$ denotes the scalar pressure. 
The second term is negligible in low-$\beta$ approximation.

The particle trajectories ${\dot {\boldsymbol{Z}} }^{\text{zow}}$ are treated as if they are crawling on a specific flux surface and given as follows: 
\begin{subequations}\label{eq:zoworbit}
\begin{alignat}{3}
 \dot{ \boldsymbol{X} } =& v \xi { \boldsymbol{b} }  + \boldsymbol{v}_{E} + \hat {\boldsymbol{v}}_{m}, 
 \\ 
 \dot{ v }      =& \frac{ -e_{a} }{ m_{a} v  } \boldsymbol{v}_{m} \cdot \nabla \psi \frac{ d \Phi }{ d \psi },
 \\ \nonumber
 \dot{ \xi }    =& - \frac{ 1 - {\xi}^{2}  }{ 2 B } \bigg ( v \boldsymbol{b} \cdot \nabla B \bigg)  \nonumber 
 \\
                & - \xi ( 1 - { \xi}^{2}) \frac{ d \Phi }{ d \psi } \frac{  \boldsymbol{B} \times \nabla B  }{ 2 B^3 } \cdot \nabla \psi .
\end{alignat}
\end{subequations}
Note that the radial magnetic drift ${\boldsymbol v}_{m} \cdot \nabla \psi$ is still kept in the time evolution of velocity $\dot v$. 
Even though the $\dot{\psi} \partial f_{1}/\partial \psi$ term is neglected in the LHS of Eq.\eqref{eq:zow-f1}, the source/sink term $S_0 \propto \dot{\psi}$ in the RHS is the same as Eq.\eqref{eq:S0_f3d} in the global model.
The $\boldsymbol{E} \times \boldsymbol{B}$ drift is defined as
\begin{equation} \label{eq:vE}
 \boldsymbol{v}_{E} \equiv \frac{ ~ d \Phi ~}{ d \psi} \frac{~ \boldsymbol{B}  \times \nabla \psi ~}{ B^2}
\end{equation}
and the magnetic drift is defined as
\begin{equation} \label{eq:vm-0}
 \boldsymbol{v}_{m} \cdot \nabla \psi \equiv \frac{ m_{a}v^2 }{2 e_{a} B^3} \bigg (  1+\xi^2 \bigg) \boldsymbol{B} \times \nabla B  \cdot \nabla \psi.
\end{equation}
The radial virtual drift velocity $\dot{\psi}$ in the local model is evaluated from 
$ \nabla \psi\cdot $ product of Eq. \eqref{eq:vm-0}, and the tangential magnetic drift $\hat{\boldsymbol v}_m$ is defined as in Eq.\eqref{eq:hat_v_m},
\begin{equation*} 
 \hat{\boldsymbol{v}}_{m} \equiv \boldsymbol{v}_{m} -\dot{\psi} \boldsymbol{e}_{\psi}.
\end{equation*}
The Jacobian of ZOW in phase space becomes
\begin{equation*}
 \mathcal{J} = \frac{ 2 \pi B^{*}_{\parallel} v^{2} }{ B } \frac{ G + \iota I }{ B^{2} } \simeq 2 \pi v^{2} \frac{ G + \iota I }{ B^{2} }.
\end{equation*}

The guiding-center equations of motion Eq.\eqref{eq:zoworbit} does not include the radial drift term $ \dot \psi$ and disobeys Hamiltonian.
As a result, $\dot{\boldsymbol{Z}}^{\text{zow}}$ is compressible on 4-dimensional phase space where
\begin{equation}\label{eq:gneq0}
  \mathcal{G} = \nabla_z \cdot  \dot{\boldsymbol Z }^{\text{zow}}  =\frac{1}{\mathcal{J}}\frac{\partial}{\partial Z_i}\cdot
  (\mathcal{J}\dot{Z}_i)\neq 0.
\end{equation}
Here, $\nabla_z$ represents the divergence in the phase space.
Following \eqref{eq:zoworbit} and \eqref{eq:gneq0}, the variation of phase-space volume along the guiding-center trajectories is
\begin{align} \label{eq:zow-J}
  \nabla_z \cdot  \dot{\boldsymbol Z }^{\text{zow}} 
                                 & = \frac{ m v^{2} ( 1 + {\xi}^{2} ) }{2eB(G+\iota I)} \Bigg \{  \frac{3}{B} \frac{ \partial B}{ \partial \psi } \Bigg ( I \frac{ \partial B }{ \partial \zeta } - G \frac{ \partial B }{ \partial \theta } \Bigg ) \nonumber 
                                 \\  &+ \Bigg ( G \frac{\partial ^2 B }{\partial\psi \partial \theta} - I \frac{\partial ^2 B }{\partial\psi \partial \zeta}  \Bigg ) \Bigg \}.
\end{align}
This term affects the balance equation of particle number, parallel momentum, and energy, which will be discussed in Sec. \ref{section:Drift-kinetic_Equation}.

\subsection{Zero Magnetic Drift (ZMD) Model} %OK
The zero magnetic drift (ZMD) model is similar to ZOW.
It follows Eq.\eqref{eq:zoworbit} but it excludes all the magnetic drift term in  $\dot{ \boldsymbol X } $. 
The particle trajectories of ZMD is given as the following:
\begin{subequations}\label{eq:zmdorbit}
\begin{alignat}{3}
 \dot{ \boldsymbol{X} } =& v \xi { \boldsymbol{b} }  + \boldsymbol{v}_{E}, 
 \\ 
 \dot{ v }      =& \frac{ -e_{a} }{ m_{a} v } \boldsymbol{v}_{m} \cdot \nabla \psi \frac{ d \Phi }{ d \psi },
 \\
 \dot{ \xi }    =& - \frac{  1 - {\xi}^{2} }{ 2 B} \bigg ( v \boldsymbol{b} \cdot \nabla B \bigg)   \nonumber
                \\ & - \xi ( 1 - { \xi}^{2}) \frac{ d \Phi}{ d \psi } \frac{  \boldsymbol{B} \times \nabla B }{ 2 B^3 } \cdot \nabla \psi.
\end{alignat}
\end{subequations}
Following the ZMD 4-dimensional guiding-center orbit, the incompressibility of the phase-space volume $\mathcal{G} = 0$ is still retained. 
%Then, the moment equations \eqref{eq:m0}, \eqref{eq:m1-parallel} and \eqref{eq:m2-kinetic} are simplified for ZMD.

\subsection{DKES-like Model} \label{sec:DKES-like}% OK
The DKES-like  model takes a further approximation on ZMD, that is, the mono-energetic assumption $\dot v = 0$. 
Then, the DKES-like model is reduced to be a 3-dimensional problem, in which $\dot{\boldsymbol{Z}}^{\text{dkes}} = d/dt(\psi, \theta, \zeta)$
on the LHS of the drift-kinetic equation.
Following the trajectory Eq.\eqref{eq:zmdorbit} and the mono-energetic particle approximation $\dot v = 0$, 
the phase space volume is not conserved:
\begin{equation} \label{eq:dles-J}
   \nabla_z \cdot \dot { \boldsymbol{Z}}^{ \text{dkes} }    = \frac{ 3 ( 1 + {\zeta}^2 ) }{ 2 B^3 \mathcal{J} } \bigg (  G \frac{ \partial B }{ \partial \theta} - I \frac{ \partial B }{ \partial \zeta } \bigg ) \frac{d \Phi}{d \psi}.
\end{equation}
In order to maintain $\mathcal{G} = 0$, the electric potential $\nabla \Phi$ is replaced by
\begin{equation}
 \nabla \Phi \simeq \nabla \Phi \frac{B^2}{ \langle B^2 \rangle }
\end{equation}
and the incompressible $\boldsymbol{E} \times \boldsymbol{B}$ drift is denoted as
\begin{equation} \label{eq:imcompressible}
 \hat{ \boldsymbol v }_{E} \equiv \frac{ \boldsymbol{E} \times \boldsymbol{B} }{ \langle B^2 \rangle }.
\end{equation}
In summary, the guiding-center trajectory in the DKES-like model is given as follows:
\begin{subequations}\label{eq:deksorbit}
\begin{alignat}{3}
 \dot{ \boldsymbol{X} } =& v \xi  { \boldsymbol{b} }  + \hat{\boldsymbol{v}}_{E}, 
 \\ 
 \dot{ v }      =& ~0,
 \\
 \dot{ \xi }    =& - \frac{ (1 - {\xi}^{2} )v }{ 2 B}  \boldsymbol{b} \cdot \nabla B,
\end{alignat}
\end{subequations}
and the particle trajectory conserves the phase space volume, $ \mathcal{G} = \nabla_z \cdot \dot { \boldsymbol{Z}}^\text{dkes}=0$.  

In the original DKES code, the collision operator is simplified by the Lorentz pitch-angle scattering operator
\begin{equation} \label{eq:pitch-angle_scattering}
 \mathcal{L} f_a = \frac{ \nu_{ab} }{2} \frac{ \partial  }{ \partial \xi } ( 1 - {\xi}^2 ) \frac{ \partial  }{ \partial \xi } f_a,
\end{equation}
where the particle does not change the velocity either by guiding-center motion or by collision. 
However, in the series of simulations in this paper, all models use the same linear collision operator to benchmark neoclassical transport. 
The linear collision operator includes the energy scattering term and field-particle part to maintain the conservation property of Fokker-Plank operator\cite{Satake_2008}. 
Between the original DKES and the DKES-like in the simulation, the effects of different collision operators appear in a quasi-symmetric geometry because of the conservation
of momentum. It is essential to evaluate neoclassical transport as discussed in Sec. \ref{sec:large_Er}.

\subsection{Two-weight $\delta f$ Scheme}\label{sec:delta_f_scheme}
The two-weight $\delta f$\cite{Satake_2008}\cite{Hu_1994} scheme is employed to solve the global and local drift-kinetic models in Section \ref{sec:Global} - \ref{sec:DKES-like}. 
%In this section, the compressibility of phase space $\mathcal G$ and the approximations of each model are taken into account in the weight function. Then, the balance equations of particle number, parallel momentum and energy are investigated for each models. 
%The requirement of adaptive source-sink term $S_1$ is explained which is essential for obtaining a steady-state solution in some models. 
Let us briefly explain the two-weight $\delta f$  scheme in the case of  $\mathcal G \neq 0$.
%\subsection{Two-weight $\delta f$ scheme}
The weight functions ${\mathfrak{w}}$ and ${\mathfrak{p}}$ are given as follows: 
\begin{subequations} \label{eq:w_p=gf}
\begin{alignat}{2}
   f_{1}  ({\boldsymbol Z} )    &=g ({\boldsymbol Z} )  {\mathfrak{w}}({\boldsymbol Z} ) \label{eq:f1=gw}
\\
   f_{M}  ({\boldsymbol Z} )    &=g ({\boldsymbol Z} )  {\mathfrak{p}}({\boldsymbol Z} ) \label{eq:fM=gp}
\end{alignat}
\end{subequations}
where $g( \boldsymbol Z)$  is the marker distribution function. 
Following Eq.\eqref{eq:DKE0-1}, an operator includes the total derivative along the particle trajectory and the test-particle collision is defined as
\begin{align}\label{eq:Df_1/Dt}
 \frac{ D f_{1} }{ Dt } &\equiv \frac{ \partial f_{1} }{ \partial t } + \dot{\boldsymbol Z} \cdot \frac{ \partial f_{1} }{ \partial \boldsymbol Z } - \mathcal{C}_T ( f_{1} ) \nonumber
 \\
 &= \mathcal{S}_{0} + \mathcal{S}_{1} +\mathcal{C}_F
\end{align}
Here, we employ the linearized collision operator decomposed into the test-particle part $ \mathcal{C}_T$ and the field-particle part $\mathcal{C}_F$\cite{Satake_2008}\cite{Z_Lin_1995}. The former is implemented by the Monte Carlo method and the latter is constructed so as to satisfy the conservation properties of particle number, parallel momentum, and energy for like-species collisions. In the ion calculation, the ion-electron collision is neglected because of the large mass-ratio $m_e/m_i\ll 1$, while the electron-ion collision is approximated by the pitch-angle scattering operator Eq. \eqref{eq:pitch-angle_scattering} with stationary background, that is, Maxwellian ions.

According to Eqs.\eqref{eq:j-variable} and \eqref{eq:d-k-1-0}, the drift-kinetic equation of marker distribution $g( \boldsymbol Z)$ is obtained  
\begin{equation} \label{eq:Dg/Dt}
 \frac{ D g }{ Dt } %\equiv  \frac{ \partial g }{ \partial t } + \frac{d \boldsymbol Z}{ dt} \cdot \frac{ \partial g }{ \partial \boldsymbol Z }  
 = - g~\mathcal{G}.
\end{equation}
 Eq.\eqref{eq:Df_1/Dt} is extended with Eq.\eqref{eq:f1=gw}
\begin{equation}\label{eq:extend_Df_1/Dt}
 \frac{ D f_{1} }{ Dt } = {\mathfrak{w}} \frac{ D g }{ Dt } + g \frac{ D {\mathfrak{w}} }{ Dt }.
\end{equation}
Following \eqref{eq:Df_1/Dt}, \eqref{eq:Dg/Dt}, and \eqref{eq:extend_Df_1/Dt}, the time evolution of the weight function ${\mathfrak{w}}$ is obtained
\begin{align}\label{eq:w0}
  %\frac{ D w }{ Dt } 
  \dot{{\mathfrak{w}}}&= \frac{1}{g} \frac{ D f_{1} }{ Dt } - \frac{{\mathfrak{w}}}{g} \frac{ D g }{ Dt }  
 \\ \nonumber
 &= \frac{ {\mathfrak{p}} }{ f_{M} } \bigg ( \mathcal{S}_{0} + \mathcal{S}_{1} +\mathcal{C}_{F} ( f_{M} )  \bigg ) + {\mathfrak{w}}  \mathcal{G}
\end{align}
Similarly, the time evolution of the weight function $\mathfrak{p}$ is obtained as follows: 
\begin{align}\label{eq:p0}
 \mathfrak{\dot p} = \frac{ \mathfrak{p} }{ f_{M} } \bigg ( \dot{\boldsymbol{Z}} \cdot \frac{ \partial  }{ \partial \boldsymbol Z } \bigg ) f_{M} + \mathfrak{p} \mathcal{G}.
 %\\ \nonumber
 %&= \frac{ \mathfrak{p} }{ f_{M} } \bigg ( \frac{ d \psi }{ dt} \frac{ \partial  }{ \partial \psi } + \frac{ d v }{ dt} \frac{ \partial  }{ \partial v} \bigg ) f_{a,M}  + \mathfrak{p} \mathcal{J}^{-1} \mathcal{G}.
\end{align}
The time evolution of the weights $\mathfrak{w}$ Eq.\eqref{eq:w0} and $\mathfrak{p}$ \eqref{eq:p0} include $\mathcal{G}$ which is non-zero in the ZOW model only. 
%Owing to the compressibility $\mathcal{G}$, the two-weight $\delta f$ scheme is extended. 
%And then, the phase-space volume is not conserved\cite{Hu_1994}. 
%The last term in Eqs. (35) and (36) are required. 
%Thus, the two-weight $\delta f$ scheme is applicable to the case, in which the phase-space volume is not conserved.
The last term in Eqs. (35) and (36) is required so that the two-weight $\delta f$ scheme is applicable to the case in which the phase-space volume is not conserved.
Note that $\dot{\boldsymbol Z}$ in RHS of Eq.\eqref{eq:p0} depends on the drift-kinetic models.
For the global model, it is denoted as
\begin{equation}
 \dot{\boldsymbol{Z}} \cdot \frac{ \partial f_{M} }{ \partial \boldsymbol Z } = - \mathcal{S}_{0};
\end{equation}
for ZOW and ZMD, it is denoted as
\begin{equation}
 \dot{\boldsymbol{Z}} \cdot \frac{ \partial f_{M} }{ \partial \boldsymbol Z } = \dot{v} \frac{ \partial }{ \partial v } f_{M};
\end{equation}
for the DKES-like, due to $\dot v = 0 $, $\dot \psi = 0$, and $\mathcal{G} = 0$, the weight function $\mathfrak{p}$ becomes constant as
\begin{equation}
 \mathfrak{\dot p} =0.
\end{equation}

\section{Moments of Drift-kinetic Equation}\label{section:Drift-kinetic_Equation}

In this section, the balance equations of particle number, parallel momentum, and energy are investigated for global and local models. 
The compressibility of phase space $\mathcal G$ and the approximations on guiding-center trajectories in each model are taken into account. 
The requirement of adaptive source-sink term $S_1$ is explained, which is essential for obtaining a steady-state solution in some models.

In order to take moments of Eq.\eqref{eq:d-k-1-0}, consider an arbitrary function $\mathcal{A} ( \boldsymbol{X}, v, \xi, t ) $
which is independent of the gyro-phase.
For density variable $\int d^3 v f \mathcal{A}$ in $\boldsymbol{X}$-space, the balance equation is yielded by multiplying $\mathcal{A}$ with Eq.\eqref{eq:d-k-1-0} and taking integral over the velocity-space.
%\begin{align} \label{eq:d-k-A-1}
% &\frac{ \partial }{ \partial t } \Bigg ( \int d^{3}v ~ f_{a}  \mathcal{A} \Bigg ) 
%    + \int d^{3}v ~ \frac{ \partial }{ \partial {Z}_{i} }  \Bigg ( f_{a}  \mathcal{A} \frac{ d {Z}_{i} }{ d t } \Bigg )
% \nonumber \\
% &= \int d^{3}v ~ \Bigg ( f_{a} \frac{ d \mathcal{A} }{ d t }  +  \big [ \mathcal{C}_a+ \mathcal{S}_a \big ] \mathcal{A} \Bigg ) \nonumber \\
% &+ \int d^3v ~  f_{a} \mathcal{G}  \mathcal{A}  
%\end{align}
%where the integral of velocity-space is given as 
%\begin{equation*}
% \int d^3v = 2 \pi \int d v v^2 \int d \xi \mathcal{J}
%\end{equation*}
%and 
%\begin{equation*}
%  \frac{ d \mathcal{A} }{ d t } 
%        \equiv  \Bigg (  \frac{\partial }{\partial t}
%          + \dot{  \boldsymbol{Z} }  \cdot \frac{\partial }{\partial \boldsymbol{Z}} 
% \Bigg ) \mathcal{A}.
%\end{equation*}
%By partial integral, Eq.\eqref{eq:d-k-A-1} is rewritten as
By partial integral, Eq.\eqref{eq:d-k-1-0} is rewritten as
\begin{align}\label{eq:d-k-A-1}
 & \frac{ \partial }{ \partial t } \Bigg ( \int d^{3}v ~ f_{a} \mathcal{A} \Bigg ) 
    + \nabla \cdot \Bigg ( \int d^{3}v ~ f_{a} ~ \mathcal{A}  \dot{ \boldsymbol{X} } \Bigg )
 \nonumber \\
 &= \int d^{3}v ~ \Bigg ( f_{a}  \frac{ d \mathcal{A} }{ d t }  + \big [ \mathcal{C}_a+\mathcal{S}_a \big ] \mathcal{A} \Bigg )
 \nonumber \\
 &+ \int d^3v ~  f_{a}  \mathcal{G} \mathcal{A},
\end{align}
where the integral of velocity-space is given as 
\begin{equation*}
 \int d^3v = 2 \pi \int d v v^2 \int d \xi \mathcal{J}.
\end{equation*}
Furthemore, the following equation is employed to derive Eq.\eqref{eq:d-k-A-1}
\begin{equation*}
  \frac{ d \mathcal{A} }{ d t } 
        \equiv  \Bigg (  \frac{\partial }{\partial t}
          + \dot{  \boldsymbol{Z} }  \cdot \frac{\partial }{\partial \boldsymbol{Z}} 
 \Bigg ) \mathcal{A}.
\end{equation*}

\subsection{The Particle and Energy Balance on the Local DKE Models} \label{section:particle_flux}
In order to derive the conservation law of particle number, substituting $\mathcal{A} = 1$ into Eq.\eqref{eq:d-k-A-1} yields
\begin{align}
 &\frac{ \partial }{ \partial t } \Bigg ( \int d^{3}v ~ f_{a}   \Bigg ) 
                             +  \nabla \cdot \Bigg ( \int d^{3}v~ f_{a}  \dot{ \boldsymbol{X} } \Bigg ) 
 \nonumber \\
 & = \int d^{3}v ~  \mathcal{S}_{a} + \int d^3v ~ f_{a} \mathcal{G}
\end{align}
where $\int d^3 v ~ \mathcal{C}_a =0$ is used.
The continuity equation is obtained as
\begin{align}\label{eq:m0}
  \frac{ \partial  n_{a}  }{ \partial t }
                             +  \nabla \cdot ( n_{a} \boldsymbol V_a  ) 
  = \int d^{3}v ~ \mathcal{S}_{a} + \int d^{3}v  f_{a} \mathcal{G} .
\end{align}
The density $n_a$ and the mean flow velocity $n_{a} \boldsymbol V_a$ are defined as
\begin{subequations}
\begin{alignat}{2}
&
n_a \equiv \int d^3 v ~ f_a,
\\
&  n_a \boldsymbol V_a \equiv  \int d^3 v ~ \dot{\boldsymbol X} f_a.
%&  n_a \boldsymbol V_a \equiv \int d^3 v ~ \boldsymbol v f_a =  \int d^3 v ~ \dot{\boldsymbol X} f_a.
\end{alignat}
\end{subequations}

The balance of kinetic energy is obtained by substituting $\mathcal{A}=\mathcal{K}$ into Eq.\eqref{eq:d-k-A-1},
\begin{align}\label{eq:m2-kinetic}
 &\frac{ \partial }{ \partial t } \Bigg ( \int d^{3}v ~ f_{a}  \mathcal{K} \Bigg ) 
   +  \nabla \cdot \Bigg ( \int d^{3}v ~ f_{a}  \mathcal{K}  \dot{ \boldsymbol{X} } \Bigg ) 
 \nonumber \\
 &= \int d^{3}v ~ \Bigg ( f_{a}  \frac { d \mathcal{K} }{dt}   +  \big [ \mathcal{C}_{a}+ \mathcal{S}_{a} \big ] \mathcal{K}  \Bigg ) 
 \nonumber \\
 &+ \int d^3v ~ f_{a} \mathcal{G}\mathcal{K} .
\end{align}
Here the kinetic energy $\mathcal{K}$ is defined as %Eq.\eqref{eq:K}
\begin{equation}
 \mathcal{K} \equiv \frac{1}{2} m_{a}{v_{\parallel}}^2 + \mu B = \mathcal{E} - e_{a} \Phi
\end{equation}
where $\mu$ is the magnetic momentum, $\mathcal{E}$ is the total energy, and $\Phi$ is the electrostatic potential.
The time derivative of the kinetic energy is denoted as %Eq.\eqref{eq:dot_K}.
\begin{align}\label{eq:K_dot}
  \frac { d \mathcal{K} }{dt} &= \frac{ d \mathcal{E}}{dt} -  e_{a} \frac{ d \Phi}{dt} 
  \nonumber \\
  &= \mu \frac{ \partial B ( \boldsymbol X , t ) }{ \partial t} +  e_{a} \boldsymbol{ E }^{*} \cdot  \frac { d \boldsymbol X}{dt},
\end{align}
where $ \boldsymbol{ E }^{*} $ is defined in Eq.\eqref{eq:E_ast}.
In the series of simulations, a stationary electromagnetic field approximation is employed, which are Eqs.\eqref{eq:dBdt=dphidt=0} and \eqref{eq:dEdt=0}.
Therefore, Eq.\eqref{eq:K_dot} is approximated as
\begin{equation} \label{eq:dkdt}
 \frac { d \mathcal{K} }{dt} \simeq - e_{a}\frac{ d \Phi}{dt} = - e_{a}\frac{ d \boldsymbol X }{ d t} \cdot \nabla \Phi.
\end{equation}
According to  Eqs.\eqref{eq:m0}, \eqref{eq:m1-parallel}, and \eqref{eq:m2-kinetic}, if the Liouville theorem is violated,  $\mathcal{G}$ affects the particle, momentum, and energy balance. The approximation trajectories and balance equations in the local models are presented in the following subsections.

The flux-surface-average is denoted as
\begin{equation}
  \langle \mathcal{A} \rangle \equiv \frac{ ~ \int d \theta d \zeta ~ \mathcal{J} \mathcal{A} ~ }{ \mathcal{V}'},
\end{equation}
where $\mathcal{A}$ is an arbitrary function and $\mathcal{V}'$ is defined as 
\begin{equation}
 \mathcal{V}' \equiv \frac{d \mathcal{V} }{ d \psi} = \int d \theta d \zeta ~ \mathcal{J}.
\end{equation}
The particle density from $f_{a,1}$ is denoted as 
\begin{equation}\label{eq:n1}
 \mathcal{N}_{1} \equiv   \int d^3 v ~ f_{1} ( \boldsymbol Z ).
\end{equation}
According to continuity equation Eq.\eqref{eq:m0}, the time evolution of density is
\begin{align} \label{eq:dz1_dt}
 & \frac{\partial   \left \langle \mathcal{N}_{1}  \right \rangle }{ \partial t} + \left \langle \nabla \cdot \bigg ( \mathcal{N}_{1} \boldsymbol{V} \bigg ) \right \rangle \nonumber
 \\ 
 &= \left \langle \int d^3 v ~ \mathcal{S}_{1} ~\right \rangle + 
       \left \langle \int d^3 v  ~ f_{1}~ \mathcal{G} \right \rangle .
\end{align}
After taking the flux-surface-average, the contribution of $\mathcal{S}_{0}$  is zero because the source/sink term is a Maxwellian Eq.\eqref{eq:f_M_local} with flux-surface functions $n$ and $T$.
Note here that in the global model the particle flow $ \mathcal{N}_{1} \boldsymbol{V} $ contains the radial component and $\mathcal{G}=0$. 
Then, Eq.\eqref{eq:dz1_dt} for the global model becomes
\begin{equation}\label{eq:global_n_a1}
 \frac{ \partial \langle \mathcal{N}_{1}  \rangle }{ \partial t } + \frac{ d  }{ d \mathcal{V}  } \left ( \Gamma^{\psi} \mathcal{V}' \right )= \left \langle \int d^3 v \mathcal{S}_{1} \right \rangle,
\end{equation}
where the particle flux is calculated by
\begin{equation} \label{eq:p-flux}
 \Gamma^\psi \equiv
  \left \langle \int d^3 v ~ f_{1} {\dot \psi} ~ \right\rangle,
\end{equation}
and the following identity is employed
\begin{equation}
 \left \langle \nabla \cdot \boldsymbol A \right \rangle 
 = \frac{d}{ d \mathcal{V} } \left \langle \boldsymbol A \cdot \nabla \mathcal{V} \right \rangle 
 = \frac{1}{  \mathcal{V}' }\frac{d}{ d \mathcal{\psi} } \left \langle \mathcal{V}' \boldsymbol A \cdot \nabla \psi  \right \rangle.
\end{equation}
The finite $d(\Gamma^{\psi} \mathcal{V}')/d \mathcal{\mathcal{V}}$ term is a corollary of global simulation in which the actual radial particle flux across a flux surface is solved. Therefore, it is essentially required to include the particle source to obtain a steady-state solution.
On the other hand, in the three local models, the $\mathcal{N}_{1} \boldsymbol{V}$ term has only the tangential component to the flux surface. Therefore, the $ \langle \nabla \cdot ( \mathcal{N}_{1} \boldsymbol{V}) \rangle $ term vanishes in ZOW, ZMD and DKES-like models. However, for ZOW, the artificial source/sink term $\mathcal{S}_{1}$ is required because of the compressibility $\mathcal{G} \neq 0$ \cite{Matsuoka2015},
\begin{align} \label{eq:particle_number_ZOW}
 \frac{\partial  \left \langle \mathcal{N}_{1} \right \rangle }{ \partial t} 
 = \left \langle \int d^3 v ~ \mathcal{S}_{1} ~\right \rangle + 
       \left \langle \int d^3 v  ~ f_{1}~ \mathcal{G} \right \rangle.
\end{align}
According to Eq.\eqref{eq:zow-J}, the last term in Eq.\eqref{eq:particle_number_ZOW} is estimated as $\mathcal{O}(\delta^2)$.
For ZMD and DKES-like, the particle density $\mathcal{N}_{1}$ is constant naturally without $\mathcal{S}_1$, as pointed out by Landreman,\cite{Landreman_2014}
\begin{align} \label{eq:particle_number_ZMD_DKES-like}
 \frac{\partial  \left \langle \mathcal{N}_{1} \right \rangle }{ \partial t} 
 = 0.
\end{align}

The energy balance equation \sout{of energy} for each model is derived similarly, as follows. 
The energy flux is introduced as
\begin{equation} \label{eq:energy-flux}
 \boldsymbol{Q} \equiv \int d^3 v f_{1}  \mathcal{K} \dot{\boldsymbol{X}},
\end{equation}
and the flux-surface-average of radial energy flux is defined as
\begin{equation}\label{eq:radial_q-flux}
 Q^{\psi} \equiv \left \langle \int d^3 v f_{1} \mathcal{K} \dot{\psi} \right \rangle.
\end{equation}
The pressure perturbation on flux surface is given as
\begin{equation}\label{eq:p1}
 P_{1} \equiv \frac{2}{3} \int d^3 v f_{1} \mathcal{K}.
\end{equation}
According to balance of kinetic energy Eq.\eqref{eq:m2-kinetic}, the time evolution of $P_{1}$ is rewritten as
\begin{align}\label{eq:m2-kinetic-1}
 & \frac{3}{2} \frac{ \partial }{ \partial t } \left \langle P_{1} \right  \rangle 
   + \left \langle \nabla \cdot \boldsymbol Q \right \rangle \nonumber
 \\ \nonumber 
 &= \left \langle \int d^{3}v ~ f_{1} ~ \frac { d \mathcal{K} }{dt} \right \rangle 
 + \left \langle \int d^{3}v ~  \mathcal{S}_{1} ~ \mathcal{K} \right \rangle
 \\ 
 &+ \left \langle \int d^3v ~ f_{1} ~ \mathcal{G} ~\mathcal{K}~ \right \rangle.
\end{align}
In Eq.\eqref{eq:m2-kinetic-1}, the contribution from $\mathcal{S}_0$ vanishes again.
The energy exchange by collision is omitted because we neglect the ion-electron collision and the electron-ion collision is approximated by pitch-angle scattering in the simulations. 
In the RHS of Eq.\eqref{eq:m2-kinetic-1}, the time evolution of kinetic energy is approximated as
\begin{align}
 &\left \langle \int d^{3}v ~ f_{1} ~ \frac { d \mathcal{K} }{dt} \right \rangle \nonumber
 \\ 
 &\simeq e E_{\psi}  \left \langle \int d^{3}v ~\dot{\psi} ~ f_{1}  \right \rangle
 =  e E_{\psi} ~ \Gamma^{\psi}, 
\end{align}
which represents the work done by the radial current.
For the global model, the finite $\langle \nabla \cdot \boldsymbol Q \rangle = d ( Q^{\psi} \mathcal{V}' ) / d \mathcal{V} $ remains as in Eq.\eqref{eq:global_n_a1}. Therefore, an energy source $\mathcal{S}_1 \mathcal{K} $ is essentially required to reach a steady-state. On the other hand, the radial energy flux $Q_a^{\psi}$ vanishes in the local models.
For the ZOW and ZMD models, $\mathcal{S}_{1} \mathcal{K} $ is required to satisfy the balance equation of energy because of $d \mathcal{K} / dt $ and $\mathcal{G}$
\begin{align}\label{eq:en_ZOW}
\frac{ \partial }{ \partial t } \left \langle P_{1} \right \rangle
  &=  \left \langle \int d^{3}v ~  \mathcal{S}_{1} ~ \mathcal{K} \right \rangle  \nonumber
 \\  
 & + e E_{\psi} ~ \Gamma^{\psi} + \left \langle \int d^3v ~ f_{1} ~ \mathcal{G} ~\mathcal{K}~ \right \rangle
\end{align} 
where $\mathcal{G}$ appears only in the ZOW model.
Eq.\eqref{eq:en_ZOW} indicates that ZMD cannot maintain the conservation law on energy when $E_{\psi} \neq 0$,
even if it holds the constant particle number in Eq.\eqref{eq:particle_number_ZMD_DKES-like}.
Finally, DKES-like maintains the energy balance without $\mathcal{S}_{1} \mathcal{K}$ because of $d \mathcal{K} /dt = 0 $ and $\mathcal{G} = 0 $. 

Recently, Sugama has derived another type of ZOW model\cite{sugama_2016} in which guiding-center variables are 
chosen as $(\boldsymbol{X},v_\parallel,\mathcal{K})$ and the tangential magnetic drift is defined as
\begin{equation}\label{eq:zow_sgm}
\hat{\boldsymbol{v}}_m=\boldsymbol{v}_m-\frac{(\boldsymbol{v}_m\cdot\nabla\psi)}{|\nabla\psi|^2}\nabla\psi.
\end{equation}
In this model, the magnetic moment $\mu$ is allowed to vary in time so that the kinetic energy $\mathcal{K}$ is conserved.
It is shown that the new local model satisfies both particle and energy balance relations without source/sink term. 
Although such a conservation property is desirable as a drift-kinetic model, we employ Matsuoka's ZOW model here for two reasons. First, the definition of tangential magnetic drift as in Eq. \eqref{eq:zow_sgm} requires the geometric factor $|\nabla\psi|^2$ on each marker's position, which will increase the computation cost. 
Second, it is necessary to find a modified Jacobian with which the phase-space volume conservation is recovered in this local model. 
%The other new numerical scheme is necessary to solve the differential equation, which is given as Eq. (84) in Ref.\cite{sugama_2016}.
%We adopt the source/sink term in ZOW and ZMD models after the verification.
%The verification shows that the source/sink term does not affect the long-term time average value of neoclassical fluxes after the simulation reaches a quasi-steady state.
To obtain such a modified Jacobian, another differential equation as Eq. (84) in Ref. 18 is required to be solved.
Instead, in this paper, we adopt the source/sink term in ZOW and ZMD models after the verification as discussed in Appendix \ref{AppendixA}. 
The verification shows that the source/sink term does not affect the long-term time average value of neoclassical fluxes after the simulation reaches a quasi-steady state.

%%%%%%%%%%%%%%%%%%%%%%%%%%%%%%%%%%%%%%%%%%%%%%%%%%%%%%%%%%%%%%%%%%%%%%%%%%%%%%%%%%%%%%%%%%%%%%%%%%%
\subsection{The Parallel Momentum Balance and Parallel Flow} \label{sec:parall_monentum_balance}
The parallel momentum balance equation is derived from Eq.\eqref{eq:d-k-A-1} with  $\mathcal{A} = m_{a} v_{\parallel}$, \cite{sugama_2016}
\begin{align}\label{eq:m1-parallel}
 &\frac{\partial}{ \partial t } ( n_{a} m_{a} V_{a, \parallel} ) + \boldsymbol{b} \cdot ( \nabla \cdot \boldsymbol{P}_{a} ) 
 \nonumber \\
 &= n_{a} e_{a} E_{\parallel} + F_{\parallel,a} 
 + \int d^{3}v ~ \mathcal{S}_{a}  m_{a} v_{\parallel} 
 \nonumber \\
 &+ \int d^3v ~ f_{a}  \mathcal{G}  m_{a} v_{\parallel},
\end{align}
where $E_{\parallel} = \boldsymbol b \cdot\boldsymbol E$ and $\boldsymbol{P}_{a}$ is the pressure tensor.
The parallel friction of collision $F_{a, \parallel}$ is given as
\begin{equation}
 F_{a, \parallel} \equiv \boldsymbol b \cdot  \sum_{b\neq a}\boldsymbol{F}_{ab}=\sum_{b\neq a}\int d^3 v ~ \mathcal{C}_{ab}(f_a ,f_b) m_{a} {v}_{\parallel}.
\end{equation}
%\begin{subequations}
%\begin{alignat}{4}
%& \boldsymbol P_{a} \equiv \boldsymbol P_{CGL,a} + { \boldsymbol \Pi }_{2,a},
%\\
%& \boldsymbol P_{CGL,a} \equiv \int d^3 v ~ [  ( m_{a} v_{\parallel}^2 \boldsymbol b \boldsymbol b + \mu B  ( \boldsymbol I - \boldsymbol b \boldsymbol b ) ] f_{a}, \label{eq:P_cgl}
%\\
%& \boldsymbol \Pi_{2,a} \equiv \int d^3 v ~ m_{a} v_{\parallel} \bigg ( \dot{  \boldsymbol X }_{\perp} \boldsymbol b + \boldsymbol b \dot{  \boldsymbol X }_{\perp} \bigg ) f_{a} \label{eq:pi_2},
%\\
%\text{and} \nonumber
%\\
%&F_{a, \parallel} \equiv \boldsymbol b \cdot  \sum_{b\neq a}\boldsymbol{F}_{ab}=\sum_{b\neq a}\int d^3 v ~ \mathcal{C}_{ab}(f_a ,f_b) m_{a} {v}_{\parallel}.
%\end{alignat}
%\end{subequations}
In order to derive Eq.\eqref{eq:m1-parallel}, the expression of the time derivative of the parallel velocity $\dot{v}_{\parallel}$ is required.
For the global model, it is given as
\begin{equation}\label{eq:dot_v_para}
  \dot{v}_{\parallel} = -\frac{ 1 }{ m }   \boldsymbol b \cdot \left( \mu  \nabla B - e \boldsymbol{E}^{*} \right)  +  {v}_{\parallel} \dot { \boldsymbol X } \cdot \boldsymbol \kappa
\end{equation}
following the particle orbit Eqs.\eqref{eq:f3d-o1} and \eqref{eq:kappa}.
We substitute Eq.\eqref{eq:dot_v_para} into the parallel momentum equation Eq.\eqref{eq:d-k-A-1}.
The pressure tensor $\boldsymbol{P}$ includes the diagonal component, the Chew-Goldbeger-Low (\text{CGL}) tensor $\boldsymbol{P}_{\text{CGL}}$, and the $ \boldsymbol{\Pi}_{2} $ term, the viscosity tensor $\boldsymbol{\Pi}_{2}$. 
See Appendix \ref{AppendixB} for the derivation.
According to the $\delta f$ method, the viscosity tensors become
\begin{subequations}
\begin{alignat}{2}
 & \boldsymbol {b} \cdot \nabla \cdot \boldsymbol P_{\text{CGL}}   \label{eq:p_cgl_1} \nonumber
 \\ &= 
  \boldsymbol {b} \cdot \nabla \cdot \left[ \int d^3 v ~ \left ( ~ ( ~ m v_{\parallel}^2 ~ \boldsymbol b \boldsymbol b + \mu B ~ ( \boldsymbol I - \boldsymbol b \boldsymbol b ) \right ) f_{1} \right ]  , 
\\
 & \boldsymbol {b} \cdot \nabla \cdot \boldsymbol \Pi_{2}   \label{eq:pi_2_f1} \nonumber
 \\ &=  
  \boldsymbol {b} \cdot \nabla \cdot \left[ \int d^3 v ~ m v_{\parallel} ~ \bigg (  \dot{\boldsymbol X}_{\perp}  \boldsymbol b + \boldsymbol b  \dot{\boldsymbol X}_{\perp}  \bigg ) f_{1} \right ] ,
\end{alignat}
\end{subequations}
where $ f_{a,0} $ is an even function but $ v_{\parallel} \dot{\boldsymbol X }_{\perp} $ is an odd function. 
Therefore, $f_{a,0}$ does not contribute to $\nabla \cdot \boldsymbol \Pi_{2}$.
According to Eq.\eqref{eq:p_cgl_1}, $\nabla \cdot \boldsymbol P_{\text{CGL}} $ does not explicitly depend on the  approximations in $\boldsymbol v_{m}$ and $\boldsymbol v_{E}$.
%Following Eq.\eqref{eq:m1-parallel}, the flux-surface-average of the parallel momentum balance equation becomes 
Multiplying Eq.\eqref{eq:m1-parallel} with $B$, the flux-surface-average of the parallel momentum balance equation becomes 
\begin{align}\label{eq:m1-parallel_f1}
 &\left \langle \frac{\partial}{ \partial t } ( n m V_{\parallel} B ) \right \rangle + \left \langle \boldsymbol{B} \cdot \nabla \cdot ( \boldsymbol{P}_{\text{CGL}} + \boldsymbol{\Pi}_{2}) \right \rangle 
 \nonumber \\
 &= \left \langle  n e E_{\parallel} B \right \rangle  + \left \langle  F_{\parallel} B \right \rangle  + \left \langle B \int d^{3}v ~ \mathcal{S}_{1} ~ m v_{\parallel} \right \rangle  
 .
\end{align}

For the ZOW model, the parallel momentum balance equation is calculated with $ \dot{\boldsymbol X}_{\perp}  = \boldsymbol v_{E} + \boldsymbol {\hat{v}}_{m} $ and the time derivative of parallel velocity
\begin{align}\label{eq:dot_v_para-zow}
 \dot{v}_{\parallel} 
 &= -\frac{ \mu }{ m }  \boldsymbol b \cdot \nabla B + {v}_{\parallel} \boldsymbol v_{E} \cdot \frac{ \nabla_{\perp} B }{ B } 
 \nonumber \\
 &= -\frac{ \mu }{ m }  \boldsymbol b \cdot \nabla B 
 + {v}_{\parallel}  \dot { \boldsymbol X }_{\perp} \cdot \boldsymbol \kappa 
 + {v}_{\parallel} \left( \frac { \dot { \psi } } { B } \frac{ \partial B }{ \partial \psi } \right),
\end{align}
following the particle orbit Eq.\eqref{eq:zoworbit}.
Then, the parallel momentum balance equation becomes
\begin{align}\label{eq:m1-parallel_f1-ZOW}
 &\left \langle \frac{\partial}{ \partial t } ( n m V_{\parallel} B ) \right \rangle + \left \langle \boldsymbol{B} \cdot \nabla \cdot ( \boldsymbol{P}_{\text{CGL}} + \boldsymbol{\Pi}_{2,\text{ZOW}}) \right \rangle 
 \nonumber \\
 &=  \left \langle  F_{\parallel} B \right \rangle  
 + \left \langle B \int d^{3}v ~ \mathcal{S}_{1} ~ m v_{\parallel} \right \rangle
 \nonumber \\ 
% &+ \left \langle B \int d^{3}v ~ m {v}_{\parallel} \left( \frac { \dot { \psi } } { B } \frac{ \partial B }{ \partial \psi }  \right) f_{1}   \right \rangle  
% \\ \nonumber
 &+ \left\langle B \int d^3v ~ f  ~ \mathcal{G} ~ mv_{\parallel}\right\rangle .
\end{align}
For the ZOW model, the $ \boldsymbol{P}_{\text{CGL}} $ term is the same form  as Eq.\eqref{eq:p_cgl_1} and the $ \boldsymbol{\Pi}_{2} $ term, Eq.\eqref{eq:pi_2_f1}, is rewritten as
\begin{align} \label{eq:pi_2_ZOW}
 & \langle \boldsymbol {B} \cdot \nabla \cdot \boldsymbol \Pi_{2, \text{ZOW} } \rangle 
% \\
% &= \left \langle \nabla \cdot \int d^3 v ~ m_{a} v_{\parallel} ~ \bigg [ ( \boldsymbol v_{E} + \hat {\boldsymbol v}_{m} )  \boldsymbol b + \boldsymbol b (\boldsymbol v_{E} + \hat {\boldsymbol v}_{m} ) \bigg ] f_{a,1} \nonumber
% \right \rangle
  \nonumber \\ 
 &=  \left \langle \boldsymbol {B} \cdot \nabla \cdot \bigg[ m n {V}_{\parallel}( \boldsymbol b \boldsymbol v_{E} + \boldsymbol v_{E} \boldsymbol b ) \bigg ]  \right \rangle \nonumber
  \nonumber \\
 &+ \left \langle \boldsymbol {B} \cdot \nabla \cdot \left [ \int d^3 v ~ m v_{\parallel} ~ \bigg (  \hat {\boldsymbol v}_{m} \boldsymbol b + \boldsymbol b  \hat {\boldsymbol v}_{m} \bigg ) f_{1} \right ] 
 \right \rangle 
  \nonumber \\
 &+ \left \langle \int d^{3}v ~ m {v}_{\parallel} B \left( \frac { \dot { \psi } } { B } \frac{ \partial B }{ \partial \psi } \right) f_{1}   \right \rangle  
\end{align}
where $\hat {\boldsymbol v}_{m}$ is defined by Eq.\eqref{eq:hat_v_m}. Eq.\eqref{eq:pi_2_ZOW} shows that $\langle \nabla \cdot \boldsymbol \Pi_{2, \text{ZOW} } \rangle  $ of the ZOW model includes not only the $\boldsymbol E \times \boldsymbol B$ drift but also the partial magnetic drift.  
In the ZOW model, there is an extra term of viscosity in Eq.\eqref{eq:m1-parallel_f1-ZOW},
\begin{align}\label{eq:extra_viscosity_ZOW}
 \left \langle  \int d^{3}v ~ m {v}_{\parallel} B \left( \frac { \dot { \psi } } { B } \frac{ \partial B }{ \partial \psi }  \right) f_{1}   \right \rangle
\end{align}
which comes from the last term of Eq.\eqref{eq:dot_v_para-zow} and is estimated as $\mathcal O (\delta^2 )$.
Actually, the $\partial B/\partial \psi$ is $\mathcal{O} (\delta)$ terms in MHD-equilibrium of helical devices, and Eq.\eqref{eq:extra_viscosity_ZOW} becomes $\mathcal O (\delta^3 )$.
%This is excluded from the $ \boldsymbol \Pi_{2, \text{ZOW} } $ in order to maintain the symmetry of $\boldsymbol \Pi_{2, \text{ZOW} }$.
The symmetry of $ \boldsymbol \Pi_{2, \text{ZOW} }$ is broken because of the third term in Eq.\eqref{eq:pi_2_ZOW}.
Furthermore, there is an additional term on the RHS in Eq.\eqref{eq:m1-parallel_f1-ZOW},
\begin{equation}\label{eq:G_ZOW}
 \left\langle B \int d^3v ~ f  ~ \mathcal{G} ~ mv_{\parallel}\right\rangle
\end{equation}
which is estimated as $\mathcal O (\delta^2 )$.
The effect of Eq.\eqref{eq:G_ZOW} on the parallel flow will be discussed in Sec.\ref{sec:incompressibility} below.
Following the order of magnitude, the contribution of Eq.\eqref{eq:pi_2_ZOW} and \eqref{eq:G_ZOW} are comparable in the parallel momentum equation Eq.\eqref{eq:m1-parallel}.   
The parallel electric field $E_{\parallel}$ and its contribution to the parallel momentum balance are neglected in the local models for simplicity.

For the ZMD model, the parallel momentum balance equation is calculated with $ \dot{\boldsymbol X}_{\perp}  = \boldsymbol v_{E}$ and the time derivative of parallel velocity
\begin{align}\label{eq:dot_v_para-zmd}
 \dot{v}_{\parallel} 
 &= -\frac{ \mu }{ m }  \boldsymbol b \cdot \nabla B + {v}_{\parallel} \boldsymbol v_{E} \cdot \frac{ \nabla_{\perp} B }{ B } 
 \nonumber \\ 
 &= -\frac{ \mu }{ m } \boldsymbol b \cdot \nabla B 
 + {v}_{\parallel} \dot { \boldsymbol X }_{\perp} \cdot \boldsymbol \kappa, 
\end{align}
following the particle orbit Eq.\eqref{eq:zmdorbit}.
If the scalar pressure is assumed as a function of $p = p (\psi)$, $\nabla_{\perp} B / B \cdot \boldsymbol v_{E} $ is rewritten as $\dot { \boldsymbol X }_{\perp} \cdot \boldsymbol \kappa $, according to Eq.\eqref{eq:kappa-app-1}.
Then, the parallel momentum balance equation becomes
\begin{align}\label{eq:m1-parallel_f1-ZMD}
 &\left \langle \frac{\partial}{ \partial t } ( n m V_{\parallel} B ) \right \rangle + \left \langle \boldsymbol{B} \cdot \nabla \cdot ( \boldsymbol{P}_{\text{CGL}} + \boldsymbol{\Pi}_{2,\text{ZMD} } ) \right \rangle 
 \nonumber \\
 &=  \left \langle  F_{\parallel} B \right \rangle  
 + \left \langle B \int d^{3}v ~ \mathcal{S}_{1} ~ m v_{\parallel} \right \rangle.
\end{align}
Equation \eqref{eq:pi_2_f1} for ZMD is rewritten as 
\begin{align}\label{eq:ZMD_Pi}
 &\langle \boldsymbol {B} \cdot \nabla \cdot   \boldsymbol \Pi_{2, \text{ZMD}  } \rangle  
 = \left \langle \boldsymbol {B} \cdot \nabla \cdot \left [ m n V_\parallel
   ( \boldsymbol b \boldsymbol v_{E} + \boldsymbol v_{E} \boldsymbol b ) \right] \right\rangle
\end{align}
where $\boldsymbol v_{m}$ does not exist in $\boldsymbol \Pi_{2}$.
This equation shows that ZMD maintains not only $\mathcal{G} = 0 $ but also the symmetry of $\langle \boldsymbol {B} \cdot \nabla \cdot   \boldsymbol \Pi_{2, \text{ZMD}}  \rangle$.  
%%%%% END of ZMD

%%%DKES
For the DKES model, the parallel momentum balance equation is calculated with $ \dot{\boldsymbol X}_{\perp}  = \hat {\boldsymbol v}_{E}$ from Eq.\eqref{eq:imcompressible} and the time derivative of parallel velocity
\begin{align}\label{eq:dot_v_para_dkes}
 \dot{v}_{\parallel} = -\frac{ \mu }{ m }  { \boldsymbol b } \cdot \nabla B,
\end{align}
following the particle orbit Eq.\eqref{eq:deksorbit}.
Then, the parallel momentum balance equation becomes
\begin{align}\label{eq:m1-parallel_f1-dkes}
 &\left \langle \frac{\partial}{ \partial t } ( n m V_{\parallel} B ) \right \rangle + \left \langle \boldsymbol{B} \cdot \nabla \cdot ( \boldsymbol{P}_{\text{CGL}} + \boldsymbol{\Pi}_{2,\text{DKES} } ) \right \rangle 
 \nonumber \\
 &=  \left \langle  F_{\parallel} B \right \rangle
 %- \left \langle m n V_{\parallel} \hat {\boldsymbol v}_{E} \cdot \boldsymbol \kappa   \right \rangle
 \nonumber \\
 &+ \left \langle B \int d^{3}v ~ \mathcal{S}_{1} ~ m v_{\parallel} \right \rangle.
\end{align}
With the incompressible $\boldsymbol E \times \boldsymbol B$ flow, Eq.\eqref{eq:pi_2_f1} is rewritten as
\begin{align}\label{eq:DKES_Pi}
   &\langle \boldsymbol {B} \cdot \nabla \cdot \boldsymbol \Pi_{2,\text{DKES} } \rangle  
   \nonumber \\ 
   &= \left \langle { \boldsymbol B  } \cdot \nabla \cdot 
   \left [ \frac{m n V_\parallel}{\langle B^2\rangle} ( { \boldsymbol b} \boldsymbol E \times \boldsymbol B + \boldsymbol E \times \boldsymbol B { \boldsymbol b} ) \right ] \right \rangle 
   \nonumber \\
   & + \left \langle B n m V_{\parallel} \hat {\boldsymbol v}_{E} \cdot \boldsymbol \kappa  \right \rangle .
\end{align}
DKES maintains $\mathcal{G} = 0$ and the symmetry is broken in viscosity $\langle \boldsymbol {B} \cdot \nabla \cdot \boldsymbol \Pi_{2,\text{DKES} } \rangle$.

The viscosity tensors are different among the ZOW, ZMD, and DKES-like models because of the approximation of incompressible $\boldsymbol E \times \boldsymbol B $ drift. 
The effect of incompressibility is discussed in Sec.\ref{sec:large_Er} below.
%%%%% END of DKES

%%%%

For the parallel momentum balance in all of the global and local models, the constraint imposed on the source/sink term $ \mathcal{S}_{1} $ is that its contribution to parallel momentum should vanish;
\begin{align}
 \int d^{3}v ~ \mathcal{S}_1 m v_{\parallel}   = 0.\label{eq:para_mom_src}
\end{align}
In fact, unlike the particle or energy balance relation, the drift-kinetic simulation reaches a steady state of parallel flow without any additional source/sink term. 
Note that the parallel momentum source vanishes not by flux-surface averaging, but is set to be zero anywhere on a flux surface.
The effect of parallel friction $F_{\parallel}$ and the finite-$\mathcal{G}$ terms on the parallel momentum are discussed in the next section.

%%%%%%%%%%%%%%%%%%%%%%%%%%%%%%%%%%%%%%%%%%%%%%%%%%%%%%%%%%%%%%%%%%%%%%%%%%%%%%%%%%%%%%%%%%%%%%%%%%%%%%%%%%%%%%%%%%%%%%%%%%%%
\section{Simulation Result and Discussion} \label{sec:result}
A series of simulations are carried out to benchmark the local and the global drift-kinetic models.
We compare the neoclassical radial particle flux $\Gamma^\psi_a$ Eq.\eqref{eq:p-flux}, radial energy flux Eq.\eqref{eq:radial_q-flux}, and the flux-surface average parallel mean flow multiplied by $B$,
\begin{equation} \label{eq:vb}
 \langle {V}_{a,\parallel} B \rangle \equiv
  \Bigg \langle \int d^3 v ~ f_{a,1}  v_{a,\parallel}  B( \psi, \theta, \zeta )  \Bigg \rangle.
\end{equation}
To see the radial fluxes and the heat fluxes in the units [1/m$^2$s] and [W/m$^2$], respectively, these are redefined as
\begin{equation*}
\Gamma_a \equiv \frac{dr}{d\psi}\Gamma_a^\psi,\quad Q_a\equiv \frac{dr}{d\psi}Q_a^\psi,
\end{equation*}
%where $r=a\sqrt{\psi/\psi_{edge}}$ and $a$ is the effective minor radius of the plasma boundary, $\psi=\psi_{edge}$. 
where $r=a\sqrt{\psi/\psi_{edge}}$ and $a$ is the effective minor radius of the plasma boundary, $\psi=\psi_{edge}$. $a$ and $\psi_{edge}$ are given from VMEC MHD equilibrium calculation code\cite{VMEC_1991}.  
Note that in the local models even though  $f_1$ does not contribute to radial fluxes in the particle and energy balance equations in Sec.\ref{section:particle_flux}, $\Gamma_a$ and $Q_a$ are evaluated by the virtual radial displacement $\boldsymbol{v}_m \cdot \nabla r$-term in the local approximations.

The plasma parameters are given as TABLE \ref{tb:para}. 
Two types of normalized ion collisionality $\nu_{i}^*$ are given in the table : 
$\nu_{i,PS}^*\equiv qR_{ax}\nu_{ii}/v_{thi}=1$ represents the Plateau - Pfirsch–Schl\"uter boundary and $\nu_{i,B}^* \equiv \nu_{i,PS}^* /(r/R_{ax})^{1.5}=1 $ is the Banana-Plateau boundary.
For LHD, the inward-shift configuration is employed, in which the neoclassical radial transport is expected to be suppressed compared to that in a standard configuration. 
For W7-X, the magnetic geometry is adjustable by the coil current system. Here, the standard configuration \cite{Gieger_PPCF(2015)_014004} in the zero-$\beta$ limit is employed. 
For HSX, the quasi-helically symmetric configuration is employed. 
The magnetic field configurations of both W7-X and HSX are chosen so as to reduce the radial guiding center excursion of trapped particles, while W7-X %the former
also aims at reducing the bootstrap current\cite{Gieger_PPCF(2015)_014004}\cite{HSX_1995} 
The artificial density and temperature profiles are given in the LHD and W7-X investigations so that the plasmas are in $1 / \nu$ regime around $|E_r| \sim 0$. The HSX kinetic profile is the diagnostic data from HSX experiment.\cite{Briesemeister_2013}  
Compared to the other devices, the collisionality of the HSX plasma is high in terms of $\nu_{iB}^*$ because of very low $T_i$.
In TABLE \ref{tb:para}, the ambipolar $E_r$ of the LHD and HSX simulations are shown, which have been evaluated by GSRAKE and DKES/PENTA, respectively.

In the following benchmarks, there are three types of DKES models, namely DKES, DKES-like, and DKES/PENTA.
 % See Fig.\ref{fig:p-flux-HSX}.
First, DKES is the original code with the pitch angle scattering collision operator.
Thus, it does not guarantee the conservation of momentum. 
Second, DKES-like is the solver of Eq.\eqref{eq:deksorbit} with the $\delta f$ method and the linearized collision operator as ZOW and ZMD. 
The test-particle portions of collision operator include both the pitch-angle and energy scattering terms.
The field-particle term maintains the conservation of particle numbers, parallel momentum, and energy in the simulation.\cite{Satake_2008} 
The third model, DKES/PENTA, is the numerical result from DKES and with momentum correction by Sugama-Nishimura method.\cite{penta_theory_2002}\cite{spong_2005} 
For LHD, local models are also benchmarked with GSRAKE code\cite{Beidler_2001}, which solves the mono-energy and the ripple-averaged drift-kinetic equations. 
GSRAKE is similar to DKES but the magnetic field spectrum in GSRAKE is approximated.\cite{Beidler_2001}
It should be emphasized that the $\boldsymbol E \times \boldsymbol B$ drift term in GSRAKE is compressible, although this point has not been clearly mentioned in previous studies. \cite{Beidler_2001}\cite{Beidler_nf2011}
The original GSRAKE code is made so that it can include the tangential magnetic drift term. 
However, the term is omitted in the present benchmarks because the magnetic drift term is found to make the simulation result unstable\cite{Satake_2006}.

\begin{table}
\centering
\caption { Simuation parameters on each configurations.}
\begin{tabular}{llll}
\hline
                  & LHD           & W7-X          & HSX           \\ 
                  \hline
r/a               & 0.7375          & 0.7500        & 0.3100        \\
$\iota$           & 0.740          & 0.886        & 1.051       \\
$R_{ax}/a$           & 3.60/0.64       & 5.51/0.51    & 1.21/0.126       \\ 
\hline
$n_{i}$ [$10^{18}/m^3$] & 3.10       & 0.406    & 3.83     \\
$T_{i}$ [$keV$]     & 0.891        & 0.350        & 0.061        \\
%$V_{i}$ [$m/s$]     & 4.1320E+05    & 2.5892E+05    & 1.0783E+05    \\
$T_{e}$ [$keV$]     & 0.891        & 0.350        & 0.544      \\
$B_{ax}$ [$T$]     & 2.99        & 2.77        & 1.00       \\ 
\hline
${\nu}^{*}_{i,B}$    & 0.0368        & 0.0910        & 17.3       \\ 
${\nu}^{*}_{i,PS}$  & 0.0017       & 0.0017        & 0.101     \\ 
\hline 
\hline
ambipolar $E_{r}$ [kV/m]   & -1.73        & N/A      & 3.47       \\ 
$\mathcal{M}_p$     & -0.015        & N/A             & 0.95       \\ 
\hline 
\end{tabular}
\label{tb:para}
\end{table}

\subsection{Effect of  $ \bold E \times  \bold  B $ Compressibility} \label{sec:large_Er} %OK
\begin{figure}
   \includegraphics[width=0.45\textwidth]{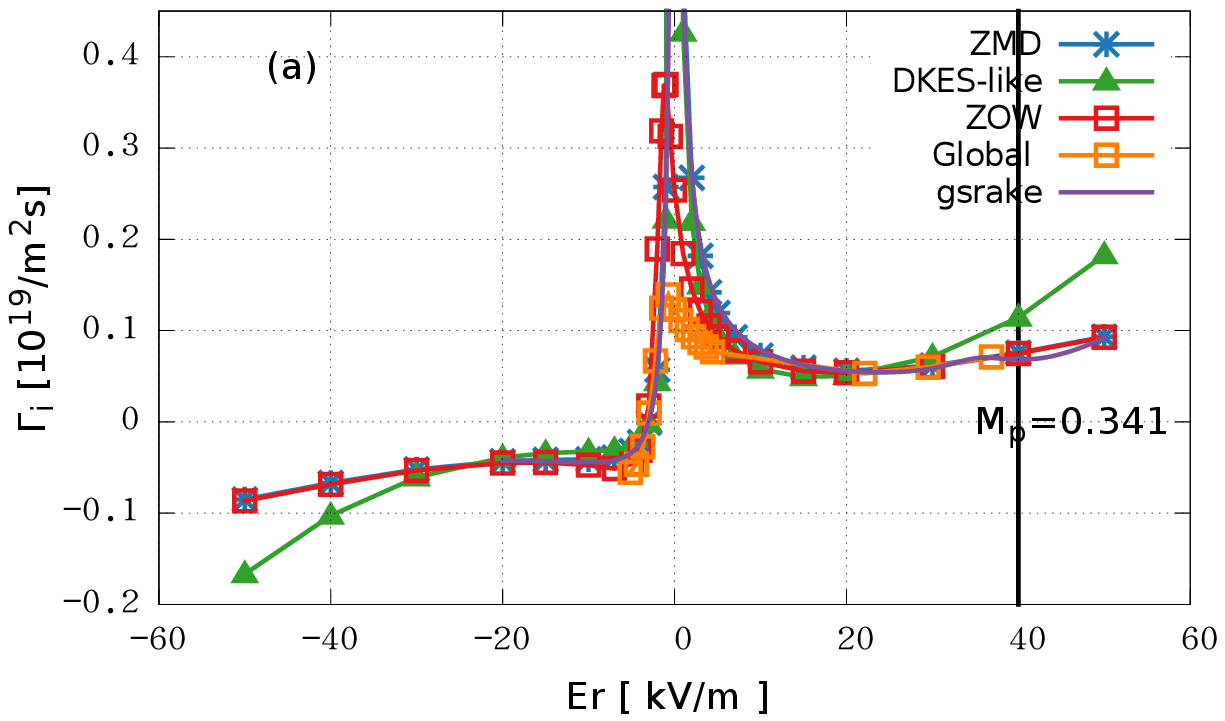} 
   \includegraphics[width=0.45\textwidth]{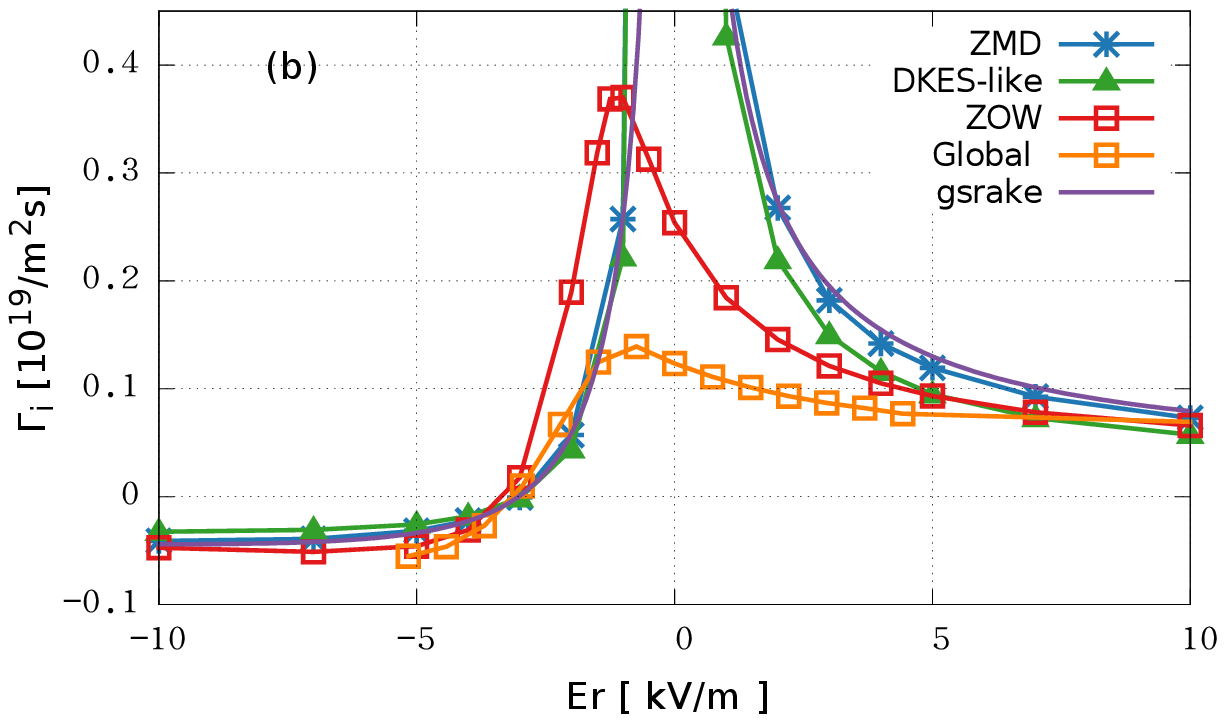} 
   \includegraphics[width=0.45\textwidth]{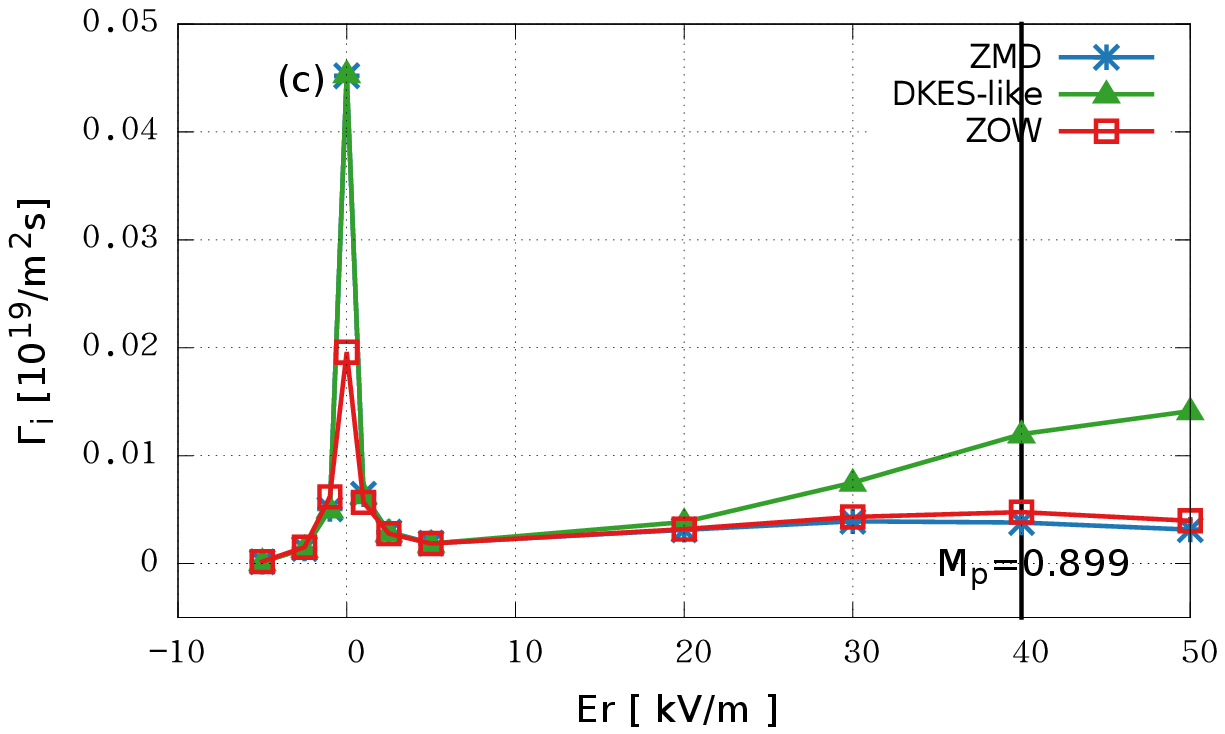} 
   \includegraphics[width=0.45\textwidth]{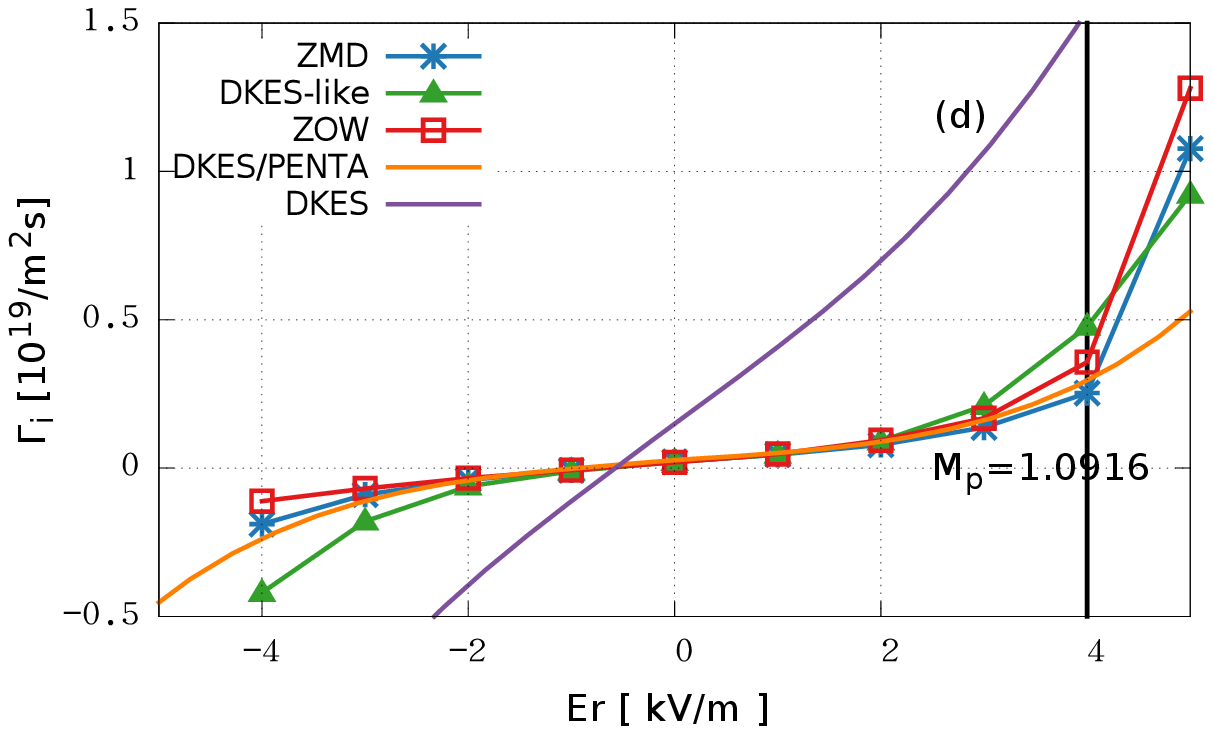}
\caption{\raggedright Ion particle fluxes $\Gamma_i$ of (a) LHD, (c) W7-X, and (d) HSX , respectively. (b) is an enlarged view of (a) around $E_r \sim 0$. The multiple numerical results of DKES model with the different collision operators are shown in (d). The vertical line shows the value of poloidal Mach number $\mathcal{M}_p$ defined in Eq.\eqref{eq:Mp}.}
\label{fig:p-flux}
\end{figure}

\begin{figure}
   \includegraphics[width=0.45\textwidth]{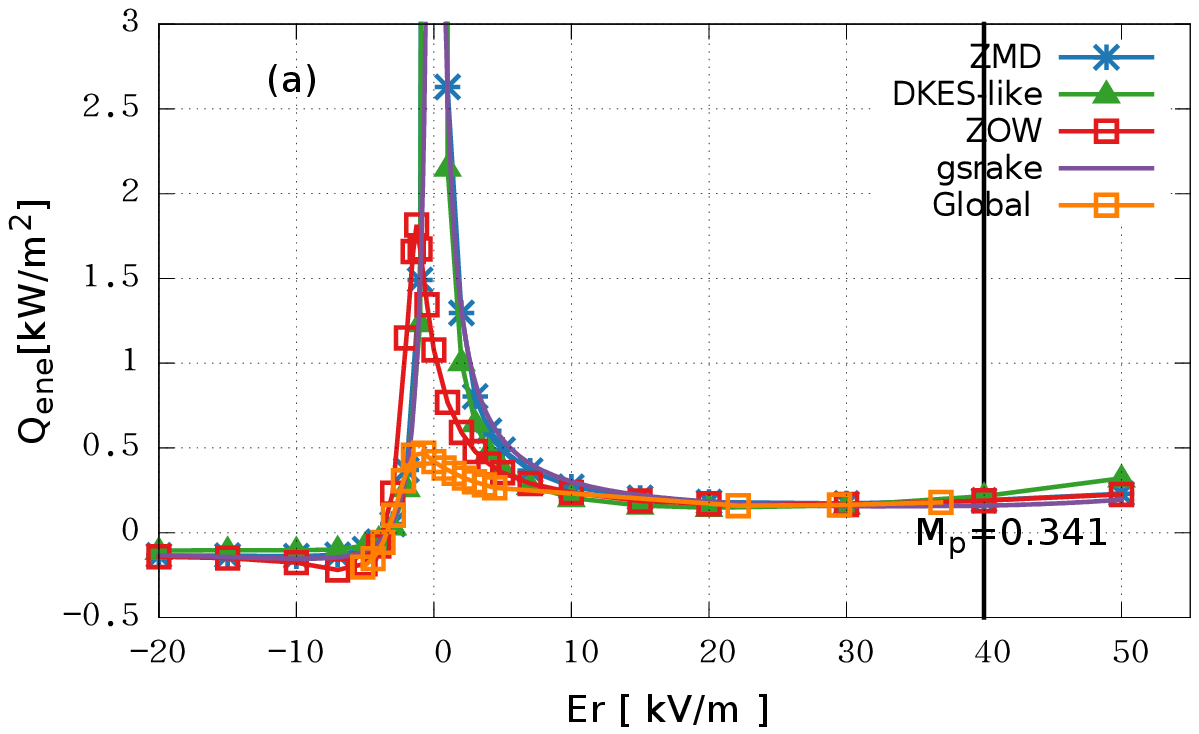} 
   \includegraphics[width=0.45\textwidth]{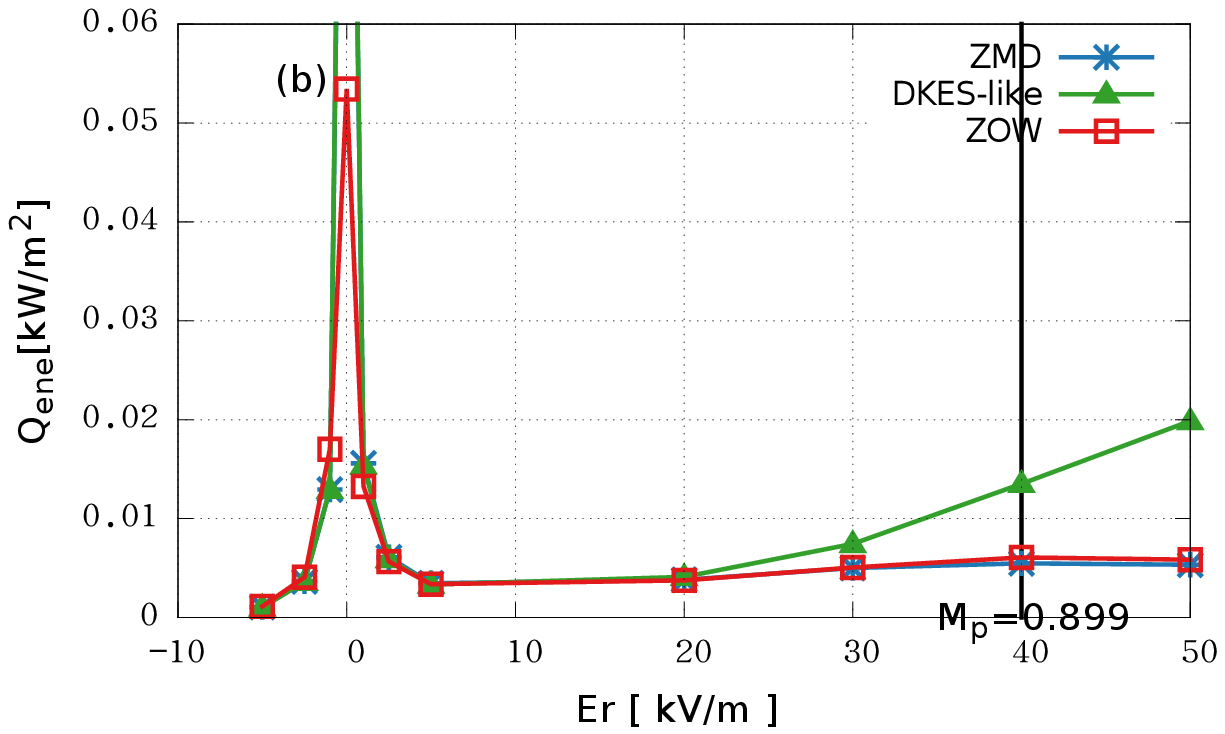} 
   \includegraphics[width=0.45\textwidth]{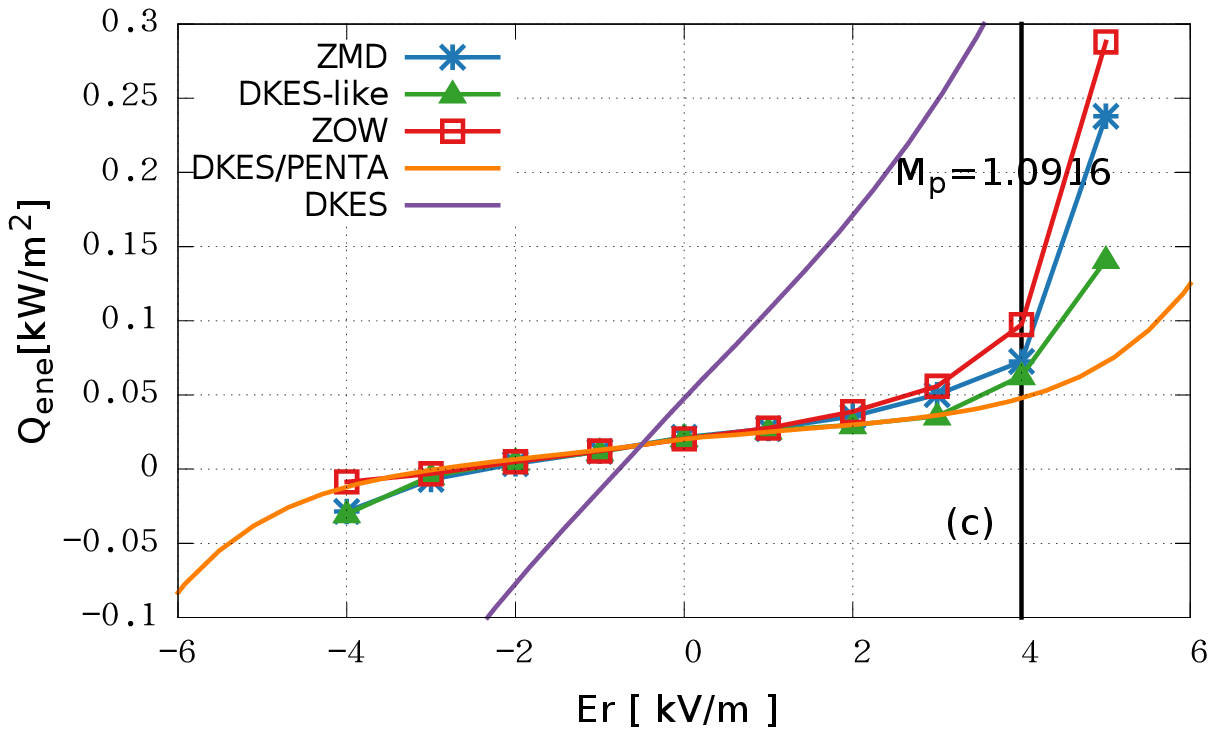}
 \caption{\raggedright Ion energy flues $Q_i$ of (a) LHD, (b) W7-X, and (c) HSX, respectively.}
 \label{fig:q-flux}
\end{figure}

\begin{figure}
   \includegraphics[width=0.45\textwidth]{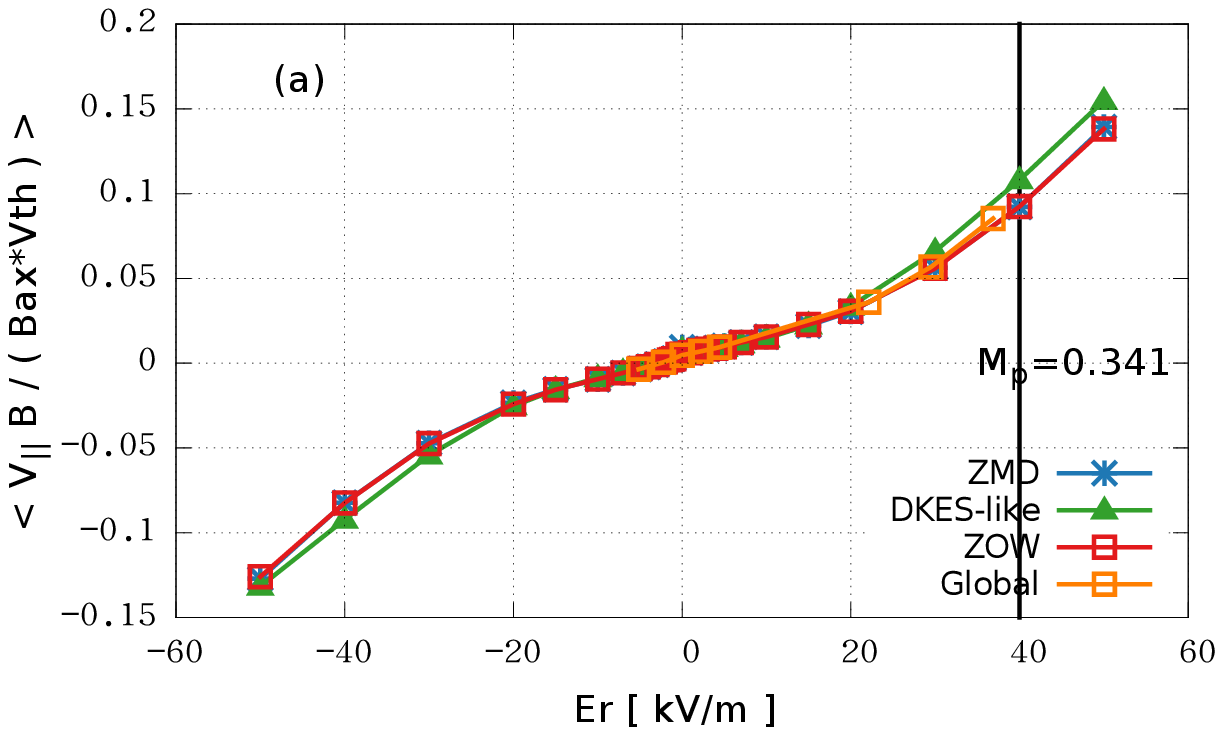} 
   \includegraphics[width=0.45\textwidth]{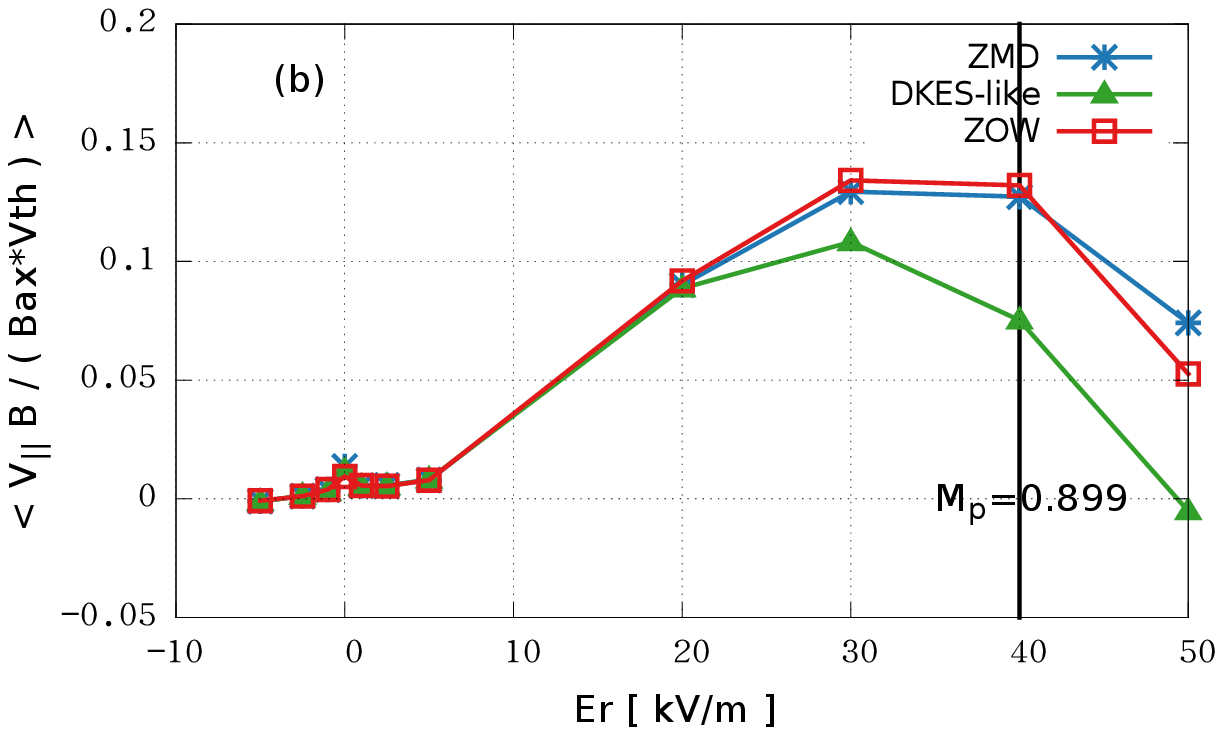} 
   \includegraphics[width=0.45\textwidth]{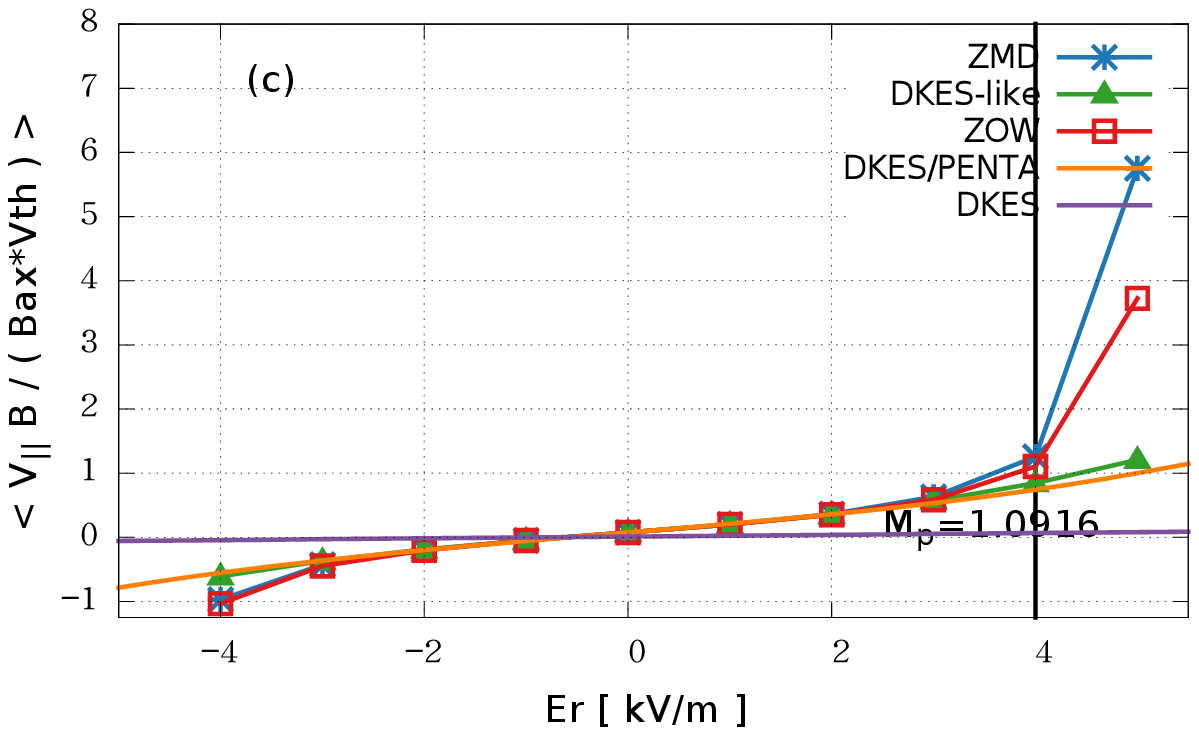}
   \includegraphics[width=0.45\textwidth]{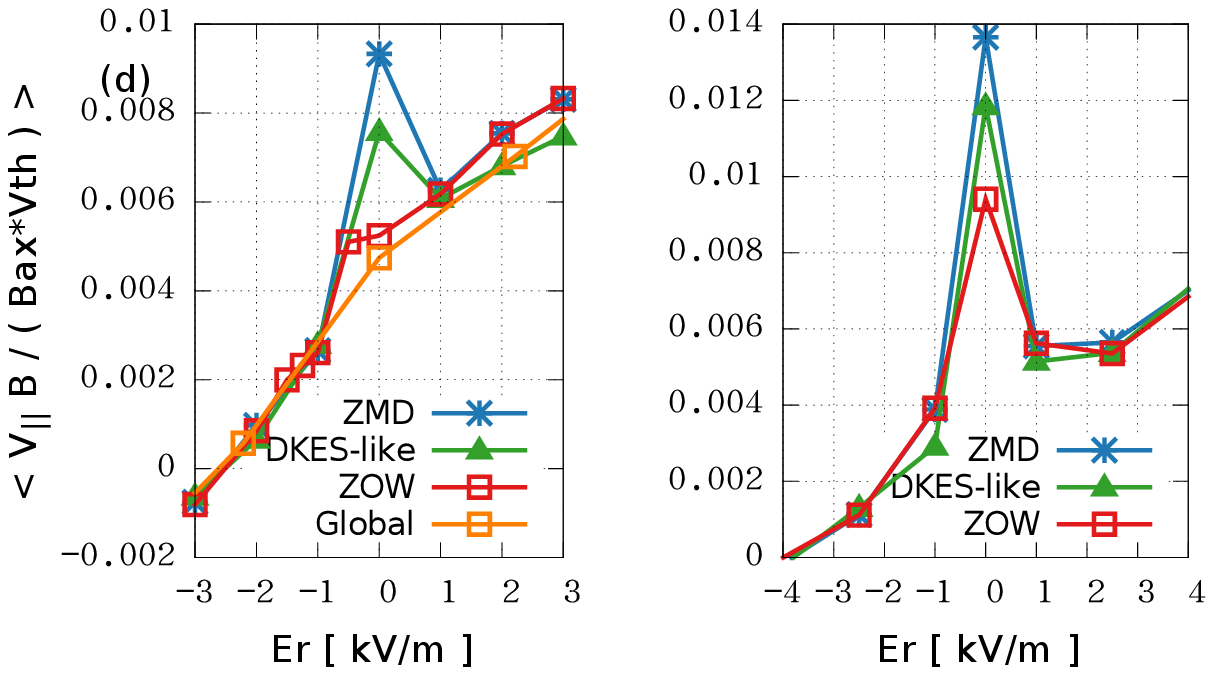}  
 \caption{\raggedright Ion parallel flow of (a) LHD, (b) W7-X, and (c) HSX, respectively. (d) presents the enlarged details around $E_r \sim 0$ for the LHD and W7-X cases. The vertical axis represents the parallel Mach number $M_\parallel$ as defined in Eq. \eqref{eq:m_para}.}
 \label{fig:vb}
\end{figure}

The radial electric field $E_r$ is given as a parameter in this series of investigations. 
In Figs. \ref{fig:p-flux} and \ref{fig:q-flux}, the ion radial particle and energy fluxes among the different approximations are presented on LHD, W7-X, and HSX,  respectively.
The figures of parallel flow simulation are shown in Fig. \ref{fig:vb}.
The global simulations are carried out for LHD only because the global simulation requires much more computational resources than the local to reach a steady-state solution of $\langle V_{i,\parallel}B\rangle$. 
In Figs. \ref{fig:p-flux} - \ref{fig:vb}, the good agreements appear among the local models in $\Gamma_i$, $Q_i$, and $\langle V_{i,\parallel}B\rangle$ if the radial electric field amplitude is moderate in terms of the poloidal Mach number, that is, $0\ll |M_p|\ll 1$.

Let us first focus on the difference which appears on the neoclassical fluxes at large-$E_r$ values.
When the amplitude of $E_r$ rises, the discrepancies increase between DKES-like and the other local models. 
As shown in Figs. \ref{fig:p-flux}(a), \ref{fig:q-flux}(a), and \ref{fig:vb}(a), the LHD radial and parallel fluxes of the ZOW, ZMD, and GSRAKE models agree with the global model well. 
Thus, the discrepancies comes from the incompressibility approximation of the $\boldsymbol E \times \boldsymbol B$ drift on  DKES-like according to Eq.\eqref{eq:imcompressible}. 
According to Figs. \ref{fig:p-flux}-\ref{fig:vb}, the $\boldsymbol E\times \boldsymbol B$ compressibility effect is expected to be significant when $|M_p|>0.4$.

The $E_r$-dependence of $\Gamma_i$, $Q_i$, and $\langle V_{i,\parallel}B\rangle$ found in the HSX case need more explanations. 
%First, in Fig.\ref{fig:p-flux}(d), except for the original DKES model, the other models including the DKES/PENTA model agreement when the poloidal Mach number is less than $1$, $\mathcal{M}_{p} \ll 1$.
First, in Fig.\ref{fig:p-flux}(d), all the cases, except for the original DKES, show a good agreement.
%when the poloidal Mach number is less than $1$, $\mathcal{M}_{p} \ll 1$.
The disagreement between the DKES model and the others is also found in the ion energy flux Fig. \ref{fig:q-flux}(c) and parallel flow Fig. \ref{fig:vb}(c).
Recall that our DKES-like simulation uses the collision operator which ensures the conservation of parallel momentum in ion-ion collisions.
The simulation result suggests that the momentum conservation property of the collision operator is essential for neoclassical transport calculation on quasi-symmetric devices like HSX. 
Secondly, as $E_r$ increases, the neoclassical fluxes of all the models disagree with one another. 
As in the LHD and W7-X cases, the $\boldsymbol E\times \boldsymbol B$ compressibility is supposed to be the main cause of the disagreement.
However, it should be pointed out that the ion parallel flow in HSX becomes supersonic at $\mathcal{M}_p > 1$ as shown in Fig.\ref{fig:vb}. 
Here, the parallel Mach number is defined as 
\begin{equation}\label{eq:m_para}
\mathcal{M}_\parallel\equiv \frac{\langle V_{\parallel}B\rangle}{v_{th}B_{ax}}.
\end{equation}
%where $B_{ax}$ is the magnetic field strength on the magnetic axis.
In the paper, the drift-kinetic models are constructed under the assumption $\mathcal{M}_{\parallel} \ll 1$ because we just takes the zeroth order distribution as the Maxwellian without the mean flow. 
See Eqs.\eqref{eq:fff} and \eqref{eq:f_M_local}.
The parallel flow dependence on $E_r$ in HSX is contrastive to that in W7-X, in which parallel mean flow remains very slow compared to thermal velocity, as in Fig. \ref{fig:vb}(b).
%This is because the neoclassical viscosity does not contribute to damp the plasma flow in quasi-symmetric devices like HSX and tokamaks\cite{Ivan_2015}\cite{Helander_2008}. 
%Consequently, the large flow in the direction of symmetry, and the large parallel mean flow, may appear. 
Both HSX and W7-X configurations aim at reducing radial neoclassical flux. 
%However, W7-X is not a quasi-symmetric system, and its magnetic configuration is chosen to reduce the parallel neoclassical flow, too. 
However, the magnetic configuration of W7-X is chosen to reduce the parallel neoclassical flow, too.
This leads to the different dependence of parallel flow on $E_r$ in these two devices.
Note also that $T_e \gg T_i$ in HSX \cite{Lore_2010} while $T_i = T_e$ in LHD and W7-X cases.
In such a $T_e\gg T_i$ plasma, $\mathcal{M}_p$ of $ \boldsymbol E\times \boldsymbol B$ flow by ambipolar-$E_r$ can be $\mathcal{O}(1)$ because of the slow ion thermal velocity $v_{th,i}$. 
For example, under the ambipolar condition, $\mathcal{M}_p \simeq -0.015$ and $E_{r} \simeq -1.73$ kV/m on LHD by GSRAKE, while $\mathcal{M}_p \simeq 0.95$, and $E_{r} \simeq 3.47$ kV/m on HSX by DKES/PENTA.
Such a large $M_p$ with the quasi-symmetric configuration of HSX results in $\mathcal{M}_{\parallel} \sim\mathcal{O}(1)$. 
When $\mathcal{M}_{\parallel}> 1$, all the drift-kinetic models violate the assumption of the slow-flow-ordering.
Therefore, although $\mathcal{M}_p\sim \mathcal{O}(1)$ $\boldsymbol E\times \boldsymbol B$ flow is allowed in ZOW and ZMD models,
the validation of the drift-kinetic models at $\mathcal{M}_{\parallel} \sim \mathcal{O}(1)$ has to be reconsidered by taking account of the centrifugal force and potential variation along the magnetic field lines\cite{sugama-horton_pop1997b}. This problem is beyond the scope of the present study.

%\clearpage
%%%%%%%%%%%%%%%%%%%%%%%%%%%%%%%%%%%%%%%%%%%%%%%%%%%%%%%%%%%%%%%%%%%%%%%%%%%%%%%%%%%%%%%%%%%%%%%%%
\subsection{Effect of Magnetic Drift and Collisionality} \label{sec:magnetic_drift} 
Let us turn to the simulations around $E_r=0$.  
In Figs. \ref{fig:p-flux}-\ref{fig:q-flux},  there are the very large peaks of $\Gamma_i$ and $Q_i$ at $E_r=0$ in the LHD and W7-X cases by the ZMD and the DKES-like models. 
On the contrary, the global and the ZOW models show the reduction of radial fluxes at $E_r\simeq 0$ and the peaks shift to negative-$E_r$ side. 
This tendency, which has been found in the previous study\cite{Matsuoka2015}, greatly modifies the neoclassical transport in $1/\nu$-regime, especially in the LHD case.
For the HSX case, however, such a peak at $E_r=0$ is not found in the ZMD and DKES-like models. 
What causes the reduction of $\Gamma_i$ and $Q_i$ in ZOW, and what makes the configuration dependence?
Here we consider the problem by analytical formulation.

In a simple stellarator/heliotron magnetic configuration like LHD, the amplitude of the magnetic field is given approximately
\begin{equation} \label{eq:B_approx}
  | \boldsymbol B | \approx B_{0} [ 1 - \epsilon_{h} \cos ( l \theta - m \zeta ) -  \epsilon \cos \theta ],
\end{equation}
where ${\epsilon}_{h}$ and $\epsilon$ are helical and toroidal magnetic field modulations, respectively. 
$l$ is the helical field coil number and $m$ is the number of toroidal periods.
Once a particle is trapped by the helical ripples, its orbit drifts across the magnetic surface and contributes to the radial flux. 
The estimation of particle flux is roughly given as\cite{miyamoto2012plasma} 
 \begin{equation}\label{eq:nVr}
 \Gamma \sim -\left\langle  \int \frac{ \nu_{ \text{eff} } }{ ( \nu_{ \text{eff} })^2 + ( \omega_{h} + \omega_{E} )^2 } V_{\perp}^2 \frac{ \partial f_{M} }{ \partial r } d^3v\right\rangle .
\end{equation}
Here, $\nu_{ \text{eff} }$ is the effective collision frequency of trapped particle and defined as $\nu_{ \text{eff} } \equiv \nu/ \epsilon_{h}$.
$\omega_h$ and $\omega_E$ represent the poloidal precession frequency of the trapped particles by the magnetic drift and $\boldsymbol E\times \boldsymbol B$ drift, respectively. 
$V_\perp$ denotes the radial drift velocity.
For trapped particles, they are estimated as \cite{Beidler_PPCF1995}
\begin{equation}
\begin{split}\label{eq:vperp}
V_\perp\sim \frac{v_d}{\epsilon_tB_0}\frac{\partial B}{\partial \theta}&\sim v_d\frac{\epsilon}{\epsilon_t},\\
\omega_h\sim \frac{v_d}{\epsilon_tB_0}\frac{\partial B}{\partial r},& \quad\omega_E\sim \frac{E_r}{rB_0},
\end{split}\end{equation}
where $v_d\equiv {\mathcal{K}}/{eB_0R_0}$ and $\epsilon_t=r/R_0$.
If $( \nu_{ \text{eff} })^2 \gg ( \omega_{h} + \omega_{E} )^2$, then Eq.\eqref{eq:nVr} indicates that the particle transport is inversely proportional to the collision frequency. The diffusion coefficient in $1/\nu$-regime is approximated as\cite{Beidler_PPCF1995}
\begin{equation} \label{eq:diffusion_coefficient}
 D_{h}  \approx{ \epsilon_{h} }^{1/2} ( \Delta_{h} )^2 \nu_{ \text{eff} }
 \sim  { \epsilon_{h} }^{3/2} \bigg ( \frac{ T }{ e B_0 R_0 } \frac{ \epsilon }{ \epsilon_t }\bigg )^2 \frac{1}{ \nu },
\end{equation}
where $\Delta h=V_\perp/\nu_{\text{eff}}$ is the estimation of the radial step size of helically trapped particles.
Approximating $\omega_h \rightarrow  0$ in Eq. \eqref{eq:nVr} corresponds to ZMD and DKES models.
Then, around $E_r=0$, $\Gamma_i$ shows $1/[\nu_{\text{eff}}(1+x^2)]$-type dependence where $x=(\omega_E/\nu_{\text{eff}})^2$. 
The $\omega_E$ is common for all the particles on a flux surface so that it makes a strong resonance at $\omega_E=0$. 
Once the finite $\omega_h$ is considered, the  peak of $\Gamma_i$ appearing at the poloidal resonance condition $\omega_h+\omega_E=0$ becomes blurred because of $\omega_h$ dependence on $\boldsymbol{v},\theta,$ and $\zeta$. 
This explains the difference between the ZOW and the ZMD models in the LHD case.

\begin{figure}   \includegraphics[width=0.45\textwidth]{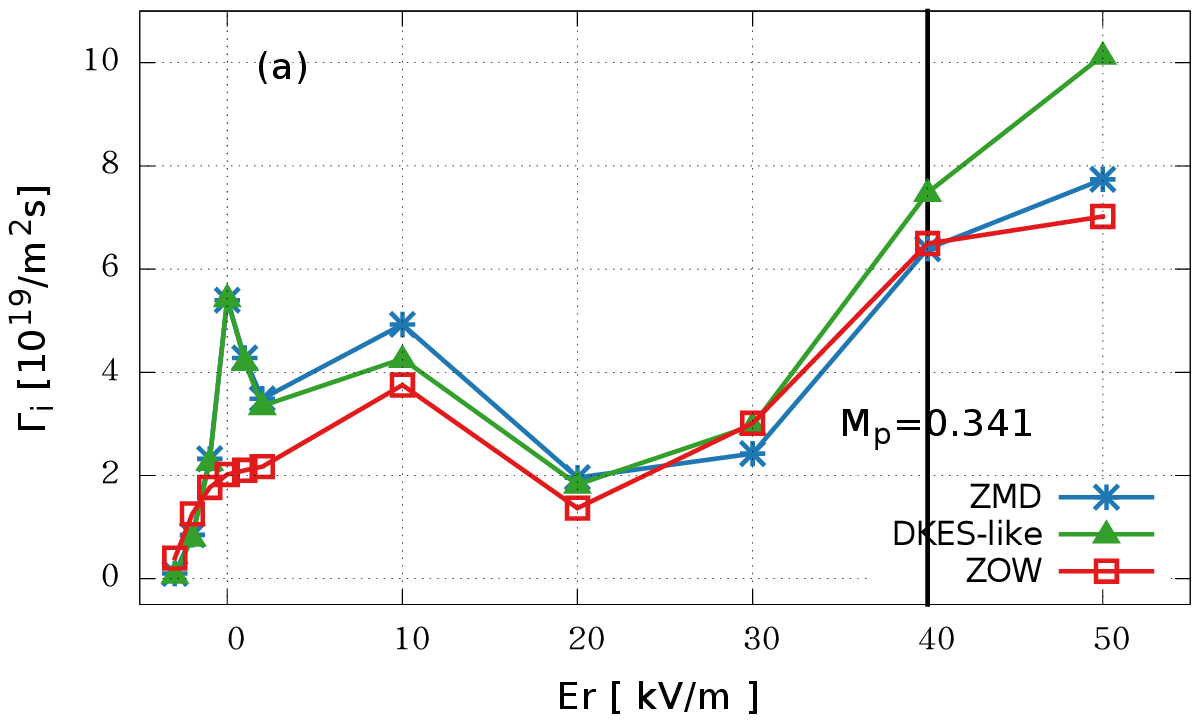} 
   \includegraphics[width=0.45\textwidth]{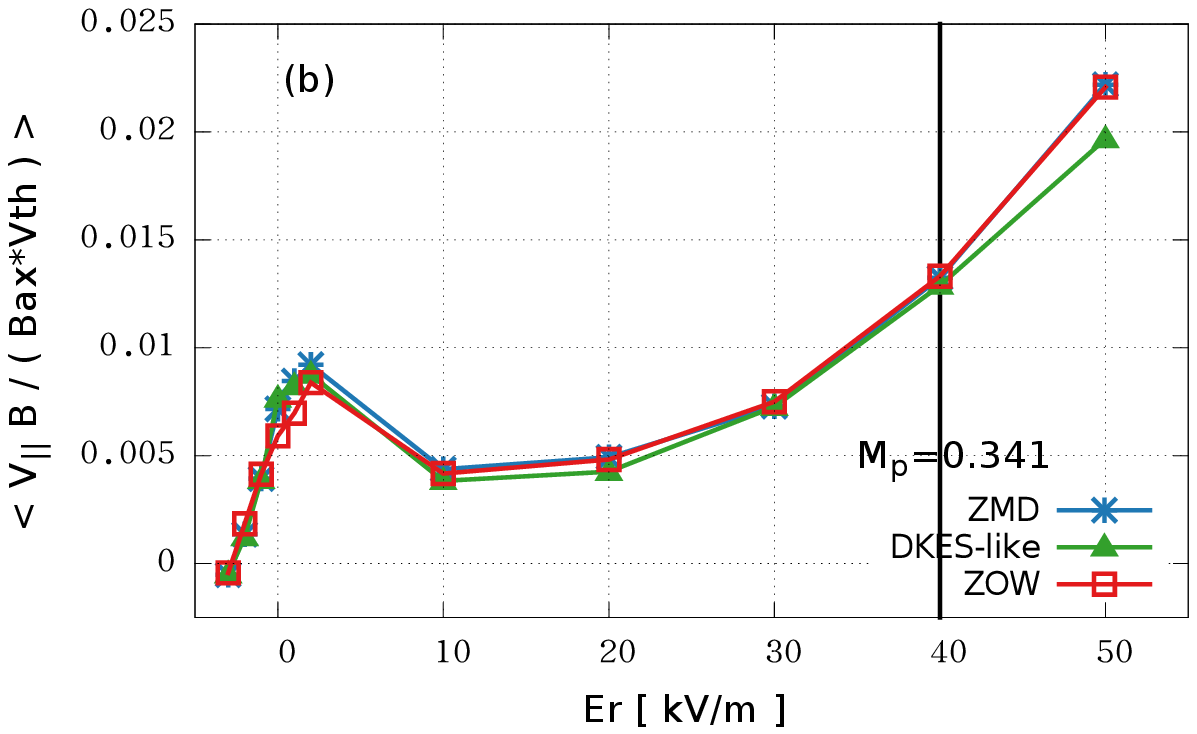} 
 \caption{\raggedright (a) The radial particle flux and (b) ion parallel flow of high collision frequency test of LHD. The normalized collision frequency is 10 times higher than $\nu^*$ on LHD in Table \ref{tb:para}.}
 \label{fig:LHD_n10}
 \end{figure}

The analytic model of the $1/\nu$-type diffusion infers that the strong resonance of trapped-particles at $E_r=0$ in ZMD and DKES-like models is damped by Coulomb collisions.
To demonstrate this, the 10 times larger density simulations are carried out for the LHD case as shown in Fig. \ref{fig:LHD_n10}. 
It is found that the strong peak in $\Gamma_i$ and $\langle V_{i,\parallel}B\rangle$ at $E_r=0$ in ZMD and DKES-like calculations are diminished, and the difference from the ZOW result is small.
It is concluded that the tangential magnetic drift  is more 
important for neoclassical transport calculation in the lower collisionality case and when $|\omega_E| < |\omega _h|$.

Quasisymmetric HSX can be regarded as the $\epsilon\rightarrow 0$ limit of Eq. \eqref{eq:B_approx}.\cite{HSX_1995}
The bounce-average radial drift $\langle V_\perp\rangle$ vanishes in the quasisymmetric limit $\epsilon/\epsilon_t=0$ so that HSX shows the low radial particle transport at $E_r\simeq0$ as in Fig.\ref{fig:p-flux}(d) in all local models.
The radial flux is of comparable level to that in equivalent tokamaks. 
However, it should be noted that the collisionality of the present HSX case is in plateau-regime. 
Then, the discussion on the radial transport level in HSX using Eq. \eqref{eq:nVr} is inadequate.
Following the previous benchmark study on local neoclassical simulations\cite{Beidler_nf2011}, there are tiny magnetic ripples in the actual HSX magnetic field made by the discrete modular coils, which causes $1/\nu$-type diffusion coefficient at very low-collisionality, $\nu_{PS}^*<10^{-3}$, in the DKES calculation.
Therefore, we benchmarked the local drift-kinetic models in HSX with $100$ times smaller plasma density 
($\nu_{PS}^*\simeq 1.0\times 10^{-3}$) at $E_r=0$. 
The results are shown in Table \ref{tb:HSX}.
The radial flux in very low-collisionality regime in HSX shows discrepancy among ZOW, ZMD, and DKES-like models, as found in the LHD and W7-X cases. 
Though the $1/\nu$-regime appears from lower $\nu^*$ value in HSX than LHD, the effect of the tangential magnetic drift on neoclassical transport appears in the same way.

Concerning the W7-X case, the magnetic field spectrum is much more complicated than the simple model Eq. \eqref{eq:B_approx}. It is generally expressed in a Fourier series as follows:
\begin{equation}
  B  ( \psi,\theta, \zeta  ) = B_{0} \sum_{ m, n } b_{m,n}  ( \psi  ) \cos ( m\theta - 5n \zeta ).
\end{equation}
Compared with LHD, W7-X has good modular coil feasibility to adjust $b_{m,n}$ \cite{Grieger1992} where the helical $b_{1,1}$ and toroidal $b_{1,0}$ magnetic field modulations are equal to $\epsilon_h$ and $\epsilon$ respectively in Eq.\eqref{eq:B_approx}. 
One of the neoclassical optimizations is performed by the reduction of average toroidal curvature $b_{1,0}/\epsilon_t\sim 0.5$\cite{Gieger_PPCF(2015)_014004} compared to that in LHD, $\epsilon/\epsilon_t\simeq 1$. 
According to Eq. \eqref{eq:diffusion_coefficient}, this partially explains the smallness of $1/\nu$-regime transport in W7-X.
However, the magnetic spectrum of W7-X contains other Fourier components which are comparable to $b_{1,0}$ and $b_{1,1}$.
Thus, the simple analytic model, such as Eqs. \eqref{eq:B_approx} and \eqref{eq:diffusion_coefficient}, is insufficient to explain its optimized neoclassical transport level.

\begin{figure}
   \includegraphics[width=0.45\textwidth]{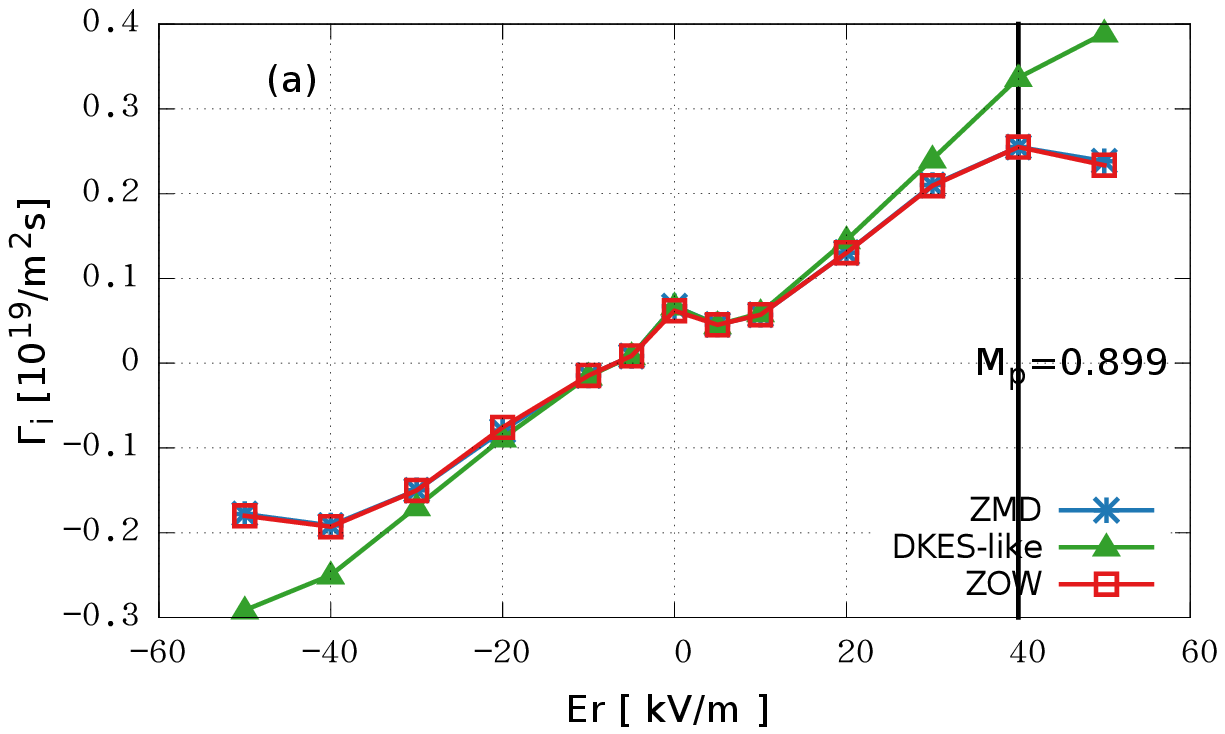} 
   \includegraphics[width=0.45\textwidth]{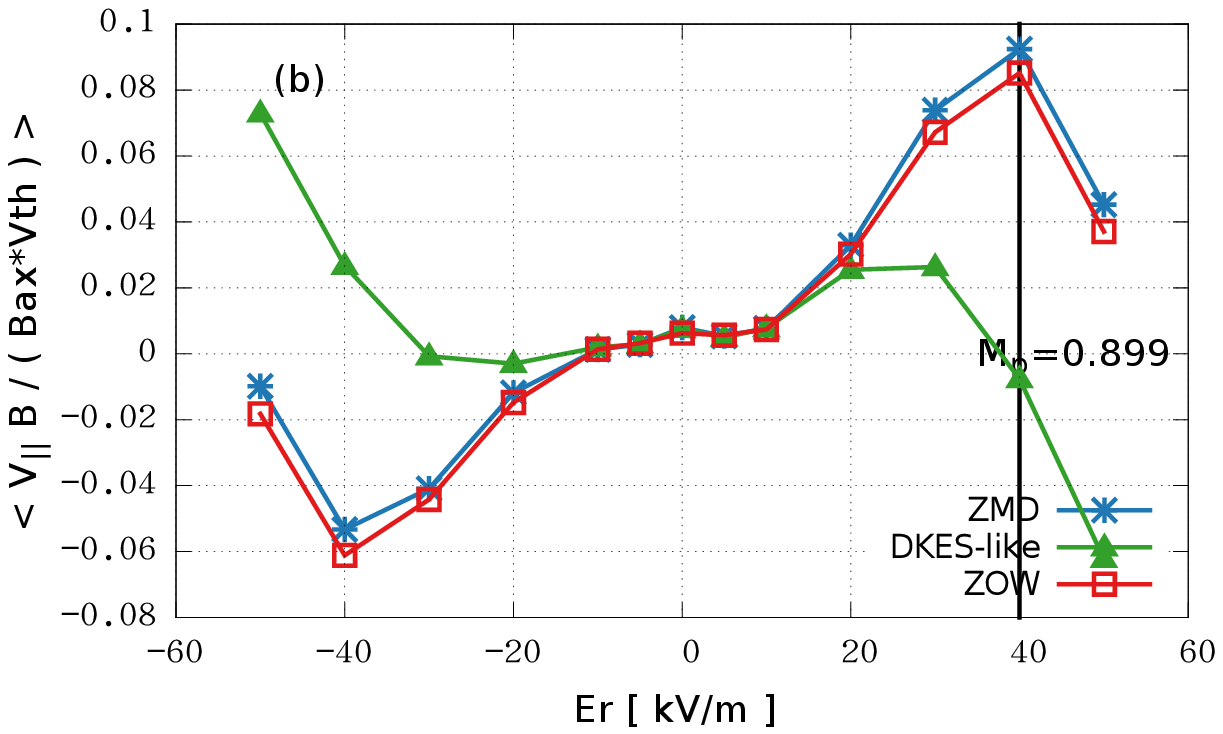} 
 \caption{\raggedright (a) The radial particle flux and (b) ion parallel flow of higher collision frequency test on W7-X. The normalized collision frequency is 10 times higher than $\nu^*$ of W7-X in Table \ref{tb:para}. }
 \label{fig:W7X}
\end{figure}

The quasi-isodynamic concept of the neoclassical optimized stellarator configuration is as follows: the trapped particles in the toroidal magnetic mirrors $b_{0,1}$ precess in the poloidal direction while their radial displacements are small and return to the same flux surface after they circulate poloidally.
The trapped particle trajectory in quasi-isodynamic W7-X configuration has been analyzed using the second adiabatic invariant\cite{Gori_Lotz_Nuhrenberg-ISPP17}
\begin{align}
\mathfrak{J}_{\parallel} & 
=   \int dl~ v_{\parallel} 
= \int d\zeta ~ \frac{ \sqrt{ ( 2 \mathcal{K} - 2 \mu B )/m } }{ \boldsymbol B \cdot \nabla \zeta } 
\nonumber \\
& \propto \int d\zeta \frac{ \sqrt{B_{ref}-B} }{B^2} ,
\end{align}
where $B_{ref}$ represents the magnetic field strength at the reflecting point of a trapped particle. 
Deeply-trapped particles move along the $\mathfrak{J}_{\parallel}=$const. surfaces. Then, if the constant-
$\mathfrak{J}_{\parallel}$-contours on a poloidal cross-section are near a flux-surface function and if the contours are closed, the radial transport  of the trapped particles are suppressed. 
However, the standard configuration, which we investigate, is not fully optimized as is the quasi-isodynamic configuration. 
The $\mathfrak{J}_{\parallel}=$ constant surfaces in the standard configuration have small deviation from the flux surfaces\cite{Gori_Lotz_Nuhrenberg-ISPP17}.
Therefore, in the limit $\omega_E+\omega_h=0$, the deeply-trapped particles drift radially along the $\mathfrak{J}_{\parallel}$ contours.
Consequently, the radial flux in W7-X solved with ZMD and DKES-like models shows the strong peak at $E_r=0$.
As expected from the form of Eq. \eqref{eq:nVr}, either by increasing the collision frequency or by taking account of finite $\omega_h$ as in the ZOW model results in decreasing the radial transport at $E_r=0$. 
In Fig. \ref{fig:W7X} we have examined the radial and parallel flux in 10 times larger density W7-X plasma than those in Figs. \ref{fig:p-flux}(c) and \ref{fig:vb}(b).
As found in the LHD case, the difference among the ZOW, ZMD, and DKES-like models at $E_r=0$ diminished in the higher collisionality W7-X case.  
It is worthwhile to note that it has already been pointed out that the improvement of collisionless particle confinement in W7-X configuration is realized not only in quasi-isodynamic geometry but also by enhancing the poloidal magnetic drift in finite-$\beta$ W7-X plasma because $\partial b_{0,0}/\partial r \propto \omega_h$ increases as the plasma-$\beta$.\cite{Yokoyama_NF2001}

Concerning the parallel flows, Fig.\ref{fig:vb} shows that all models agree with each other well at $ 0 \ll \mathcal{M}_p \ll 1 $.
Compared to the radial flux, the magnetic drift does not influence the parallel flow strongly at $E_r \sim 0$, even in the 
low-collisionality LHD and W7-X cases. 
%On the other hand, the $\boldsymbol E \times \boldsymbol B$ compressibility strongly affects the calculations of parallel flow at large-$|E_r|$.   
On the other hand, the discrepancies of parallel flows at large-$|\mathcal{M}_p|$ appear as clearly as that of the radial flux.

In the simulations, steady-state solution of parallel flow is obtained when the parallel momentum balance relation Eq.\eqref{eq:m1-parallel_f1} is satisfied. 
As explored in Sec. \ref{sec:parall_monentum_balance}, in the parallel momentum balance relation, the differences among the drift-kinetic models includes four parts:
(1) the explicit difference of the tangential drift velocities in $\langle  \boldsymbol B \cdot \nabla \cdot \boldsymbol{\Pi}_{2}\rangle$,
(2) the implicit difference of $\langle \boldsymbol B \cdot \nabla \cdot \boldsymbol{P}_{\text{CGL}}\rangle$ through $f_1$,
(3) the extra term Eq.\eqref{eq:extra_viscosity_ZOW} which breaks the symmetry of $\Pi_{2}$ in ZOW, 
and (4) the term \eqref{eq:G_ZOW}\label{sec:incompressibility} related to $\mathcal G = \nabla_z \cdot \dot{\boldsymbol Z}^{\text {ZOW} } \neq 0$. 
$\langle  \boldsymbol B \cdot \nabla \cdot \boldsymbol{\Pi}_{2}\rangle$ in DKES-like and ZMD models do not contain $\hat{\boldsymbol v}_{m}$. 
These models disagree with each other gradually as $E_r$ increases. 
This indicates that the discrepancy between Eqs.\eqref{eq:ZMD_Pi} and \eqref{eq:DKES_Pi} on the compressibility of $\boldsymbol E \times \boldsymbol B$ affects the evaluation of parallel flow. 
Meanwhile, the ZMD and ZOW tendencies are similar in the wide range of $E_r$ in Figs.\ref{fig:vb}.
As a result, the two extra parallel-viscosity terms appearing in the ZOW model do not influence the parallel flow.
In Fig.\ref{fig:vb}(d), there are small peaks at $E_r = 0$.
When $E_r = 0$, the poloidal resonance leads to the extra large radial fluxes in Fig.\ref{fig:p-flux}(a) and \ref{fig:p-flux}(c).
Equations \eqref{eq:p-flux} and \eqref{eq:nVr} suggest that $f_1$ becomes very large at the resonance.
However, the resonance occurs on trapped particles, which cannot contribute to parallel flow.
The influence of resonance is passed to the passing particles via collisions to change the momentum balance through $\langle \boldsymbol B \cdot \nabla \cdot \boldsymbol{P}_{\text{CGL}}\rangle$.
The parallel flows peak at $E_r=0$ is much less than the radial flux peaks because it is driven by this indirect mechanism.

In summary, as long as the collisionality is low enough to present the $1/\nu$-type diffusion
at the condition $|\omega_E|<|\omega_h|$, the ZMD and DKES-like models, which ignore the 
tangential magnetic drift term, tend to overestimate the neoclassical flux at $E_r\rightarrow 0$
in all three helical configurations in this work. 
The ZOW model reproduces the similar trend as the global simulation in which the finite $\omega_h$ term results in reducing the $1/\nu$-type diffusion.
%The unphysical peaky dependence of neoclassical flux in the ZMD and DKES-like models at $E_r=0$ becomes more significant as the plasma collisionality is lower.
The strong poloidal resonance $\omega_E=0$ without $\omega_h$ term in these local models results in the strong modification in the perturbed distribution function $f_1$, and it indirectly affects the evaluation of parallel flow $\langle V_\parallel B\rangle$, too.

%%%%%%table
\begin{table}[]
\centering
\begin{tabular}{ll}
\hline
Model         & $\Gamma$ [$1/m^2s$]  \\  
                  \hline 
ZOW            & $1.72 \times 10^{15}$      \\
ZMD            & $2.15 \times 10^{16}$      \\
DKES-like      & $2.10 \times 10^{16}$      \\ 
\hline 
\end{tabular}
\caption {The particle flux of HSX at $E_r = 0$ with $0.01$ times density than that in Table \ref{tb:para}.}
\label{tb:HSX}
\end{table}

\subsection{Electron Neoclassical Transport and Bootstrap Current}
\begin{figure}
   \includegraphics[width=0.45\textwidth]{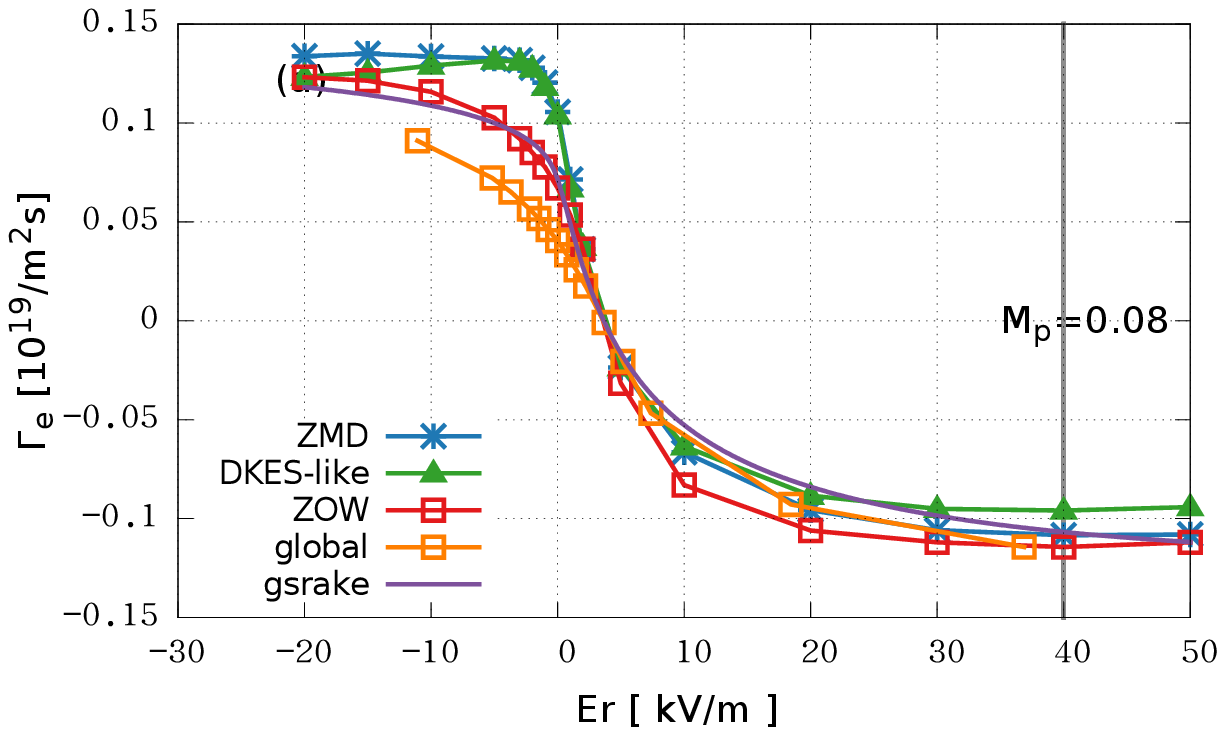}
   \includegraphics[width=0.45\textwidth]{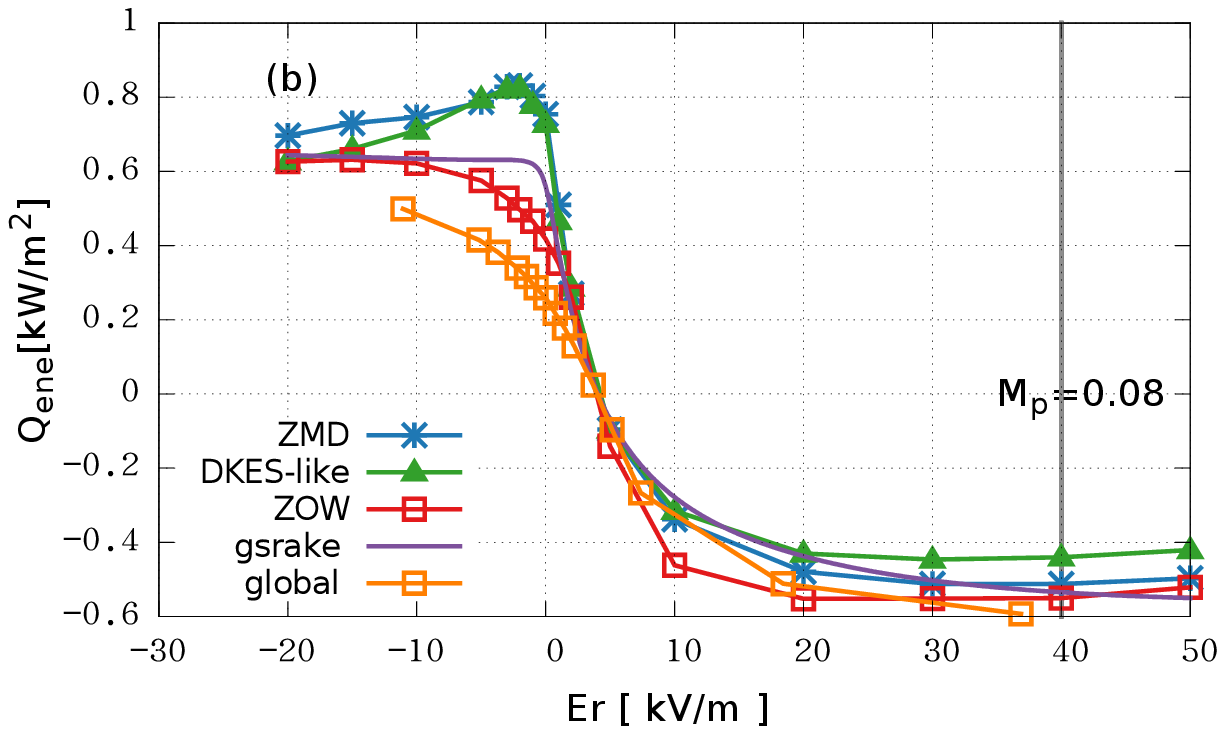} 
   \includegraphics[width=0.45\textwidth]{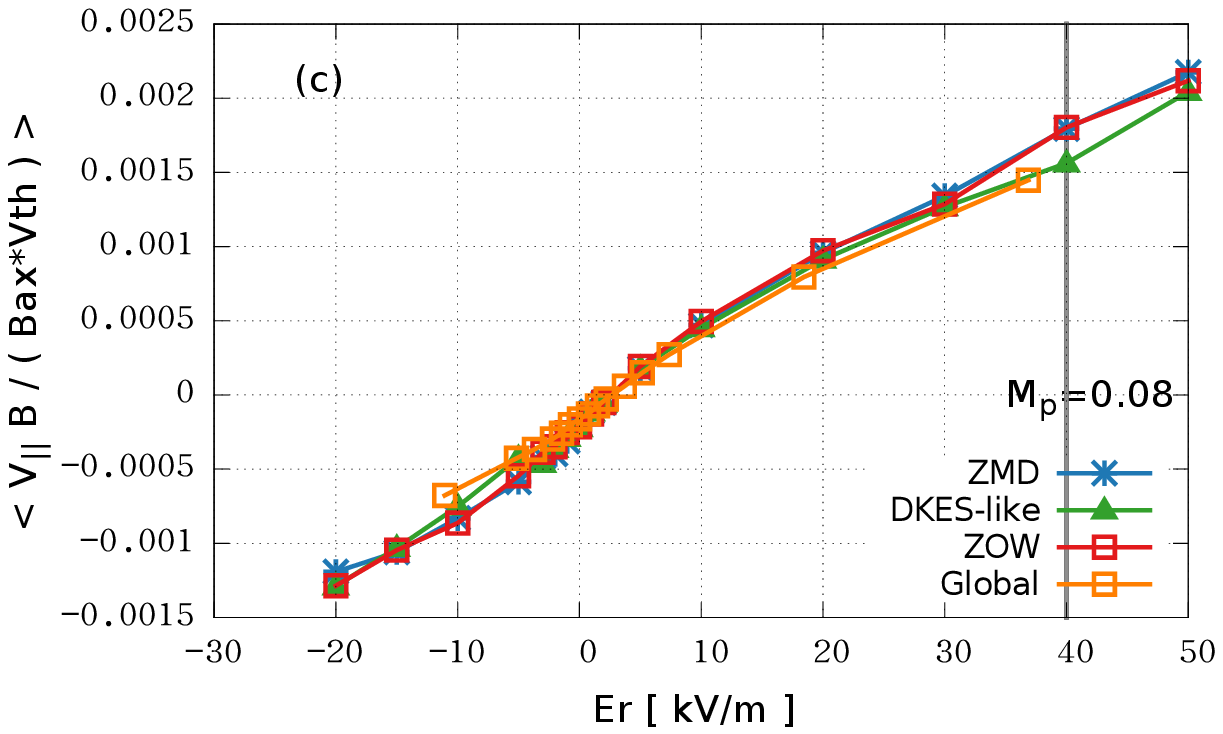}  
 \caption{\raggedright (a) The electron radial particle flux, (b) the energy flux, and (c) the parallel flow in the LHD case which are shown in Table \ref{tb:para}. }
 \label{fig:LHD_e}
\end{figure}

In order to benchmark the bootstrap current calculation at ambipolar condition among the local models, the electron neoclassical transport simulations were carried out for the LHD case. 
The results are shown in Figs. \ref{fig:LHD_e}.
In the entire range of $E_r$, it is found that the differences of  $\Gamma_e$, $Q_e$, and $\langle V_{e,\parallel} B\rangle$ between the two groups, i.e., (global, ZOW) and (ZMD, DKES-like), are smaller than those in the ion calculations.
As the electron thermal velocity is much faster than the ions, the poloidal Mach number for electrons is always regarded as $\mathcal{M}_{p,e} \sim \mathcal{O}(\delta)$. 
Therefore, the $\boldsymbol E\times \boldsymbol B$-compressibility is not important for the electron calculation. 
Moreover, compared with Fig.\ref{fig:p-flux}(a), Fig.\ref{fig:LHD_e}(a) does not present any obviously unphysical peak of the radial particle transport at $E_r \simeq 0$. 
There is the same feature in the energy flux. ( See Fig.\ref{fig:p-flux}(a) and \ref{fig:LHD_e}(b).) 
Even though the normalized collision frequencies $\nu_{*,B}$ (or $\nu_{*,PS}$) are the same in the ion and electron simulations, it seems that the collision effect is stronger in electrons than ions to blur the tangential magnetic drift effect around $E_r=0$. 
Note that the precession drift frequency by the magnetic drift is also the same order between ions and electrons. 
See Eq.\eqref{eq:vperp}.
The difference of the tendency at $E_r\simeq 0$ between ions and electrons is considered as follows.
The collision frequency of particle species $a$ is proportional to \cite{Braginskii_1965}
\begin{equation}
 \nu_{a} \propto \frac{ ( e_a )^4  n_{a} }{ m_a^2  v_a^3 }.
\end{equation}
For the LHD simulations, the temperature is set as $T_e = T_i$.
Thus, the ratio of collision frequency between the electron and ion is 
\begin{equation} \label{eq:nu_00}
 \frac{ \nu_e }{ \nu_i } \propto \bigg ( \frac{  m_i }{ m_e } \bigg )^{ \frac{1}{2} } \gg 1
\end{equation}
because of $Z_i=Z_e=1$ in this work.  
On the other hand, the normalized collision frequency $\nu_{*}(=\nu_{*,PS})$ is defined as 
$\nu_{*,a} \equiv q R_{ax} \nu_{a}/v_{th,a}$.
Therefore, the ratio between the normalized electron and ion collision frequency is 
\begin{equation}\label{eq:normalized_nu_01}
 \frac{ \nu_{*,i} }{\nu_{*,e}} \propto \bigg ( \frac{  m_i }{ m_e } \bigg )^{ \frac{1}{2} } ~ 
 \frac{ v_{th, e} }{ v_{th, i} }.
\end{equation}
Eqs. \eqref{eq:nu_00} and \eqref{eq:normalized_nu_01} suggest that
 $\nu_{e} \gg \nu_{i}$ though $\nu_{*,e} =\nu_{*,i}$.
In Eq.\eqref{eq:nVr}, it is not the normalized collision frequency but the real collision frequency that appears in the form $\nu_{a, \text{eff}}=\nu_{a}/\epsilon_h$.
$\omega_h$ and $\omega_E$ are the same order between ions and electrons so that the ratio between these terms in the denominator of Eq. \eqref{eq:nVr},
\begin{equation*}	
(\omega_h+\omega_E)^2/\nu_{\text{eff}}^2
\end{equation*}
is smaller for electrons than that for ions. 
Therefore, the finite-$\omega_h$ effect in the ZOW  model is not as important for electrons than as for ions.

\begin{figure*}
\centering
\begin{minipage}[b]{.45\textwidth}
   \includegraphics[width=1\textwidth]{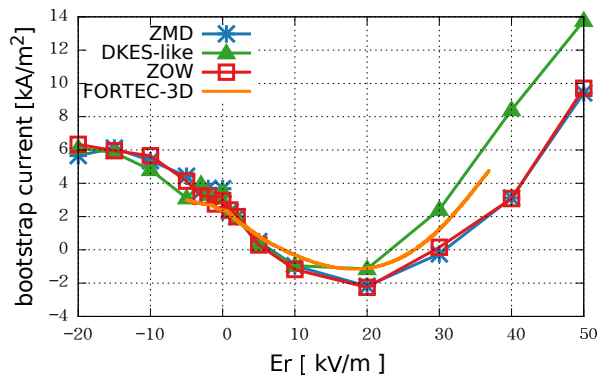} 
 \caption{\raggedright The bootstrap current in the LHD case by combining Fig.\ref{fig:vb}(a) with Fig.\ref{fig:LHD_e}(c).}
 \label{fig:BC}
\end{minipage}\qquad
\end{figure*}

The LHD bootstrap current is investigated among the drift-kinetic models. 
In Fig.\ref{fig:BC}, the bootstrap current is estimated by ion and electron parallel flows as
\begin{equation}
 J_{BC} = e ( Z_i\langle v_{i, \parallel} B \rangle n_{i} - \langle v_{e, \parallel} B \rangle n_{e}) /B_{ax}.
\end{equation}
It is found that the discrepancy of bootstrap current among the models increases when $E_r$ rises.
This indicates that the gap mainly comes from the effect of $\boldsymbol E \times \boldsymbol B$ compressibility on the ion parallel flow as it is found in Fig.\ref{fig:vb}(a). 
The local drift-kinetic models are divided into two groups, DKES-like and the others.
In the following discussion, the two extra terms in the ZOW model, Eqs.\eqref{eq:extra_viscosity_ZOW} and \eqref{eq:G_ZOW}, are ignored because it is found that the difference caused from these two terms is negligible among the ZMD and the ZOW models. 
Neglecting the $ n e E_{\parallel} B $ term in Eq.\eqref{eq:m1-parallel_f1}, the parallel momentum balance in a steady-state is written as
\begin{align} %\label{eq:tensor_and_friction01}
   \langle  \boldsymbol{B} \cdot \nabla \cdot ( \boldsymbol{P}_{\text{CGL}} + \boldsymbol{\Pi}_{2} ) \rangle_a
     = \langle B  F_{a,\parallel} \rangle.
\end{align}
The friction $ F_{\parallel}$ is estimated as follows:
For ion, the friction between ions and electrons is ignored because of large mass ratio.
And, the parallel momentum balance depends only on $\boldsymbol{P}_{\text{CGL}}$ and $ \boldsymbol{\Pi}_{2}$:
\begin{align}\label{eq:tensor_and_friction01}
   \langle  \boldsymbol{B} \cdot \nabla \cdot ( \boldsymbol{P}_{\text{CGL}} + \boldsymbol{\Pi}_{2} ) \rangle_{i}  = 0.
\end{align}
For electrons, not only the viscosity but also the electron-ion parallel friction $F_{ei,\parallel}$ are considered.
And the parallel friction is approximated by
\begin{equation} \label{eq:friction_model}
 \langle F_{ei,\parallel} \rangle \simeq \nu_{ei,\parallel} m_{e} n_{e} ( V_{i,\parallel} - V_{e,\parallel}  ),
\end{equation}
where $\nu_{ei,\parallel}$ is the parallel momentum-transfer frequency. 
The friction force acting on ions is ignored, $F_{ie,\parallel}=-F_{ei,\parallel}$ so that the total parallel momentum is not conserved in the simulation. 
Moreover, as explained in section \ref{sec:delta_f_scheme}, the electron-ion collision in the simulation is simplified by the pitch-angle scattering operator Eq.\eqref{eq:pitch-angle_scattering} where ion mean flow is ignored.
Therefore, in the present simulation models, the electron parallel momentum balance is approximated as
\begin{equation}\label{eq:tensor_and_friction02}
   \langle  \boldsymbol{B} \cdot \nabla \cdot ( \boldsymbol{P}_{\text{CGL}} + \boldsymbol{\Pi}_{2} ) \rangle_{e}  
   = -\nu_{\parallel,ei} m_{e} n_{e} V_{e,\parallel} B.
\end{equation}
In Eqs.\eqref{eq:friction_model} and \eqref{eq:tensor_and_friction02},  
the viscosity $\boldsymbol \Pi_{2}$ is directly influenced by the treatment of the guiding center motion tangential to the flux surface.
See Eqs.\eqref{eq:pi_2_ZOW}, \eqref{eq:ZMD_Pi}, and \eqref{eq:DKES_Pi}.
$J_{BS}$ in the DKES-like model deviates from that in the ZOW and the ZMD model. 
This shows that the incompressible-$\boldsymbol E \times \boldsymbol B$ assumption in $\boldsymbol{\Pi}_{2}$ mainly causes the difference in parallel momentum balance. Meanwhile, the contribution of the tangential magnetic drift $ \hat {\boldsymbol{v}}_{m} $ is minor in the parallel momentum balance equation because the difference is negligible between the ZMD and the ZOW models in Fig.\ref{fig:BC}.
It should be noted that the approximation in the $F_{\parallel, ei}$ in our simulation is valid when $ |V_{\parallel,e}| \gg |V_{\parallel,i}|$. 
Actually, the electron and ion parallel flows can become comparable. 
For a more quantitative evaluation of bootstrap current, the effect should be considered when ion mean flow dominates the bootstrap current, for example, when $J_{BS}$ is at $E_r > 30 kV/m$ in Fig.\ref{fig:BC}. 
The present work is to investigate neoclassical transport among the local drift-kinetic models so that the rigorous treatment of the parallel friction is left for future work.

In this section, the dependence of neoclassical transport on radial electric field was studied. 
The obvious difference appears at $E_r\simeq 0$ or $\mathcal{M}_{p} \sim 1$ among the drift-kinetic models. For the practical application on helical devices, it is important for evaluating the neoclassical fluxes at the ambipolar condition. The LHD ambipolar condition is investigated by searching the $E_r$ value where $Z_i \Gamma_i = \Gamma_e$. 
As shown in Table \ref{tb:ambi}, the ambipolar-$E_r$ values from different models are located between $-2.6$  and $-1.5$  [$kV/m$]. 
The amplitude of electric field, radial flux, and bootstrap current at the ambipolar condition are obtained by the interpolation as shown in Table \ref{tb:ambi}. 
The ambipolar-$E_r$ magnitude of the ZMD model is close to the DKES-like and GSRAKE magnitudes, while the ZOW model predicts closer $E_r$ to the global simulation.
Around the ambipolar condition, the bootstrap current amplitudes are just minor differences among the drift-kinetic models. 
Owing to $T_i \sim T_e$, the ambipolar condition is on the ion-root. In the present case, the finite $E_r$ on the 
ion-root is sufficient to suppress the poloidal resonance but insufficient to make an obvious gap by 
the $\boldsymbol E \times \boldsymbol B$ compressibility. 
The present case does not show any obvious advantage of the ZOW model compared to the other local models. 
If the tangential magnetic drift $\hat{v}_{m}$ increases or if the plasma collisionality is lower, the ZOW model will perhaps
be more reliable than the other models in predicting the ambipolar-$E_r$, bootstrap current, and radial fluxes.
The result of the ZOW model is close to the global simulation values so that the code requires less computation resources than the global.
For example, in the LHD case, the ZOW model takes about $20\%$ computational resources compared to a global calculation with the same number of radial flux surfaces.
In local simulation, one can choose a proper time step size according to the local parameters. 
On the other hand, in a global code, the time step size is a common parameter for all the markers. 
The step size must be small enough to resolve the fast guiding-center motion in the core, but it is much too fine
for the markers in the low-temperature peripheral region.
Another advantage of local simulation is fewer time steps to finish a calculation than a global one.
For a local model, the calculation can be stopped after the time evolution converges on a single flux surface.
For a global model, the calculation has to be continued untill the whole the plasma reaches a steady state. 

\begin{table*}[]
\centering
\begin{tabular}{llllll}
\hline
                  & $E_{r}$ [$kV/m$]          & $\Gamma$ [$10^{19}/m^{2}s$]        & $J_{BS}$  [$kA/m^{2}$]    & $Q_i$[$kW/m^2$]  & $Q_e$[$kW/m^2$] \\ 
                  \hline 
Global    & -2.34        &  0.057             & 2.88     & 0.272  & 0.343 \\ 
ZOW       & -2.59        &  0.089             & 3.23     & 0.614  & 0.515 \\ 
ZMD       & -1.55        &  0.123             & 3.22     & 0.880  & 0.819 \\ 
DKES      & -1.66        &  0.125             & 3.55     & 0.595  & 0.807 \\ 
GSRAKE    & -1.73        &  0.089             & N/A      & 0.519  & 0.628 \\ 
\hline 
\end{tabular}
\caption { The ambipolar conditions obtained from each model in the LHD case.}
\label{tb:ambi}
\end{table*}
%\twocolumngrid

\section{summary}\label{sec:summary}
A series of neoclassical transport benchmarks have been presented among the drift-kinetic models in
helical plasmas. The two-weight $\delta f$ scheme is employed to carry out the calculations of particle flux, energy flux, and parallel flow. The $\delta f$ formulation in this work allows the violation of Liouville's theorem in
a local drift-kinetic approximation as in the ZOW model. 
The treatments of the convective derivative term $(\boldsymbol v_E+\boldsymbol v_m)\cdot\nabla f_{a,1}$ are 
different among the local drift-kinetic models.
For example, the ZOW model maintains the tangential magnetic drift $\hat{ \boldsymbol v }_{m}$ which results in  
the compressible phase-space flow, $ \mathcal{G} \neq 0$. 
On the contrary, in the ZMD and the DKES models, the magnetic drift is completely neglected, but instead the phase-space volume is conserved.
%It is found that the finite $\mathcal{G}$ term effect term is an order of magnitude higher correction in the particle, parallel momentum, and energy balance equations.
The finite $\mathcal{G}$ term in ZOW brings $\mathcal{O}( \delta^2 )$-correction in the particle, parallel momentum, and energy balance equations.
The simulation results have demonstrated that the ZOW and the ZMD models agree with each other well in the wide range of $E_r$ value.
This indicates that the $\mathcal{O}( \delta^2 )$-correction term is negligible in neoclassical transport calculation.
The only exception is around $\boldsymbol v_E\simeq 0$, where the ZMD and DKES-like models show the very large peaks of neoclassical flux. 
Owing to the tangential magnetic drift $\hat{ \boldsymbol v }_{m}$, the ZOW simulation evaluates the radial fluxes and parallel flows around $E_r \simeq 0$ which are much more smoothly dependent on $E_r$ and similar to those obtained from the global calculations.

Effects of the tangential magnetic drift $\hat{v}_{m}$ because stronger under the following conditions.
First, according to the simulations, the tangential magnetic drift $\hat{ v }_{m} $ is more obvious in LHD than W7-X and HSX. 
In W7-X and HSX, the magnetic configuration is chosen so as to reduce the radial drift of trapped particles and remains the neoclassical transport in $1/\nu$-regime.
This reduces the peak value of $\Gamma_i$ at the poloidal 
resonance, $\omega_E+\omega_h=0$ in Eq. \eqref{eq:nVr}, and results in the small gap between the ZMD and the ZOW models in
these machines compared to LHD.
Second, the effect is obvious in the low collisional plasma. 
At $E_r \simeq 0$, the tangential magnetic drift is required to avoid the poloidal resonance. 
Otherwise, the artificially strong $1/\nu$-type neoclassical transport will occur. 
Third, the ZOW, ZMD, and DKES-like models agree with one another in a series of electron simulations. 
The discrepancies occur more clearly on the ions. 
This suggests that the conventional local drift-kinetic models are sufficient for electron simulation. 

The difference in the treatment of the $\boldsymbol E\times \boldsymbol B$ drift term has also been found to cause a large error in neoclassical transport calculation. 
%The gap between the DKES-like and the other models grows with the magnitude of poloidal Mach number when $|M_p|>0.4$.
%The influence of the radial electric field on neoclassical transport is miscalculated in the DKES-like model because the incompressibility of $\boldsymbol E \times \boldsymbol B$ drift is assumed.
%Note here that the poloidal Mach number depends on the particle mass, $\mathcal{M}_{p,a} \propto v_{E}/v_{th,a} \sim \sqrt{ m_{a} }$, %if the temperatures are the same among the particles species.
%The fuels are Deuterium and Tritium in a fusion device. 
%It is possible for Helium from the D-T reaction and Tungsten to escape from the wall. 
%The heavier species has larger $\mathcal{M}_p$ values. 
The assumption of incompressible $E \times B$ drift in the DKES-like model results in the miscalculation of the neoclassical transport for the larger poloidal Mach number of $\mathcal{M}_{p} > 0.4$. Due to the mass dependency of $\mathcal{M}_{p} > \propto v_E/v_{th,a} \sim \sqrt{m_a}$, the heavier ion $\mathcal{M}_{p}$ such as He and W increases.
%Therefore, if the incompressible-$\boldsymbol E\times \boldsymbol B$ approximation is employed, the parameter window of  will be narrower.
Therefore, the parameter window in which the incompressible-$\boldsymbol E\times \boldsymbol B$ approximation is valid will be narrower for heavier species.

Regarding the practical application, the neoclassical flux and bootstrap current are evaluated at the ambipolar condition. 
The ion-root usually exists when $T_i \simeq T_e$; the electron-root appears when $T_i \ll T_e$\cite{yokoyama_CPP_2010}.
The peak of $\Gamma_i$ at $E_r=0$ is an artifact of the ZMD and the DKES-like models.
It suggests that the $T_e/T_i$ is the threshold of transition between the ion-root and the electron-root. 
Therefore,  the magnitude of $T_e/T_i$ will be less/lower in the global and the ZOW models than in the ZMD and the DKES models. 
The neoclassical transport varies drastically if the ambipolar-$E_r$ switches from an ion-root to an electron-root.
Therefore, the introduction of tangential magnetic drift term in a local code plays an importance role for the investigation of the ambipolar-root transition.
Figure \ref{fig:BC} indicates that the $\hat{v}_{m}$ term slightly affects the bootstrap current evaluation.
Furthermore, the sign of the bootstrap current may change when the ambipolar-$E_r$ transits from a negative to a positive root.
This will be also related to the study on the bootstrap current effect on MHD equilibrium. 

On the basis of the present study, the particle flux, energy flux, and bootstrap current of FFHR-d1 will be studied in the future.
The investigation will be carried out by iteration between the MHD equilibrium and the bootstrap current calculations in order to collect data for the design of FFHR-d1.
The FFHR-d1 magnetic configuration is similar to LHD so that the present study on an LHD configuration provides useful insight on the magnetic drift effect on the neoclassical transport in FFHR-d1. 
The effect of the bootstrap current on the MHD equilibrium will play a more important role in FFHR-d1 than that in present LHD operations because the central $\beta$ will be about $ 5 \% $\cite{goto_nf2015}. 

It is found that the $\hat{v}_m$ term does not only decrease the height of the peak of $\Gamma_i$ but also changes the value of $E_r$ at which $\Gamma_i(r,E_r)$ peaks. The approximated amount of the shift in $E_r$ in LHD can be estimated by the bounce-averaged poloidal precession drift\cite{Beidler_2001} of thermal ions as in Eq.\eqref{eq:vperp}.
The bounce-averaged magnetic drift for deeply-trapped particles is approximated as
\begin{equation}\begin{split}\label{eq:bave}
\omega_h & \sim \frac{ v_d }{\epsilon_t B_0 } \frac{ \partial B_{2,10} (\rho) }{ \partial r } 
\left\langle \cos ( m \theta - n \zeta )\right\rangle_b 
\\ & \sim - \frac{ 4v_d}{ a }
\end{split}\end{equation}
where $\rho\equiv r /a$ and $\langle\cdots\rangle_b$ denotes the bounce-average over a particle trajectory trapped 
in a helical magnetic ripple. The radial dependence of the helical component is approximated as $B_{2,10} (\rho) \simeq 2 (a/R_0) B_0 \rho^2 $ according to the tendency found in the MHD equilibrium for LHD plasma. 
In Eq.\eqref{eq:bave}, $(\theta, \zeta) = (0,\pi/10)$ is chosen because this is the bottom position of both toroidal and helical ripples.  Substituting the parameters $B_0, a, \epsilon_t$, and $v_d$ for the LHD case, the shift of the $\Gamma_i$-peak is estimated as
\begin{equation} \label{eq:Er-shift}
   E_r \simeq - \frac{ 4 T_i \rho}{ e_i R_0}
\end{equation}
at which poloidal resonance $\omega_E+\omega_h=0$ occurs.
Eq.\eqref{eq:Er-shift} agrees with the tendency of the peak shift in $\Gamma_i$ from the ZOW and the global models, which are Figs.8-10 in 
Matsuoka et al.\cite{Matsuoka2015} 
Since high-temperature discharge $T_i>10 keV$ is planned in FFHR-d1, it is anticipated that 
the peak of $\Gamma_i$ in the ZOW model will appear more negative-$E_r$ which can be close to the ion-root $E_r$ 
value. In such a case, the difference between the ZOW and ZMD models becomes significant in evaluating the
neoclassical transport level in the ambipolar condition.

%%%%%%%%%%%%%%%%%%%%%%%%%%%%%%%%%%%%%%%%%%%%%%%%%%%%%%%%%%%%%%%%%%%%%%%%%%%%%%%%%
\begin{acknowledgments}
The authors would like to thank Dr. J. M. Garc\'{\i}a Rega\~{n}a for the W7-X configuration data, and Mr. Jason Smoniewski for the DKES and PENTA numerical results of HSX.
The simulations are carried out by Plasma Simulator, National Institute for Fusion Science.
This work was supported in part by Japan Ministry of Education, Culture, Sports, Science and Technology (Grant No. 16K06941) and in part by the NIFS Collaborative Research Programs (NIFS16KNST092 and NIFS16KNTT035).
\end{acknowledgments}
%%%%%%%%%%%%%%%%%%%%%%%%%%%%%%%%%%%%%%%%%%%%%%%%%%%%%%%%%%%%%%%%%%%%%%%%%%%%%%%%%%
\appendix
 
\section{Source and Sink term in FORTEC-3D}\label{AppendixA}
As explained in Sec. \ref{section:particle_flux} and \ref{sec:parall_monentum_balance}, an adaptive source and sink term is introduced in the global and local FORTEC-3D codes.
Thus, the flux-surface averaged density and pressure perturbation from the $f_{1}$ part, which are defined by Eqs. \eqref{eq:n1} and \eqref{eq:p1}, become negligible compared to the background density and pressure, i.e., $\langle\mathcal{N}_{1}\rangle\ll n$ and $\langle P_{1}\rangle \ll  nT$.
Such a source/sink term is constructed according to the following considerations.

First, the source/sink term acts to reduce the flux-surface average perturbations $\langle\mathcal{N}_1 \rangle$ and 
$\langle P_1\rangle$.
It is considered that the source/sink term should \textbf{not} smoothen the spatial variation of them on the flux surface, because the non-uniform distribution reflects the compressible flow on the flux surface.
Therefore, the source-sink term is constructed to reduce $\langle \mathcal{N}_1\rangle$ and $\langle P_1\rangle$, while it maintains the fluctuation patterns on the flux surface, $\mathcal{N}_1- \langle\mathcal{N}_1\rangle$ and ${P}_1-\langle P_1\rangle$. 
Second, the source-sink term should be adaptive. 
The strength of the source-sink term is proportional to $\langle \mathcal{N}_1\rangle$ and $\langle P_1\rangle$ so that the users do not have to control the strength of the source-sink term. 
Third, the source/sink term does not contribute as a parallel momentum source as shown in Eq.\eqref{eq:para_mom_src}, because the steady-state parallel momentum balance can be found without giving an artificial source/sink term.

\begin{figure}   
   \includegraphics[width=0.9\columnwidth]{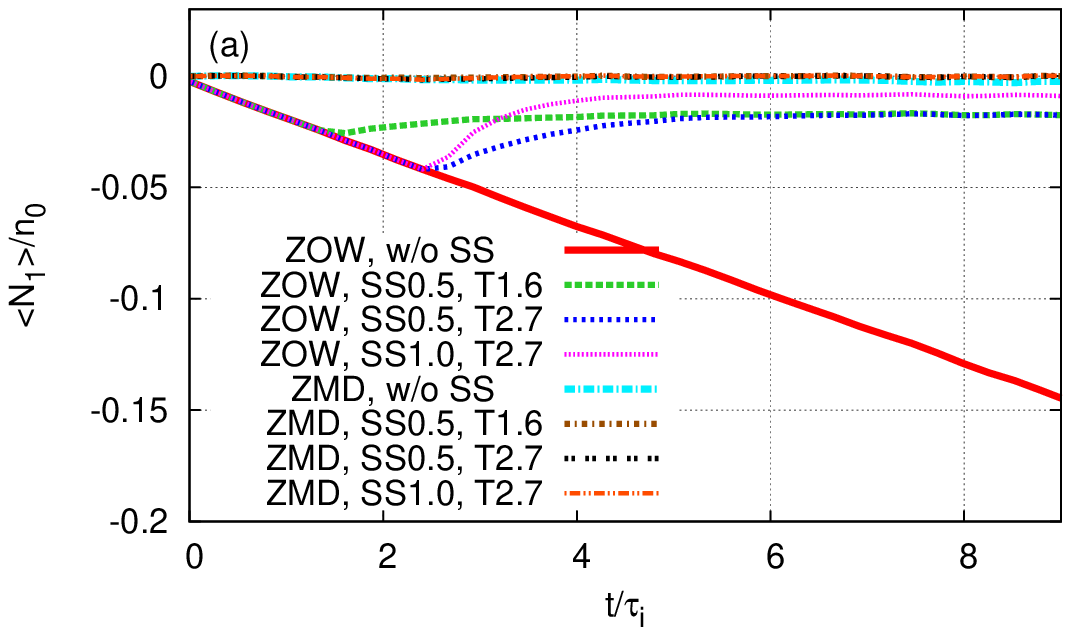} 
   \includegraphics[width=0.9\columnwidth]{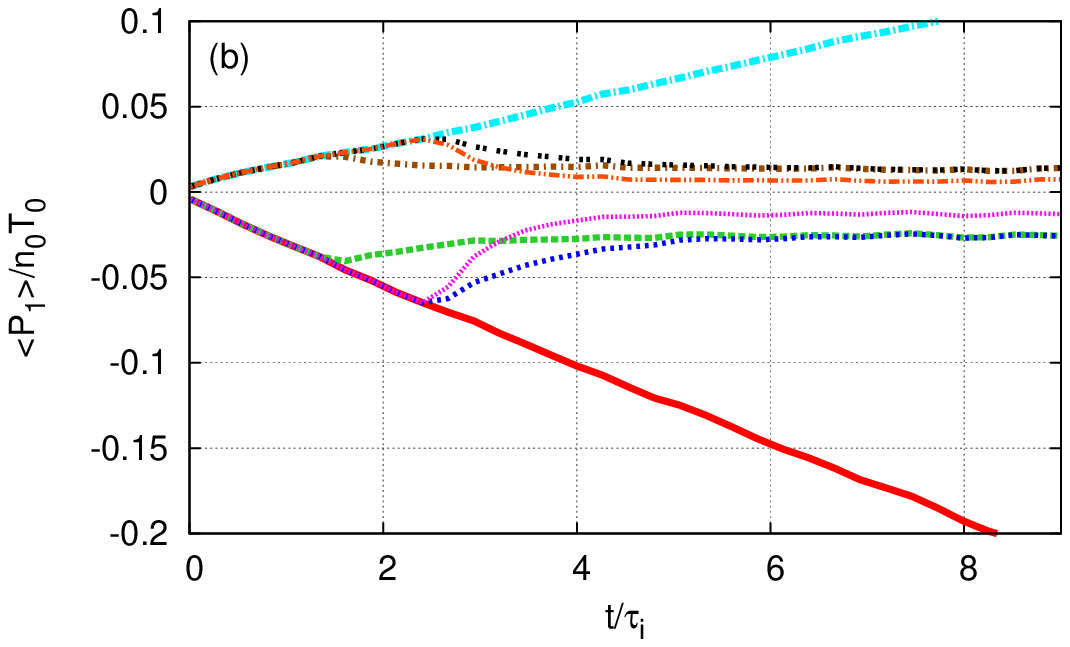} 
   \includegraphics[width=0.9\columnwidth]{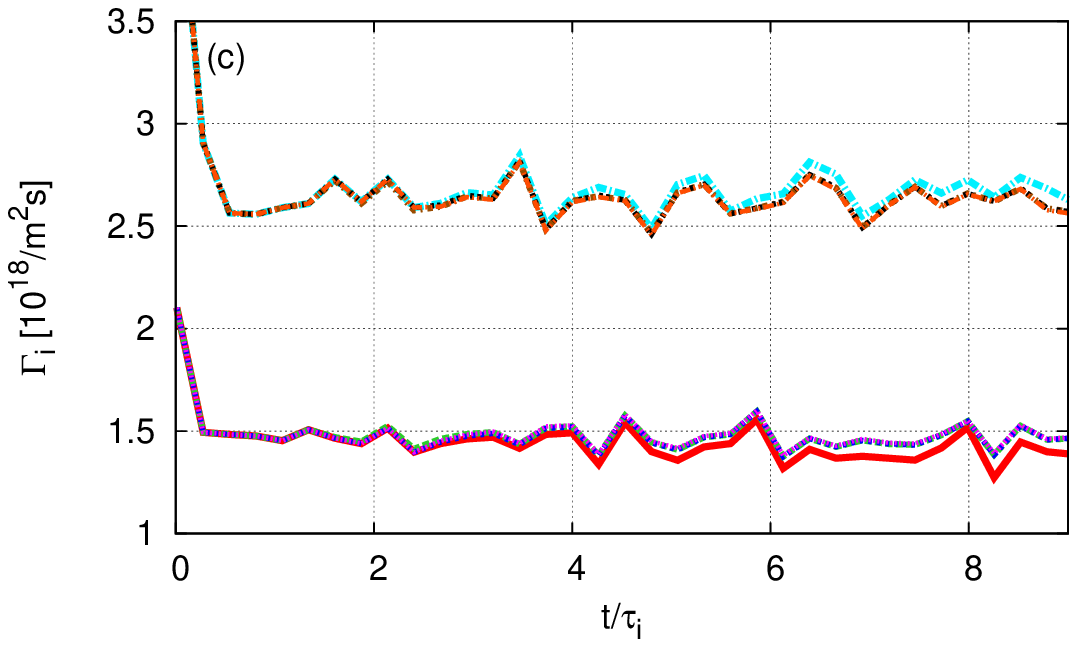} 
   \includegraphics[width=0.9\columnwidth]{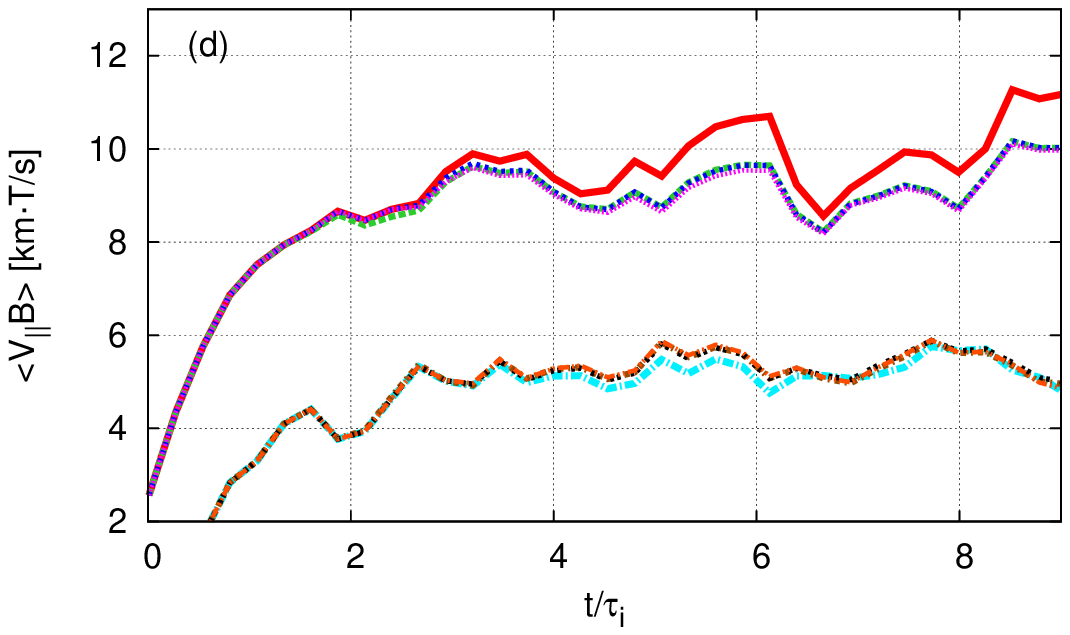} 
 \caption{\raggedright The time evolution of (a) the density perturbation $\langle \mathcal{N}_1\rangle$, (b) the pressure perturbation $\langle P_1\rangle$, (c) the neoclassical particle flux $\Gamma_i$ and (d) the parallel flow $\langle V_\parallel B\rangle$ are the LHD ion case shown in Sec. \ref{sec:large_Er}. Furthermore, Figs.(a) and (b) are normalized by background density and pressure, respectively. In Fig.(d), the parallel flows of the ZMD model are plotted offset by $-4$. The source/sink term is turned on at $t=1.6\tau_i$ or $2.7\tau_i$. The numbers after ``SS'' in the legend indicate the strength of the source/sink term, $\nu_S$.  }
\label{fig:apd1}
\end{figure}

In the drift-kinetic equation for $f_{1}$ \eqref{eq:Df_1/Dt}, the source-sink term $S_1$, which satisfies the conditions explained above, is given in the form $S_1=s(\psi,v,\xi,t)f_M$ with the following constraints:
\begin{eqnarray}\label{eq:ss_constraint}
\int d^3v~ sf_{M}&=&-\nu_S \langle \mathcal{N}_1\rangle,\nonumber\\
\int d^3v~ m_av_\parallel sf_{M}&=&0,\\
\int d^3v~ \frac{m_av^2}{2}sf_{M}&=&- \frac{3}{2} \nu_S \langle P_1\rangle,\nonumber
\end{eqnarray}
where $\nu_S$ is a numerical factor to control the strength of the adaptive source-sink term.
There is arbitrariness to make a source/sink term which satisfies Eq.\eqref{eq:ss_constraint}.  
The examples of the adaptive source/sink terms can be found in the references\cite{Landreman_2014}\cite{Jolliet_NF_2012_023026}. 
In FORTEC-3D code, the source/sink term is implemented by 
diverting the field-particle collision operator  $\mathcal{C}_Ff_M$. The field-particle operator is made 
so as to satisfy the following conservation laws for the like-particle linearized collision term\cite{satake_Comp_Phys_Comm2010},
\begin{eqnarray}\label{eq:ctcf}
\int d^3v~ \mathcal{C}_Ff_M&=&-\int d^3v~ \mathcal{C}_T(f_1),\nonumber\\
\int d^3v~ mv_\parallel \mathcal{C}_Ff_M&=&-\int d^3v~ mv_\parallel\mathcal{C}_T(f_{1}),\\
\int d^3v~ \frac{mv^2}{2}\mathcal{C}_Ff_M&=&-\int d^3v~ \frac{mv^2}{2}\mathcal{C}_T(f_{1}).\nonumber
\end{eqnarray}
By comparing Eqs. \eqref{eq:ss_constraint} and \eqref{eq:ctcf}, one can see that operator $\mathcal{C}_F$ can be 
directly used to implement the source/sink term. 
In FORTEC-3D, the source/sink term is operated in the $(\theta,\zeta)$ cells on a flux-surface which is the same as those prepared for the collision terms. 
In this simulation, $20 \times  10 ( 20 \times 20 )$ cells on a $(\theta,\zeta)$-plane are employed.
The strength of the source/sink term $\nu_S$ is varied case by case because the growth rate of $\langle\mathcal{N}_1\rangle$ and $\langle P_1 \rangle$ depends on the drift-kinetic model, magnetic configuration, and parameters such as $E_\psi$.
See  Eqs.\eqref{eq:dz1_dt} and \eqref{eq:en_ZOW}.
In most cases, the moderate strength $\nu_S=0.5\sim 1.0\times \nu_{i}$ is enough to suppress $\mathcal{N}_1$ and $P_1$ to $\mathcal{O}(10^{-2})$, where $\nu_i$ is the ion-ion collision frequency. 
As demonstrated in Fig. \ref{fig:apd1} for the ZOW and ZMD simulations in the LHD case, it is confirmed that the final steady-state solutions of the neoclassical fluxes are not affected by the strength of the source/sink term nor the timing from when the source/sink term is turned on. 
It is obvious that without the source/sink term the ZMD model does not conserve $\langle P_1\rangle$.
The $\langle\mathcal{N}_1\rangle$ and $\langle P_1 \rangle$ both continue to change in the ZOW model, as expected from the particle and energy balance relations in Sec. \ref{section:particle_flux}. 
In the series of simulations without source/sink, the neoclassical fluxes $\Gamma_i$ and $\langle V_\parallel B\rangle$ continue evolving and one cannot obtain a quasi-steady state solution.
By adopting $\nu_S= 0.5$ or $1.0$, the ZOW and ZMD models both converge to a quasi-steady state at which one can take a time average.
It is observed that the pattern of the fluctuations on the flux surface, $\mathcal{N}_1- \langle\mathcal{N}_1\rangle$ and ${P}_1-\langle P_1\rangle$, are sustained before and after turning on the source/sink term.
This scheme works well in the global, ZOW, and ZMD models. 
For the DKES-like model, the source/sink term is not necessary because it preserves the total particle number and energy ideally. 
However, the weak source/sink was given in the DKES-like model in this work to reduce the numerical error accumulation in $\mathcal{N}_1$ and $P_1$.

\section{Derivation of Viscosity Tensor}\label{AppendixB}
The parallel moment equation is derived from Eq.\eqref{eq:d-k-A-1} with $\mathcal{A} = m v_{\parallel}$,
\begin{align}\label{eq:d-k-A-A1}
 &\frac{\partial}{\partial t} \left( \int d v^3 f m v_{\parallel}  \right) + \nabla \cdot \left( \int d v^3 f m v_{\parallel} \dot{ \boldsymbol X } \right) 
  \nonumber \\
 =& \left( \int d v^3 f m  \dot{ v }_{\parallel} \right) 
 + \left( \int d v^3 f m [ \mathcal{S} + \mathcal{C} ] \right)
  \nonumber \\
 +& \int d v^3 f m v_{\parallel} \mathcal{G}.
\end{align}
With $\dot {v}_{\parallel}$ in the global model, Eq.\eqref{eq:dot_v_para}, we have the following relation
\begin{align}\label{eq:d-k-A-A2}
 \nonumber
 &\nabla \cdot \left( \int d v^3 f m v_{\parallel} \dot{ \boldsymbol X } \right) 
 - \left( \int d v^3 f m  \dot{ v }_{\parallel} \right) 
 \nonumber \\
 & = \nabla \cdot \left( \int d v^3 f m v_{\parallel}^2 \boldsymbol b \right) 
 + \int d v^3 f \boldsymbol b \cdot \left( \mu \nabla B - e_a \boldsymbol E \right)
 \nonumber \\
 & + \nabla \cdot \left( \int d v^3 f m v_{\parallel} \dot{ \boldsymbol X }_{\perp} \right) 
 - \int d v^3 f m v_{\parallel} \dot{ \boldsymbol X }_{\perp}  \cdot \boldsymbol \kappa
 \nonumber \\
 & = \boldsymbol b \cdot \left\lbrace \nabla \cdot \left( \int d v^3 f \left[ m v_{\parallel}^2 \boldsymbol b \boldsymbol b  + \mu B \left( \boldsymbol I - \boldsymbol b \boldsymbol b \right) \right] \right)
 \right\rbrace
 \nonumber \\
 & + \boldsymbol b \cdot \left\lbrace \nabla \cdot \left[ \int d v^3 f m v_{\parallel} \left( \boldsymbol b \dot{ \boldsymbol X }_{\perp} + \dot{ \boldsymbol X }_{\perp} \boldsymbol b \right) \right]
 \right\rbrace
  \nonumber \\
  & + e_a E_{\parallel} \int d v^3 f.
\end{align}
Then, Eq.\eqref{eq:m1-parallel} is obtained by rewriting Eq.\eqref{eq:d-k-A-A1},
\begin{align*}
 &\frac{\partial}{ \partial t } ( n m V_{ \parallel} ) + \boldsymbol{b} \cdot ( \nabla \cdot \boldsymbol{P} ) 
 \nonumber \\
 &= n e_{a} E_{\parallel} + F_{\parallel} 
 + \int d^{3}v ~ \mathcal{S}  m v_{\parallel} 
 + \int d^3v ~ f  \mathcal{G}  m v_{\parallel},
\end{align*}
where
\begin{subequations}
\begin{alignat}{4}
& \boldsymbol P \equiv \boldsymbol P_{CGL} + { \boldsymbol \Pi }_{2},
\\
& \boldsymbol P_{CGL} \equiv \int d^3 v ~ [  ( m v_{\parallel}^2 \boldsymbol b \boldsymbol b + \mu B  ( \boldsymbol I - \boldsymbol b \boldsymbol b ) ] f, \label{eq:P_cgl}
\\
& \boldsymbol \Pi_{2} \equiv \int d^3 v ~ m v_{\parallel} \bigg ( \dot{  \boldsymbol X }_{\perp} \boldsymbol b + \boldsymbol b \dot{  \boldsymbol X }_{\perp} \bigg ) f\label{eq:pi_2}
.
\end{alignat}
\end{subequations}
It should be noted that the $\dot{ \boldsymbol X}_{\perp} \cdot \boldsymbol \kappa$ term in Eq.\eqref{eq:d-k-A-A2} is involved in  the symmetry of the $\boldsymbol \Pi_{2}$ tensor. 
On the other hand, Eq.\eqref{eq:d-k-A-A2} is independent of the explicit form of $\dot{\boldsymbol X}_{\perp}$.

For the ZOW model, the parallel momentum balance equation is calculated with 
${\dot { \boldsymbol X }}_{\text ZOW} = v_{\parallel} \boldsymbol b + \boldsymbol{v}_{E} + \hat{\boldsymbol v}_{m} $ 
and 
\begin{align}\label{eq:dot_v_para-zow-1}
  \dot{v}_{\parallel} 
  &= -\frac{ 1 }{ m }  { \boldsymbol b } \cdot \left( \mu  \nabla B \right)  
  + {v}_{\parallel} \boldsymbol v_{E} \cdot \frac{ \nabla_{\perp} B }{ B }    
  \nonumber \\ 
  &= -\frac{ 1 }{ m }  { \boldsymbol b } \cdot \left( \mu  \nabla B \right)  
  + {v}_{\parallel} {\dot { \boldsymbol X }}_{\text ZOW} \cdot \boldsymbol \kappa   
  \nonumber\\ 
  & - {v}_{\parallel} \left( 
    {\dot { \boldsymbol X }}_{\text ZOW} \cdot \boldsymbol \kappa  
  -\boldsymbol v_{E} \cdot \frac{ \nabla_{\perp} B }{ B }
  \right).
\end{align}
Then, the last term in Eq.\eqref{eq:dot_v_para-zow-1} is rewritten as
\begin{align}\label{eq:last_zow}
   & {\dot { \boldsymbol X }}_{\text ZOW} \cdot \boldsymbol \kappa   -  \boldsymbol v_{E} \cdot \frac{ \nabla_{\perp} B }{ B }
   \nonumber \\
   & = \left( \hat{\boldsymbol v}_{m} + \boldsymbol{v}_{E} \right) \cdot \boldsymbol \kappa  -  \boldsymbol v_{E} \cdot \frac{ \nabla_{\perp} B }{ B }
   \nonumber \\
   & = \left( \hat{\boldsymbol v}_{m} + \boldsymbol{v}_{E} \right) \cdot \left( \frac{ {\nabla}_{\perp} B }{ B } 
   +  \frac{ \mu_{0} \boldsymbol J \times \boldsymbol B }{ B^2 } \right) 
   - \boldsymbol v_{E} \cdot \frac{ \nabla_{\perp} B }{ B }
   \nonumber \\
   & = \left[ \boldsymbol {v}_{m} \cdot \left( \boldsymbol I - \nabla \psi \boldsymbol{e}_{\psi} \right) \right] \cdot \left( \frac{ {\nabla}_{\perp} B }{ B } 
   + \frac{ \mu_{0} \nabla p  }{ B^2 } \right)
   +  \boldsymbol v_{E} \cdot \frac{ \mu_{0} \nabla p }{ B^2 }
   %\nonumber \\
   %& = -\frac{ {\nabla}_{\perp} B }{ B } \cdot \left( \boldsymbol {v}_{m} \cdot \nabla \psi \boldsymbol{e}_{\psi} \right) 
   \nonumber \\
   & = -\frac{ 1 }{ B } \frac{ \partial B }{ \partial \psi }  \dot{\psi} .
\end{align}
Therefore, Eq.\eqref{eq:dot_v_para-zow} is obtained.
Using this $\dot{v}_{\parallel}$  for the ZOW model, Eq.\eqref{eq:d-k-A-A2} is rewritten as 
\begin{align}
 \nonumber
 &\frac{\partial}{\partial t} \left( \int d v^3 f m v_{\parallel}  \right)
 + \boldsymbol b \cdot \left\lbrace \nabla \cdot \left( \int d v^3 f m v_{\parallel}^2 \boldsymbol b \boldsymbol b \right) 
 \right\rbrace
 %\nonumber \\
 %& + \boldsymbol b \cdot \left\lbrace \nabla \cdot \left[ \int d v^3 f \mu B \left( \boldsymbol I - \boldsymbol b \boldsymbol b \right)  + \right] 
 %\right\rbrace
 \nonumber \\
 %& + \boldsymbol b \cdot \left\lbrace \nabla \cdot \left[ \int d v^3 f m v_{\parallel} \left( \boldsymbol b \dot{ \boldsymbol X }_{\perp, {\text ZOW}} + \dot{ \boldsymbol X }_{\perp,{\text ZOW}} \boldsymbol b \right) \right]
 %\right\rbrace
 & + \boldsymbol b \cdot \nabla \cdot \left[ \boldsymbol P_{CGL} + \boldsymbol \Pi_{2.\text{ZOW}} \right] 
 \nonumber \\
 & = \left( \int d v^3 f m v_{\parallel} \mathcal{S} \right) + \left( \int d v^3 f m v_{\parallel} \mathcal{G} \right)
 ,
\end{align}
where 
\begin{align} \label{eq:B_pi_zow}
 &\boldsymbol b \cdot \nabla \cdot \boldsymbol \Pi_{2, \text{ZOW}}
 \nonumber \\
 & = \boldsymbol b \cdot \left\lbrace \nabla \cdot \left[ \int d v^3 f m v_{\parallel} \left( \boldsymbol b \dot{ \boldsymbol X }_{\perp, {\text ZOW}} + \dot{ \boldsymbol X }_{\perp,{\text ZOW}} \boldsymbol b \right) \right] \right\rbrace
 \nonumber \\
 &- \left( \int d v^3 {v}_{\parallel} \frac{ 1 }{ B } \frac{ \partial B }{ \partial \psi }  \dot{\psi} \right).
\end{align}
The second term in Eq.\eqref{eq:B_pi_zow} breaks the symmetry of the $\Pi_{2}$ tensor.

For the ZMD model, the parallel momentum balance equation is calculated with 
${\dot { \boldsymbol X }}_{\text ZMD} = v_{\parallel} \boldsymbol b + \boldsymbol{v}_{E}$ 
and 
\begin{align}\label{eq:dot_v_para-zmd-1}
  \dot{v}_{\parallel} 
  = -\frac{ 1 }{ m }  { \boldsymbol b } \cdot \left( \mu  \nabla B \right)  
  + {v}_{\parallel} \boldsymbol v_{E} \cdot \frac{ \nabla_{\perp} B }{ B }.
\end{align}
%\begin{align}\label{eq:dot_v_para-zmd-1}
%  \dot{v}_{\parallel} 
%  &= -\frac{ 1 }{ m }  { \boldsymbol b } \cdot \left( \mu  \nabla B \right)  
%  + {v}_{\parallel} {\dot { \boldsymbol X }}_{\text ZMD} \cdot \boldsymbol \kappa    
%  \\ \nonumber
%  & - {v}_{\parallel} \left( {\dot { \boldsymbol X }}_{\text ZMD} \cdot \boldsymbol \kappa    + \boldsymbol v_{E} \cdot \frac{ \nabla_{\perp} B }{ B } \right)
%  .
%\end{align}
Because of the difference of $\dot{\boldsymbol X}_{\perp}$ between ZOW and ZMD, one finds that 
\begin{align}
   & \boldsymbol v_{E} \cdot \frac{ \nabla_{\perp} B }{ B } - {\dot { \boldsymbol X }}_{\text ZMD} \cdot \boldsymbol \kappa 
   \nonumber \\  
   & =  \boldsymbol v_{E} \cdot \frac{ \nabla_{\perp} B }{ B } - \left( \frac{ {\nabla}_{\perp} B }{ B } 
  + \boldsymbol v_{E}  \cdot  \frac{ \mu_{0} \boldsymbol J \times \boldsymbol B }{ B^2 } \right)
   \nonumber \\ 
   & = 0.
\end{align}
%\begin{align}
%   & {\dot { \boldsymbol X }}_{\text ZMD} \cdot \boldsymbol \kappa    -  \boldsymbol v_{E} \cdot \frac{ \nabla_{\perp} B }{ B }
%   \nonumber \\ 
%   & = \boldsymbol \kappa \cdot \boldsymbol{v}_{E} -  \boldsymbol v_{E} \cdot \frac{ \nabla_{\perp} B }{ B }
%   \nonumber \\ 
%   & = \left( \frac{ {\nabla}_{\perp} B }{ B } 
%   + \boldsymbol v_{E}  \cdot  \frac{ \mu_{0} \boldsymbol J \times \boldsymbol B }{ B^2 } \right)
%   - \boldsymbol v_{E} \cdot \frac{ \nabla_{\perp} B }{ B }
%   \nonumber \\ 
%   & = 0.
%\end{align}
%Therefore, Eq.\eqref{eq:dot_v_para-zmd} is obtained.
%For the ZMD model, Eq.\eqref{eq:d-k-A-A2} could be rewritten as 
%\begin{align}\label{eq:d-k-A-A1-zmd}
% \nonumber
% &\frac{\partial}{\partial t} \left( \int d v^3 f m v_{\parallel}  \right)
% + \boldsymbol b \cdot \left\lbrace \nabla \cdot \left( \int d v^3 f m v_{\parallel}^2 \boldsymbol b \boldsymbol b \right) 
% \right\rbrace
% \\ \nonumber
% & + \boldsymbol b \cdot \left\lbrace \nabla \cdot \left[ \int d v^3 f \mu B \left( \boldsymbol I - \boldsymbol b \boldsymbol b \right) \right] 
% \right\rbrace
% \\ \nonumber
% & + \boldsymbol b \cdot \left\lbrace \nabla \cdot \left[ \int d v^3 f m v_{\parallel} \left( \boldsymbol b\boldsymbol{v}_{E} + \boldsymbol{v}_{E} \boldsymbol b \right) \right]
% \right\rbrace
% \\
% &= \left( \int d v^3 f m v_{\parallel} \mathcal{S}  \right).
%\end{align}
%Furthemore, $\boldsymbol \Pi_{2, \text{ZMD}}$ becomes
Therefore, $\boldsymbol \Pi_{2, \text{ZMD}}$ becomes
\begin{align}\label{eq:B_pi_zmd}
 &\boldsymbol b \cdot \nabla \cdot  \boldsymbol \Pi_{2, \text{ZMD}}
 \nonumber \\
 & = \boldsymbol b \cdot \left\lbrace \nabla \cdot \left[ \int d v^3 f m v_{\parallel} \left( \boldsymbol b \dot{ \boldsymbol X }_{\perp, {\text ZMD}} + \dot{ \boldsymbol X }_{\perp,{\text ZMD}} \boldsymbol b \right) \right] \right\rbrace
  \nonumber \\
 &= \boldsymbol b \cdot \left\lbrace \nabla \cdot \left[ n m V_{\parallel} \left( \boldsymbol b { \boldsymbol v }_{E} + {\boldsymbol v}_{E} \boldsymbol b \right) \right] \right\rbrace.
\end{align}
Note that Eq.\eqref{eq:B_pi_zmd} is equivalent to Eq.(33) in Ref\cite{Landreman_2014}.

For the DKES model, the parallel momentum balance equation is calculated with 
${\dot { \boldsymbol X }}_{\text DKES} = v_{\parallel} \boldsymbol b + \hat{\boldsymbol{v}}_{E}$ 
and  
\begin{align}\label{eq:dot_v_para-dkes-1}
  \dot{v}_{\parallel} 
  = -\frac{ 1 }{ m }  { \boldsymbol b } \cdot \left( \mu  \nabla B \right),
\end{align}
%\begin{align}\label{eq:dot_v_para-dkes-1}
%  \dot{v}_{\parallel} 
%  &= -\frac{ 1 }{ m }  { \boldsymbol b } \cdot \left( \mu  \nabla B \right)  
%  + {v}_{\parallel} {\dot { \boldsymbol X }}_{\text DKES} \cdot \boldsymbol \kappa    
%  \\ \nonumber
%  & - {v}_{\parallel} {\dot { \boldsymbol X }}_{\text DKES} \cdot \boldsymbol \kappa   .
%\end{align}
which lacks in the $ {\dot { \boldsymbol X }}\cdot \boldsymbol \kappa $ term.
%\begin{align}
%    {\dot { \boldsymbol X }}_{\text DKES} \cdot \boldsymbol \kappa   
%   = \hat{\boldsymbol{v}}_{E} \cdot \left( \frac{ {\nabla}_{\perp} B }{ B } 
%   + \mu_{0} \frac{ \boldsymbol J \times \boldsymbol B }{ B^2 } \right) 
%    \neq 0.
%\end{align}
%Therefore, Eq.\eqref{eq:dot_v_para_dkes} is obtain.
%For the DKES model, Eq.\eqref{eq:d-k-A-A2} could be rewritten as 
%\begin{align}\label{eq:d-k-A-A1-dkes}
% \nonumber
% &\frac{\partial}{\partial t} \left( \int d v^3 f m v_{\parallel}  \right)
% + \boldsymbol b \cdot \left\lbrace \nabla \cdot \left( \int d v^3 f m v_{\parallel}^2 \boldsymbol b \boldsymbol b \right) 
% \right\rbrace
% \\ \nonumber
% & + \boldsymbol b \cdot \left\lbrace \nabla \cdot \left[ \int d v^3 f \mu B \left( \boldsymbol I - \boldsymbol b \boldsymbol b \right) \right] 
% \right\rbrace
% \\ \nonumber
% & + \boldsymbol b \cdot \left\lbrace \nabla \cdot \left[ \int d v^3 f m v_{\parallel} \left( \boldsymbol b \hat{ \boldsymbol v }_{E}  + \hat{ \boldsymbol v }_{E} \boldsymbol b \right) \right]
% \right\rbrace
% \\
% &=- n m V_{\parallel}  \hat{ \boldsymbol v }_{E} \cdot \boldsymbol \kappa  
%  + \left( \int d v^3 f m v_{\parallel} \mathcal{S} \right).
%\end{align}
Therefore, $\boldsymbol \Pi_{2, \text{DKES}}$ becomes
\begin{align}\label{eq:B_pi_dkes}
 \boldsymbol b \cdot \nabla \cdot  \boldsymbol \Pi_{2, \text{DKES}}
 & = \boldsymbol b \cdot \left\lbrace \nabla \cdot \left[ \int d v^3 f m v_{\parallel} \left( \boldsymbol b \hat{ \boldsymbol v }_{E} + \hat{ \boldsymbol v }_{E} \boldsymbol b \right) \right] \right\rbrace 
 \nonumber \\
 &+ n m V_{\parallel}  \hat{ \boldsymbol v }_{E} \cdot \boldsymbol \kappa  
\end{align}
which is equivalent to Eq.(34) in Ref\cite{Landreman_2014}.
The symmetry of Eq.\eqref{eq:B_pi_dkes} is broken.
Note that in the derivations shown in Appendix \ref{AppendixB}, we use assumptions $p = p(\psi)$, $\boldsymbol J \times \boldsymbol B  = \nabla p $, and $\boldsymbol E = - \nabla \Phi (\psi)$.

In conclusion, the symmetry of viscosity tensor $\boldsymbol \Pi_{2}$ depends on the form of $\dot{ \boldsymbol X } \cdot \boldsymbol \kappa$ term in $\dot{v}_{\parallel}$ in each local model.
%\clearpage
%\bibliography{./thesis.bib}
%\bibliography{/home/data/working/jabref/all.bib}
%\bibliography{./all.bib}
%

\end{document}